\documentclass[twocolumn,aps,floatfix,superscriptaddress,prb]{revtex4-2}
\usepackage[utf8]{inputenc}
\usepackage[T1]{fontenc}
\usepackage{newtxtext}
\usepackage[smallerops]{newtxmath}
\usepackage{amsmath,bm}
\usepackage{graphicx}
\usepackage[version=4]{mhchem}
\usepackage[colorlinks,bookmarks=true,citecolor=blue,linkcolor=blue,urlcolor=blue,breaklinks=true]{hyperref}
\usepackage{xcolor}
\usepackage{comment}

\begin{document}

%%%%%%%%%%%%%%%%%%%%%%%%%%%%%%%%%%%%
\title{From dimensional reduction to tetramerization in mixed ferro-antiferro breathing pyrochlores}

\author{Sourin Chatterjee}
\affiliation{Department of Physics, Indian Institute of Technology Madras, Chennai 600036, India}

\author{Kelvin Salou-Smith}
\affiliation{CNRS, University of Bordeaux, LOMA, UMR 5798, F-33400 Talence, France}

\author{Benedikt Schneider}
\affiliation{Arnold Sommerfeld Center for Theoretical Physics, Center for NanoScience, and Munich Center for Quantum Science and Technology, Ludwig-Maximilians-Universität München, 80333 Munich, Germany}
\affiliation{Department of Physics, Indian Institute of Technology Madras, Chennai 600036, India}

\author{Imre Hagymási}
\affiliation{Institute for Solid State Physics and Optics, HUN-REN Wigner
Research Centre for Physics, P.O. Box 49, H-1525 Budapest, Hungary}

\author{Arnaud Ralko }
\affiliation{Institut N\'eel, UPR2940, Universit\'e Grenoble Alpes et CNRS, Grenoble 38042, France}
\affiliation{Department of Physics, Indian Institute of Technology Madras, Chennai 600036, India}

\author{Johannes Reuther}
\affiliation{Helmholtz-Zentrum Berlin für Materialien und Energie, Hahn-Meitner-Platz 1, 14109 Berlin, Germany}
\affiliation{Dahlem Center for Complex Quantum Systems and Fachbereich Physik,
Freie Universität Berlin, Arnimallee 14, 14195 Berlin, Germany}
\affiliation{Department of Physics, Indian Institute of Technology Madras, Chennai 600036, India}

\author{Jeffrey G. Rau}
\affiliation{Department of Physics, University of Windsor, 401 Sunset Avenue, Windsor, Ontario, N9B 3P4, Canada}
\affiliation{Department of Physics, Indian Institute of Technology Madras, Chennai 600036, India}

\author{Karlo Penc}
\affiliation{Institute for Solid State Physics and Optics, HUN-REN Wigner
Research Centre for Physics, P.O. Box 49, H-1525 Budapest, Hungary}
\affiliation{Department of Physics, Indian Institute of Technology Madras, Chennai 600036, India}

\author{Harald O. Jeschke}
\affiliation{Research Institute for Interdisciplinary Science, Okayama University, Okayama 700-8530, Japan}
\affiliation{Department of Physics, Indian Institute of Technology Madras, Chennai 600036, India}

\author{Ludovic D. C. Jaubert}
\affiliation{CNRS, University of Bordeaux, LOMA, UMR 5798, F-33400 Talence, France}
\affiliation{Department of Physics, Indian Institute of Technology Madras, Chennai 600036, India}

\author{Yasir Iqbal}
\affiliation{Department of Physics, Indian Institute of Technology Madras, Chennai 600036, India}

\begin{abstract}
We study the spin-\(1/2\) nearest-neighbor Heisenberg model on the breathing
pyrochlore lattice in the mixed ferro-antiferromagnetic regime, where one
tetrahedral sublattice is antiferromagnetic and the other ferromagnetic.  In this
regime the classical ground-state manifold is not the pyrochlore Coulomb phase,
but that of the nearest-neighbor face-centered-cubic (fcc) antiferromagnet built
from composite tetrahedral moments.  Using classical Monte Carlo simulations and
the self-consistent Gaussian approximation, we show that thermal fluctuations lift
the subextensive degeneracy of this manifold by order-by-disorder and select the
collinear Type-I state with ordering wave vector \(X=(1,0,0)\) through a strongly
first-order transition.  Quantum fluctuations modify this outcome:
pseudofermion functional renormalization group calculations for \(S=1/2\) and
\(S=1\) reveal a finite nonmagnetic window adjacent to the decoupled
antiferromagnetic-tetrahedron limit, beyond which the same \(X=(1,0,0)\) order
reappears.  Density matrix renormalization group calculations show that this window
is not featureless: the antiferromagnetic tetrahedra develop nearly ideal tetramer
correlations, with four bonds carrying
\(\langle\mathbf{S}_i\cdot\mathbf{S}_j\rangle\simeq-1/2\) and the two remaining
opposite bonds \(\simeq+1/4\), while the spin structure factor retains broad maxima
at the \(X\) points.  We account for this by deriving the third-order effective
Hamiltonian in the manifold of tetrahedral singlets and showing that reversing the
sign of the inter-tetrahedron coupling converts Tsunetsugu's dimer selection into a
uniform tetramer selection; exact diagonalization of the resulting fcc pseudospin
model confirms this.  A dynamic high-temperature expansion then traces how the local
tetrahedral magnetic excitations give way, with increasing mixed-sign coupling, to
low-energy \(X\)-centered spectral weight.  Finally, we place density-functional
parameter sets for eight structures of spin-\(3/2\) breathing chromium thiospinels in the
classical effective-fcc phase diagram and compare them with the experimental
literature.  The mapping reproduces the observed \(\mathbf{k}=(1,0,0)\) order of
\ce{CuInCr4S8}, the only member of the family whose magnetic structure is resolved
and whose lattice remains cubic.  It also accounts for the absence of long-range
order in \ce{LiGaCr4S8}, whose room-temperature and low-temperature structures
fall on opposite sides of a phase boundary, and it predicts Type-I order for
\ce{LiInCr4S8}, whose \(24\)~K transition has not been characterized
microscopically.  Mixed-sign breathing pyrochlores thus bring effective-fcc
frustration, thermal order-by-disorder, quantum suppression of dipolar order and
tetrahedral singlet formation together in a single model.
\end{abstract}

\date{\today}
\maketitle

\tableofcontents
%%%%%%%%%%%%%%%%%%%%%%%%%%%%%%%%%%%%

\section{Introduction}

Frustrated magnets on the pyrochlore lattice have long served as paradigmatic examples of how local constraints can suppress conventional magnetic order and generate collective low-energy degrees of freedom.  In the nearest-neighbor Heisenberg antiferromagnet, the condition of vanishing total spin on every tetrahedron produces an extensively degenerate classical manifold and, at low temperature, a Coulomb phase with algebraic correlations and pinch-point singularities in the spin structure factor~\cite{Moessner-1998a,Moessner-1998b}.  Closely related constraint physics underlies the phenomenology of spin-ice materials, chromium spinels, and several other three-dimensional frustrated magnets~\cite{Gardner-2010,Savary-2017,Henley-2010}.

Breathing pyrochlores enrich this setting by breaking the equivalence between the two tetrahedral sublattices.  The magnetic ions still form a network of corner-sharing tetrahedra, but the ``up'' and ``down'' tetrahedra have different bond lengths and hence different exchange couplings.  This structure is realized in the chromium spinel oxides \ce{LiGaCr4O8} and \ce{LiInCr4O8}~\cite{Okamoto13a,tanaka14-PRL113,Okamoto-2015,nilsen15-PRB91,Lee-2016,Saha-2016,wawrzynczak17,Okamoto-2017,okamoto18-JPSP87,He-2021}, in the chromium sulfides \ce{LiInCr4S8}~\cite{okamoto18-JPSP87}, \ce{LiGaCr4S8}~\cite{Pokharel-2018,Pokharel-2020}, \ce{CuInCr4S8}~\cite{Plumier-1971,Plumier-1977,Gao-2021}, \ce{CuAlCr4S8}~\cite{Sharma-2022,Gen-2024} and \ce{CuGaCr4S8}~\cite{Gen-2023,Gen-2024}, and in the selenide \ce{CuInCr4Se8}~\cite{Duda-2008,Gao-2022}.  The same structural motif also appears in effective spin-$1/2$ rare-earth and cluster magnets, including \ce{Ba3Yb2Zn5O11}~\cite{KimuraPRB14,HakuPRB16,Dissanayake-2022}, where strong breathing anisotropy has motivated proposals involving unusual multipolar or higher-rank gauge structures~\cite{Yan-2020,Gresista_PRL}.

The simplest theoretical description is the nearest-neighbor breathing-pyrochlore Heisenberg model,
\begin{equation}
    H =
    J_A \sum_{\langle ij\rangle_A} \mathbf{S}_i\cdot\mathbf{S}_j
    +
    J_B \sum_{\langle ij\rangle_B} \mathbf{S}_i\cdot\mathbf{S}_j ,
    \label{eq:breathing_hamiltonian}
\end{equation}
where $\langle ij\rangle_A$ and $\langle ij\rangle_B$ denote bonds belonging to the two inequivalent tetrahedral sublattices.  When both couplings are antiferromagnetic, the classical ground-state condition remains the vanishing of the total spin on every tetrahedron, and the Coulomb phase survives the breathing deformation~\cite{Benton15c}.  Quantum versions of the antiferromagnetic breathing model have also been studied extensively, especially in the strong-breathing limit where the physics can be understood in terms of weakly coupled tetrahedral singlets~\cite{Harris-1991,Isoda-1998,Koga-2001,Tsunetsugu-2001a,Tsunetsugu-2001b,Canals-1998,Canals-2000,Tsunetsugu-2002}. Pseudofermion functional renormalization group (pf-FRG) studies further indicate that, for purely antiferromagnetic couplings, the nonmagnetic regime extends over the full range of breathing anisotropy for both $S=1/2$ and $S=1$~\cite{Yasir2019}.

The situation is qualitatively different when the two exchanges have opposite signs.  This mixed ferro-antiferro regime is natural in chromium sulfides and selenides, where the balance between direct antiferromagnetic exchange and indirect ferromagnetic exchange can be altered by bond length, bond angle, and chemical substitution.  Benton and Shannon showed that, in this regime, the classical ground-state manifold is no longer the pyrochlore Coulomb manifold.  Instead, the ferromagnetic tetrahedra behave as composite spins living on an fcc lattice, while the antiferromagnetic tetrahedra impose the nearest-neighbor fcc antiferromagnetic constraint~\cite{Benton15c}.  The resulting manifold has an $O(L)$ degeneracy associated with lines of soft modes and an effective decoupling of antiferromagnetic planes.  By analogy with the nearest-neighbor fcc antiferromagnet, thermal fluctuations are expected to select the Type-I antiferromagnet with ordering wave vector
\begin{equation}
    \mathbf{q}_{\rm ord}=X=(1,0,0)
\end{equation}
and symmetry-related wave vectors~\cite{Gvozdikova05a,Benton15c,Schick20a}.

\begin{figure*}[!t]
\includegraphics[width=\linewidth]{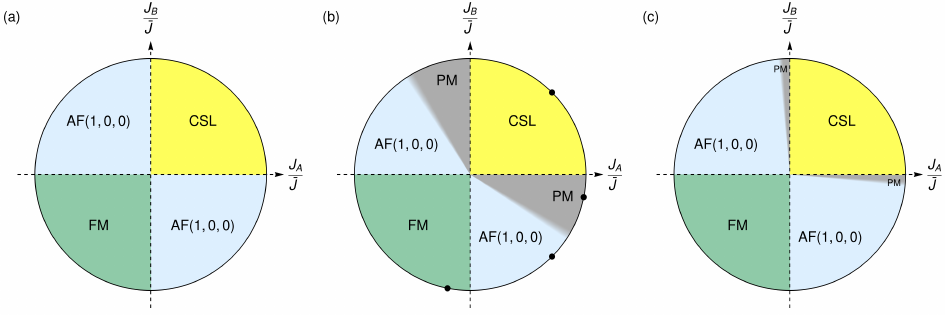}
\caption{
\textbf{Phase diagrams of the nearest-neighbor breathing-pyrochlore Heisenberg
model}, parametrized by
\(J_A=\bar{J}\cos\theta\) and \(J_B=\bar{J}\sin\theta\) with \(\bar{J}>0\)
[Eq.~\eqref{eq:polar_parametrization}].
(a) Classical phase diagram of the low-temperature phases from Monte Carlo
simulations, which agrees with Ref.~\cite{Benton15c} at \(T\to0^{+}\).  The mixed-sign quadrants are selected by
thermal order-by-disorder into the Type-I \(X=(1,0,0)\) antiferromagnet.
(b),(c) pf-FRG phase diagrams for \(S=1/2\) and \(S=1\), respectively.
Quantum fluctuations open a nonmagnetic regime near the decoupled
antiferromagnetic-tetrahedron limits, while the \(X=(1,0,0)\) ordered
phase survives at stronger mixed-sign coupling.  The black dots mark the
representative points whose susceptibility profiles are shown in
Fig.~\ref{fig:pffrg_suscep}.
}
\label{fig:PD}
\end{figure*}

This classical result is the starting point of the present work.  We ask what becomes of the dimensionally reduced effective-fcc manifold once thermal and quantum fluctuations are treated explicitly.  First, using classical Monte Carlo simulations, we show that the mixed-sign breathing pyrochlore indeed undergoes a finite-temperature order-by-disorder transition into the $X=(1,0,0)$ state.  This establishes the classical reference point and confirms that the line degeneracy of the effective-fcc manifold is thermally resolved in favor of Type-I antiferromagnetism.  We then introduce quantum fluctuations through pf-FRG.  In contrast to the classical limit, where any infinitesimal mixed-sign coupling selects an ordered ground state, pf-FRG finds a finite nonmagnetic region for both $S=1/2$ and $S=1$ when ferromagnetic coupling is added to decoupled antiferromagnetic tetrahedra.  The ordered phase reached beyond this nonmagnetic window has the same $X=(1,0,0)$ wave vector selected by the classical order-by-disorder mechanism.  The resulting phase diagrams are collected in Fig.~\ref{fig:PD}.

The absence of a pf-FRG flow breakdown in the nonmagnetic regime establishes the lack of conventional dipolar order, but it does not by itself determine what kind of paramagnet is realized.  To address this question, we use the density matrix renormalization group (DMRG) to probe real-space correlations and finite-cluster structure factors.  The DMRG results reveal enhanced tetramer correlations on the antiferromagnetic tetrahedra, while the spin structure factor retains soft maxima at the same $X$ points that control the proximate ordered phase.  We then use degenerate perturbation theory in the strong-breathing limit, following Tsunetsugu's construction of the tetrahedral singlet pseudospin model~\cite{Tsunetsugu-2001a,Tsunetsugu-2001b,Tsunetsugu-2002}, to derive an effective fcc pseudospin Hamiltonian.  Exact diagonalization of this effective model explains why the mixed-sign perturbation favors the tetramer correlations seen in DMRG.

A separate question is whether the effective-fcc reorganization is a peculiarity of the idealized nearest-neighbor model or a feature of real materials.  We address it by combining density-functional calculations of the microscopic Cr--Cr exchange network of the breathing chromium thiospinels with an explicit coarse graining over Cr\(_4\) tetrahedra, and by comparing the resulting effective-fcc parameters with what is known experimentally about the magnetic ground states of these compounds.  We find that the degeneracy of the fcc soft-mode manifold is lifted by a single combination of further-neighbor effective couplings, that this combination is small in every compound considered, and that the resulting near degeneracy accounts for the range of behavior observed across the family: commensurate Type-I order where the lattice remains cubic, incommensurate helimagnetism where a magnetostructural transition intervenes, and no long-range order where the room-temperature and low-temperature structures sit on opposite sides of a phase boundary.

The resulting picture is therefore not a simple replacement of classical order by a featureless quantum paramagnet, but a sequence of reorganizations of the same effective-fcc geometry. Thermal fluctuations select the fcc Type-I antiferromagnet. Quantum fluctuations first melt this dipolar order, but the nonmagnetic phase retains the memory of the nearby fcc manifold through soft $X$-point correlations and reorganizes its local singlet degrees of freedom into tetramerized patterns. This provides a unified framework for understanding mixed ferro-antiferromagnetic breathing pyrochlores and for interpreting the magnetic ground states and finite-temperature scattering signatures of the chromium thiospinels.

\begin{figure}
    \centering
    \includegraphics[width=0.6\linewidth]{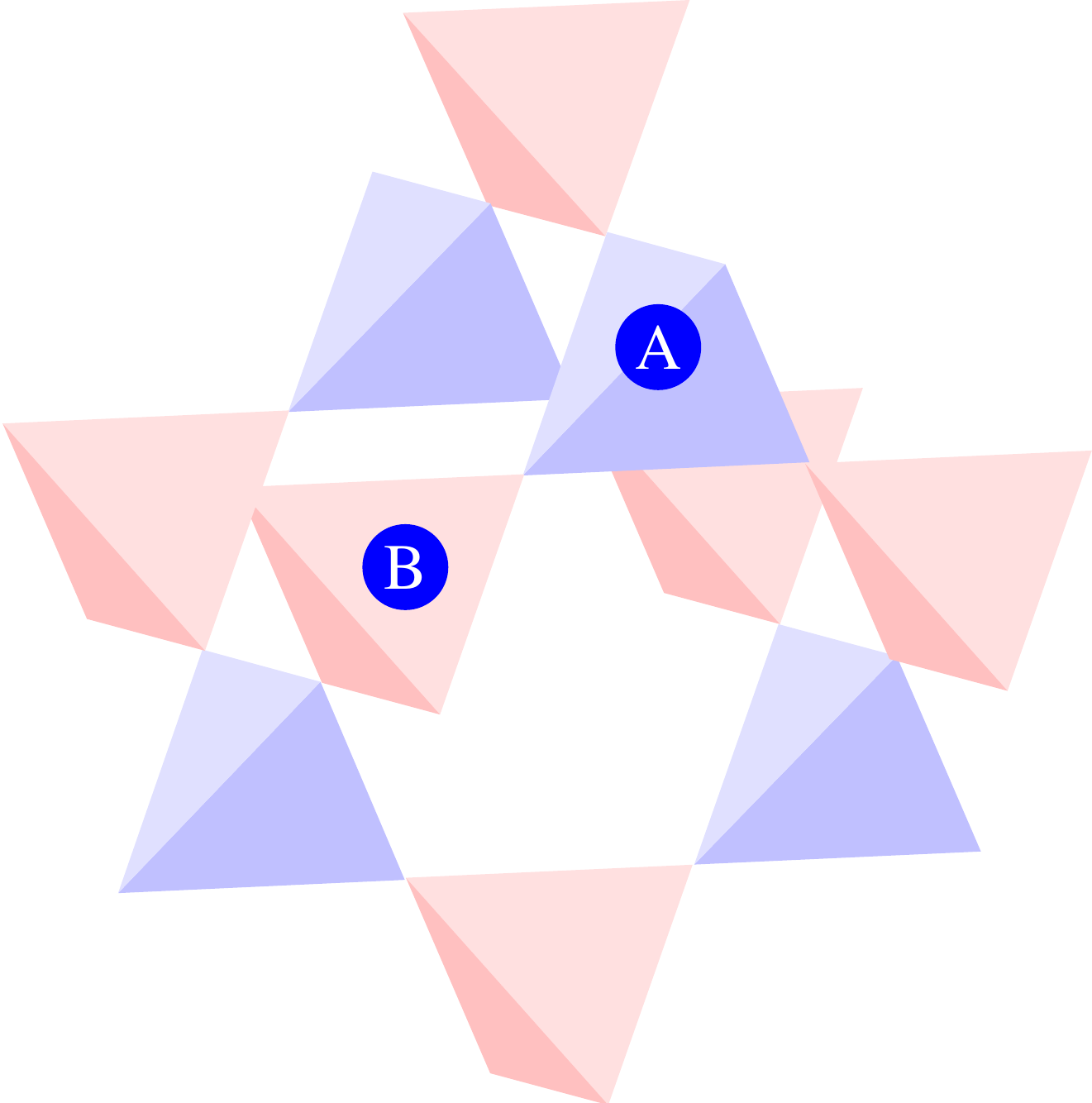}
    \caption{Breathing pyrochlore lattice, built from alternating ``up'' (\(A\)) and ``down'' (\(B\)) tetrahedra carrying the exchange constants \(J_A\) and \(J_B\) of Eq.~\eqref{eq:breathing_hamiltonian}.  The centers of the \(A\) tetrahedra, and likewise those of the \(B\) tetrahedra, form an fcc lattice.}
    \label{fig:bp_lattice}
\end{figure}
%=================================

%=================================

\section{Model and classical reference point}
\label{sec:model_classical}

We consider the nearest-neighbor Heisenberg model on the breathing pyrochlore lattice of Fig.~\ref{fig:bp_lattice}, Eq.~\eqref{eq:breathing_hamiltonian}.  The two exchange constants are parametrized as
\begin{equation}
    J_A=\bar{J}\cos\theta,\qquad
    J_B=\bar{J}\sin\theta ,
    \label{eq:polar_parametrization}
\end{equation}
with $\bar{J}>0$.  In this convention, $0<\theta<\pi/2$ corresponds to the antiferromagnetic breathing pyrochlore, $\pi<\theta<3\pi/2$ to the ferromagnet, and the two remaining quadrants to mixed ferro-antiferro coupling.  Unless otherwise stated, we focus on the quadrant
\begin{equation}
    J_A>0,\qquad J_B<0 ,
    \label{eq:mixed_sign_quadrant}
\end{equation}
where antiferromagnetic tetrahedra are coupled through ferromagnetic tetrahedra.  The opposite mixed-sign quadrant is related by interchanging the two tetrahedral sublattices.

The classical ground-state structure of Eq.~\eqref{eq:breathing_hamiltonian} can be understood by rewriting the Hamiltonian in terms of the total spin of each tetrahedron.  For a tetrahedron $t$, define
\begin{equation}
    \mathbf{M}_t=\sum_{i\in t}\mathbf{S}_i .
\end{equation}
Up to a constant, the energy is a sum of terms proportional to $\mathbf{M}_t^2$ on the two tetrahedral sublattices.  If $J_A,J_B>0$, the ground-state condition is $\mathbf{M}_t=0$ on every tetrahedron, giving the Coulomb phase.  If $J_A,J_B<0$, the energy is minimized by maximizing $|\mathbf{M}_t|$ on every tetrahedron, giving a ferromagnet.  If the two signs differ, however, the ferromagnetic tetrahedra are internally polarized, while the antiferromagnetic tetrahedra impose a zero-total-spin constraint on neighboring polarized tetrahedra.

This last case produces the effective-fcc manifold identified in Ref.~\cite{Benton15c}.  In the quadrant $J_A>0$, $J_B<0$, each ferromagnetic $B$ tetrahedron behaves, at the classical level, as a composite spin.  The centers of the $B$ tetrahedra form an fcc lattice.  The condition that each intervening $A$ tetrahedron have zero total spin becomes precisely the nearest-neighbor antiferromagnetic constraint on this fcc lattice.  Thus, the mixed-sign breathing pyrochlore and the nearest-neighbor fcc antiferromagnet share the same classical ground-state manifold.

Throughout this paper, wave vectors are quoted in reciprocal-lattice units of \(2\pi/a\), where \(a\) is the cubic lattice constant.

\begin{figure*}[!t]
    \centering
    \includegraphics[width=\textwidth]{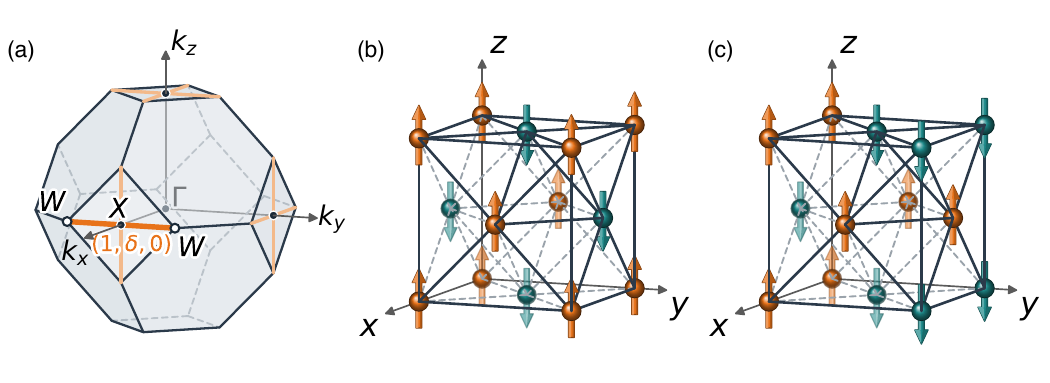}
    \caption{
    \textbf{Soft-mode line and collinear ordered states of the effective fcc
    antiferromagnet.}
    (a) First Brillouin zone of the fcc lattice, in units of \(2\pi/a\).  The
    \(X\) points (filled circles) sit at the centers of the square faces and the
    \(W\) points (open circles) at their corners.  The soft-mode line
    \(\mathbf{q}=(1,\delta,0)\) of Eq.~\eqref{eq:softline} is the diagonal of the
    \(k_x=1\) face running from \(W\) through \(X\) to \(W\) (thick orange);
    its symmetry-related partners (light orange) make every \(X\) point the
    crossing of two such lines.
    (b),(c) Type-I (AF1) and Type-III (AF3) structures on one conventional
    cubic cell of the fcc lattice formed by the centers of the ferromagnetic
    \(B\) tetrahedra, with ordering wave vectors \(X=(1,0,0)\) and
    \(W=(1,\tfrac12,0)\) of Eq.~\eqref{eq:XandW}, respectively.  Orange and
    teal arrows denote the two spin orientations; the four sites at the back
    of the cell and their bonds are shown faded and dashed.  In the Type-I
    state, ferromagnetic \((100)\) planes alternate in sign along \(x\); in the
    Type-III state, each \((100)\) plane is in addition modulated along \(y\)
    with period \(2a\), so that the \(x\)-stacking is combined with a doubling
    of the cell along \(y\).
    }
    \label{fig:af1_af3}
\end{figure*}

The fcc antiferromagnetic manifold is subextensively degenerate.  It contains lines of soft modes with wave vectors of the form
\begin{equation}
    \mathbf{q}=(1,\delta,0)
    \label{eq:softline}
\end{equation}
and symmetry-related directions.  The two commensurate wave vectors on this line,
\begin{equation}
    X=(1,0,0),
    \qquad
    W=\left(1,\tfrac{1}{2},0\right),
    \label{eq:XandW}
\end{equation}
support the collinear Type-I (AF1) and Type-III (AF3) antiferromagnets, respectively~\cite{Schick20a}; their location in the Brillouin zone and their spin configurations on the conventional fcc cell are shown in Fig.~\ref{fig:af1_af3}.

How these collinear states are built from a single wave vector deserves a comment, since a generic single-\(\mathbf{q}\) state on the soft line is a spiral rather than a collinear structure.  Writing
\begin{equation}
    \mathbf{S}_i
    =
    \mathbf{l}_1\cos(\mathbf{q}\cdot\mathbf{r}_i)
    +
    \mathbf{l}_2\sin(\mathbf{q}\cdot\mathbf{r}_i) ,
    \label{eq:single_q_parametrization}
\end{equation}
the familiar choice \(|\mathbf{l}_1|=|\mathbf{l}_2|\) with \(\mathbf{l}_1\perp\mathbf{l}_2\) gives a coplanar spiral whose moment length is uniform for any \(\mathbf{q}\).  The collinear states are obtained instead by taking \(\mathbf{l}_1\) and \(\mathbf{l}_2\) parallel and of equal length, which turns Eq.~\eqref{eq:single_q_parametrization} into a single cosine with a \(\pi/4\) phase shift.  This is the form in which the two states of Eq.~\eqref{eq:XandW} are quoted in Ref.~\cite{Gvozdikova05a},
\begin{equation}
    \mathbf{S}_i=\hat{\mathbf{e}}\cos(X\cdot\mathbf{r}_i),
    \qquad
    \mathbf{S}_i=\hat{\mathbf{e}}\cos\!\left(W\cdot\mathbf{r}_i+\tfrac{\pi}{4}\right) ,
    \label{eq:collinear_single_q}
\end{equation}
with \(\hat{\mathbf{e}}\) a common unit vector.  Such a state is collinear by construction, but its moment length is site dependent for generic \(\mathbf{q}\), so that equal moments on every site occur only at special commensurate wave vectors.  Both \(X\) and \(W\) are of this kind: on the fcc lattice \(\mathbf{q}\cdot\mathbf{r}_i\) is an integer multiple of \(\pi\) at \(X\) and of \(\pi/2\) at \(W\), so that the cosine takes the values \(\pm1\) and \(\pm1/\sqrt{2}\), respectively.  The \(\pi/4\) shift is what makes this work at \(W\): without it the cosine would vanish on half of the sites.  The spiral and the collinear state built on the same wave vector have the same classical energy, because the two parametrizations differ by a term that sums to zero on the lattice; the collinear member is selected by thermal fluctuations, through the effective biquadratic interaction generated by short-wavelength modes~\cite{Gvozdikova05a,Henley-1987}, as the quadrupolar order parameter of Sec.~\ref{sec:classical_cmc} confirms.

The \(X\) points lie at intersections of the soft-mode lines [Fig.~\ref{fig:af1_af3}(a)].  They therefore possess an enhanced fluctuation phase space and are selected by the thermal order-by-disorder mechanism in the nearest-neighbor fcc antiferromagnet~\cite{Henley-1987,Gvozdikova05a,Schick20a}.  The same reasoning implies that the mixed-sign breathing pyrochlore should order into the corresponding $X=(1,0,0)$ state in the classical finite-temperature problem~\cite{Benton15c}.

%------------------------------------------------------------------------
%------------------------------------------------------------------------

Figure~\ref{fig:PD} summarizes the global phase structure that will be developed in the following sections.  Panel~(a) shows the classical phase diagram, where the mixed-sign region is ordered by thermal fluctuations into the $X=(1,0,0)$ antiferromagnet.  Panels~(b) and~(c) show the quantum phase diagrams obtained from pf-FRG for $S=1/2$ and $S=1$.  The central result is that quantum fluctuations carve out a finite nonmagnetic regime near the decoupled antiferromagnetic-tetrahedron limit before the system enters the same $X=(1,0,0)$ ordered phase selected classically.

The rest of the paper follows the hierarchy suggested by this phase diagram.  We first show that classical thermal fluctuations indeed select the $X=(1,0,0)$ state in the mixed-sign regime.  We then demonstrate that quantum fluctuations partially melt this order into a nonmagnetic phase, determine the internal structure of that phase using DMRG, and finally explain its tetramer correlations using a strong-breathing effective pseudospin theory.

\section{Classical finite-temperature selection}
\label{sec:classical_cmc}

We begin with the finite-temperature behavior of the classical model in the mixed-sign regime.  As discussed in Sec.~\ref{sec:model_classical}, when one tetrahedral sublattice is antiferromagnetic and the other ferromagnetic, the classical ground-state manifold is equivalent to that of the nearest-neighbor antiferromagnet on the fcc lattice~\cite{Benton15c}.  This manifold contains lines of soft modes and is only subextensively degenerate.  The points \(X=(1,0,0)\), and those related to them by cubic symmetry, lie at intersections of these soft-mode lines.  They are therefore expected to be favored by thermal fluctuations, as in the Type-I ordered state of the nearest-neighbor fcc antiferromagnet~\cite{Henley-1987,Gvozdikova05a,Schick20a}.

We test this expectation using classical Monte Carlo simulations.  Throughout this section we focus on the quadrant \(J_A>0\), \(J_B<0\),
and set \(J_A=1\).  Thus \(J_A\) denotes the antiferromagnetic exchange on the \(A\)-tetrahedra, while \(J_B\) denotes the ferromagnetic exchange on the \(B\)-tetrahedra.  The simulations were carried out for systems of \(N=16L^3\) classical spins of length \(|\mathbf{S}|=1/2\), using heat-bath updates, over-relaxation sweeps, and parallel tempering.  We also compare with the self-consistent Gaussian approximation (SCGA), which describes the diffuse paramagnetic correlations but does not capture the singular thermodynamics of the ordering transition.

%------------------------------------------------------------------------
\begin{figure*}[!t]
\centering
\includegraphics[width=\textwidth]{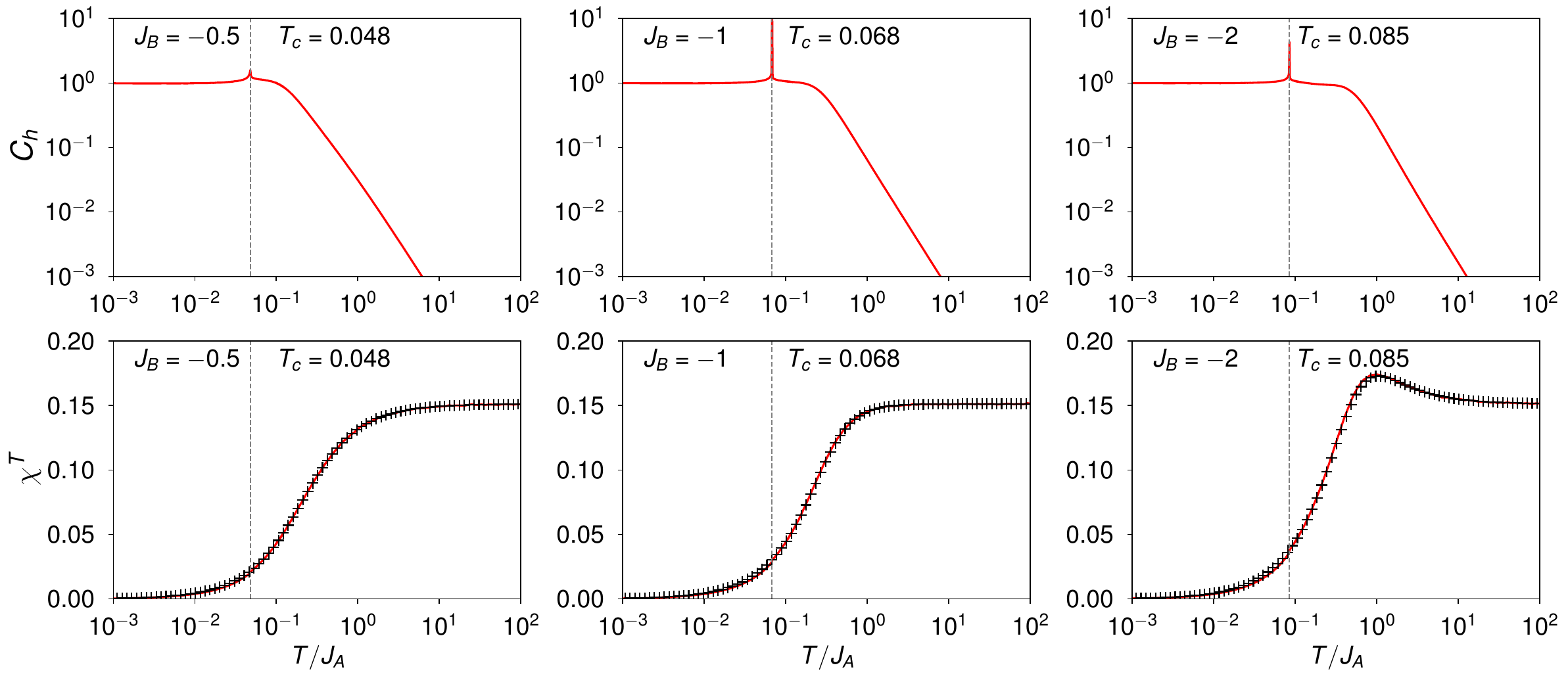}
\caption{
\textbf{Heat capacity $C_{h}$ and reduced susceptibility $\chi T$} for $J_A=1$ and $J_B=-0.5$ (left), $-1$ (middle) and $-2$ (right). We clearly see a phase transition in the heat capacity that is invisible in the magnetic susceptibility, indicating that the low-temperature order carries no magnetization. $\chi T$ displays a characteristic Curie-law crossover, with the onset of ferromagnetic fluctuations for $J_B=-2$ (see the bump). In the limit $J_B/J_A \rightarrow -\infty $, we have at intermediate temperatures, $J_A \ll T \ll |J_B|$, a paramagnet on the fcc lattice whose super-spins are the four collinear spins of each $B$ tetrahedron. The Curie constant is then 4 times bigger than for the paramagnet on the pyrochlore lattice, hence the bump. The red curves are Monte Carlo data, while the black crosses are analytical SCGA results. Monte Carlo simulations are performed with the heat-bath algorithm, over-relaxation, and parallel tempering, for classical spins of length \(|\mathbf{S}|=1/2\).
}
\label{fig:cmc_thermo}
\end{figure*}
%------------------------------------------------------------------------

\subsection{Thermodynamic signatures}

Figure~\ref{fig:cmc_thermo} shows the heat capacity \(C_h\) and the reduced uniform susceptibility \(\chi T\) for three representative values of the ferromagnetic coupling, \(J_B=-0.5,-1\), and \(-2\), with \(J_A=1\).  The heat capacity develops a sharp anomaly at a finite transition temperature,
\begin{equation}
    T_c/J_A \simeq 0.048,\quad 0.068,\quad 0.085 ,
\end{equation}
respectively.  The increase of \(T_c\) with \(|J_B|\) is consistent with the effective-fcc description.  As the ferromagnetic coupling grows, the spins within each \(B\)-tetrahedron become more rigidly aligned, and the composite tetrahedral moments order more readily through the antiferromagnetic constraints imposed by the neighboring \(A\)-tetrahedra.

The uniform susceptibility shows a much weaker signature of the transition.  This is expected, since the selected state is antiferromagnetic in the effective-fcc sense and has no net magnetization.  Away from the transition, the smooth part of \(\chi T\) is well reproduced by the SCGA.  For \(J_B=-2\), \(\chi T\) has a broad maximum at intermediate temperature.  This reflects the formation of nearly ferromagnetic \(B\)-tetrahedra before long-range antiferromagnetic correlations develop between them.  In this regime the system behaves approximately as weakly correlated composite moments on the fcc lattice.

%------------------------------------------------------------------------
\begin{figure*}[!t]
\centering
\includegraphics[width=\linewidth]{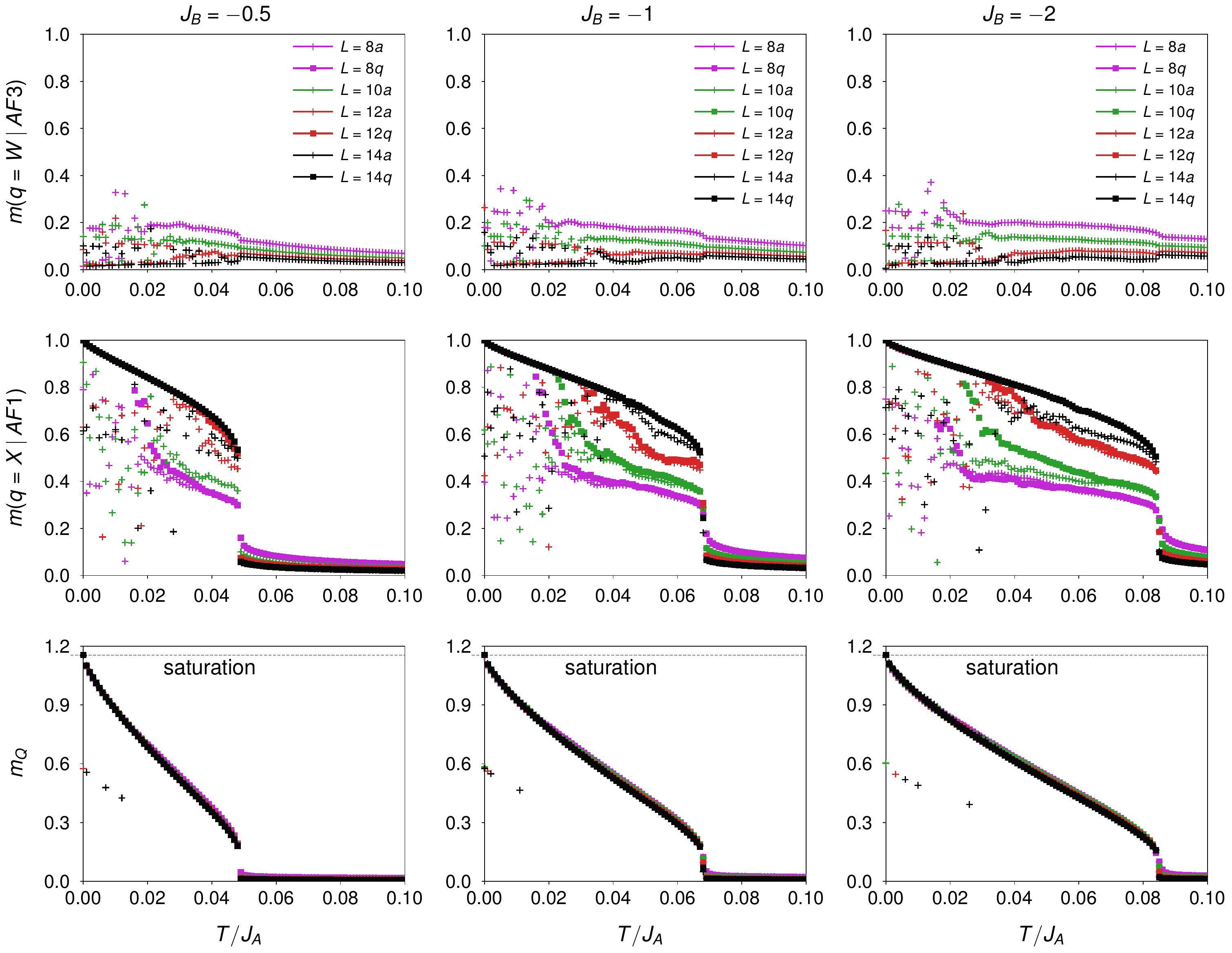}
\caption{
\textbf{Order parameters for the order-by-disorder phase transition in breathing pyrochlore} for $J_A=1$ and $J_B=-0.5$ (left), $-1$ (middle) and $-2$ (right). Systems are made of $N=16\,L^{3}$ classical spins of length $|\mathbf{S}|=1/2$ with $L\in\{8,10,12,14\}$. The quadrupolar order parameter (bottom panels) reaches saturation $m_Q\rightarrow \sqrt{4/3}$ as $T\rightarrow 0$, indicating that the spins become collinear, as expected. We then examine the dipolar order parameters at wave vectors $\mathbf{q}=X$ and $\mathbf{q}=W$, corresponding respectively to the Type-I (AF1) and Type-III (AF3) antiferromagnets on the fcc lattice~\cite{Gvozdikova05a,Schick20a}. The labels $a$ and $q$ in the legends refer to the thermalization protocol: $a$ for slow annealing from high temperature, and $q$ for a quench at temperature $T$ into the AF1 state. The fact that, for a given system size, the annealed and quenched simulations give roughly the same result (at least at intermediate temperatures) suggests that the simulations are reasonably thermalized. We notice strong finite-size effects, which are not uncommon in order-by-disorder phenomena, but the order parameter for AF3 diminishes with increasing system size, while the one for AF1 increases (for both annealed and quenched simulations). Monte Carlo simulations clearly show selection of AF1 at low temperatures, as predicted in Refs.~\cite{Benton15c,Gvozdikova05a}, through a strongly first-order transition. Since the transition is also first order on the fcc lattice, which would correspond to the limit $J_B \ll -J_A$, we can expect it to remain first order for all negative values of $J_B$ (and $J_A=+1$).
}
\label{fig:cmc_order}
\end{figure*}
%------------------------------------------------------------------------

\subsection{Order-by-disorder selection}

To identify the selected state, we monitor order parameters for two collinear fcc antiferromagnets.  The first is the Type-I state, denoted AF1, with ordering wave vector \(X=(1,0,0)\).  The second is the Type-III state, denoted AF3, associated with the \(W=(1,\tfrac{1}{2},0)\) point of Eq.~\eqref{eq:XandW}.  We also track a quadrupolar order parameter \(m_Q\), defined as the norm of the site-averaged rank-two tensor~\cite{Shannon-2010,Mizoguchi18a},
\begin{equation}
    m_Q
    =
    \Bigl[\sum_{\alpha}\bigl(Q^{\alpha}\bigr)^{2}\Bigr]^{1/2},
    \qquad
    Q^{\alpha}=\frac{1}{S^{2}N}\sum_{i}Q_i^{\alpha},
    \label{eq:mQ}
\end{equation}
where \(Q_i^{\alpha}\) denotes the five quadrupole components of spin \(i\),
\(Q_i^{3z^2-r^2}=\bigl[2(S_i^z)^2-(S_i^x)^2-(S_i^y)^2\bigr]/\sqrt{3}\),
\(Q_i^{x^2-y^2}=(S_i^x)^2-(S_i^y)^2\),
\(Q_i^{xy}=2S_i^xS_i^y\), \(Q_i^{yz}=2S_i^yS_i^z\), and \(Q_i^{zx}=2S_i^zS_i^x\), and \(S=|\mathbf{S}_i|\) is the classical spin length, so that \(m_Q\) is independent of the normalization of the spins.
It vanishes for isotropically distributed spins and saturates at \(\sqrt{4/3}\) for a perfectly collinear spin configuration.

The results are shown in Fig.~\ref{fig:cmc_order}.  The quadrupolar parameter grows rapidly below the transition and approaches saturation at low temperature, showing that the selected state is collinear. The dipolar order parameters then distinguish between the two candidate collinear states.  For all three couplings, the AF1 order parameter grows upon cooling and increases with system size.  By contrast, the AF3 order parameter remains small and is suppressed as the system size is increased. This finite-size trend identifies the selected state as the Type-I antiferromagnet at \(X=(1,0,0)\).

The transition is strongly first order, as in the nearest-neighbor fcc antiferromagnet~\cite{Diep-1989,Gvozdikova05a}.  This is visible in the sharp heat-capacity anomaly of Fig.~\ref{fig:cmc_thermo} and in the abrupt onset of AF1 order in Fig.~\ref{fig:cmc_order}. It partially explains the numerical noise, since parallel tempering cannot help thermalize below a strongly discontinuous transition. The other reason for the difficult thermalization comes from the intrinsic nature of the transition. An order-by-disorder mechanism cannot rely on an energetic selection, since the selection is entropic in origin, made across a degenerate ground-state manifold. Technically, the Metropolis argument in Monte Carlo cannot directly select the ground state, as in traditional phase transitions. It is the proper exploration of the statistical Gibbs ensemble, via thermal fluctuations, that selects AF1 over AF3. In the present model, this selection is made even more complex as it must include the collective fluctuations of 16 spins (the size of the magnetic unit cell for AF1). Confirming AF1 order in the fcc antiferromagnet had already been a tour-de-force for a four-site unit cell~\cite{Gvozdikova05a}.

In order to confirm our results, we have compared two thermalization procedures in simulations: (i) a slow annealing from high temperature down to the temperature of simulation $T$, followed by a thermalization at $T$ versus (ii) an immediate quench into the AF1 manifold followed by a thermalization at $T$. The former favors the disordered phase while the latter favors order. The two procedures find the same transition temperature, while the order parameter for AF1 overlaps well at intermediate temperatures below the transition. This overlap confirms that simulations are reasonably thermalized in this temperature regime. Interestingly, finite-size effects persist up to system sizes $N\sim 40\,000$, which indicates a fairly large correlation length. In summary, this combination of results confirms that thermal fluctuations select the AF1 state with ordering wave vector \(X\).

\subsection{Structure factor and dimensional reduction}

\begin{figure}[t]
\centering
\includegraphics[width=\columnwidth]{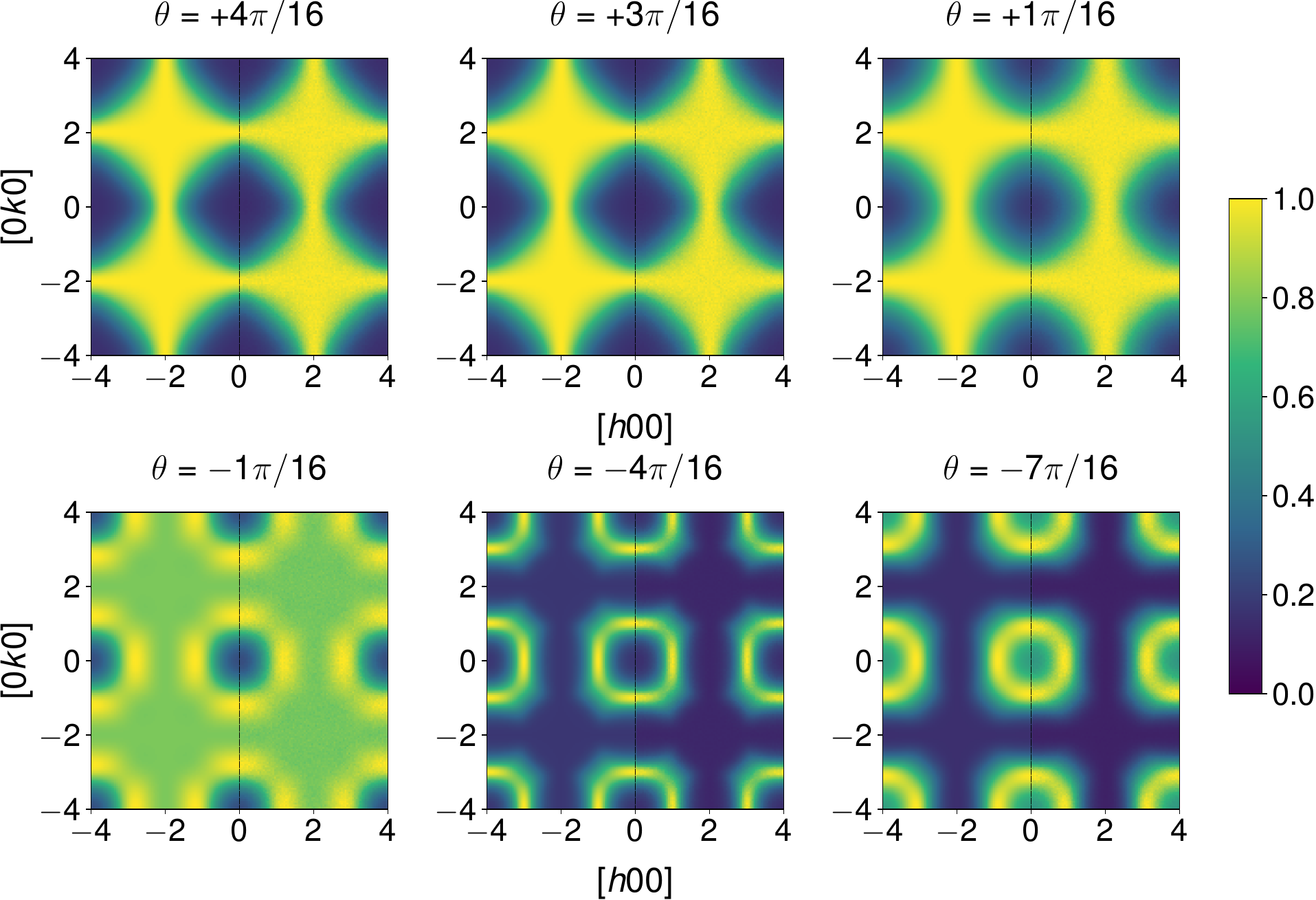}
\caption{
\textbf{Static structure factors of the breathing pyrochlore} at $T/(\bar{J}S^2)=0.5$ in the $[hk0]$ plane, for $\theta=+4\pi/16$, $+3\pi/16$, $+\pi/16$ (top row, left to right) and $\theta=-\pi/16$, $-4\pi/16$, $-7\pi/16$ (bottom row, left to right) in the parametrization of Eq.~\eqref{eq:polar_parametrization}, i.e.\ three points in the antiferromagnetic quadrant and three in the mixed-sign quadrant of Fig.~\ref{fig:PD}(a), measured by SCGA calculations and Monte Carlo simulations (the left and right parts of each panel, respectively). This confirms the square-ring patterns of the correlations above the transitions when $J_B$ and $J_A$ are of opposite signs. The small mismatch between mean-field theory and simulation is expected as $T$ approaches the transition temperature. The color scale has been normalized to 1 at the maximum of scattering for each panel.
}
\label{fig:cmc_sq}
\end{figure}

The same physics is visible in the equal-time spin structure factor,
\begin{equation}
    S(\mathbf{q})
    =
    \frac{1}{N}
    \sum_{ij}
    \left\langle
    \mathbf{S}_i\cdot\mathbf{S}_j
    \right\rangle
    e^{i\mathbf{q}\cdot(\mathbf{r}_i-\mathbf{r}_j)} ,
    \label{eq:sq}
\end{equation}
where \(\mathbf{r}_i\) is the position of spin \(i\), including the four-site pyrochlore basis.  Since the correlator is real and symmetric under \(i\leftrightarrow j\), the exponential may equivalently be replaced by a cosine.
Figure~\ref{fig:cmc_sq} compares Monte Carlo and SCGA results~\cite{Benton15c} in the \([hk0]\) plane at fixed temperature \(T/(\bar{J}S^2)=0.5\).  For \(\theta>0\), in the antiferromagnetic breathing-pyrochlore regime, the correlations have the familiar Coulomb-phase form, with pinch-point-like structures associated with the local zero-magnetization constraint on each tetrahedron.  For \(\theta<0\), in the mixed-sign regime, the scattering changes qualitatively.  The dominant features are square-ring patterns associated with the soft-mode lines of the effective-fcc manifold~\cite{Benton15c}.

The SCGA reproduces the diffuse structure of the Monte Carlo data at temperatures well above the transition.  This shows that the square-ring pattern is already present in the correlated paramagnet and is not merely a remnant of the ordered state.  It reflects the effective decoupling of antiferromagnetic planes in the mixed-sign classical manifold.  On further cooling, the intersections of these soft-mode lines are selected, leading to the \(X=(1,0,0)\) order identified in Fig.~\ref{fig:cmc_order}.

%------------------------------------------------------------------------
%------------------------------------------------------------------------

\subsection{From pinch points to diffuse patterns}

\begin{figure}[t]
\includegraphics[width=\columnwidth]{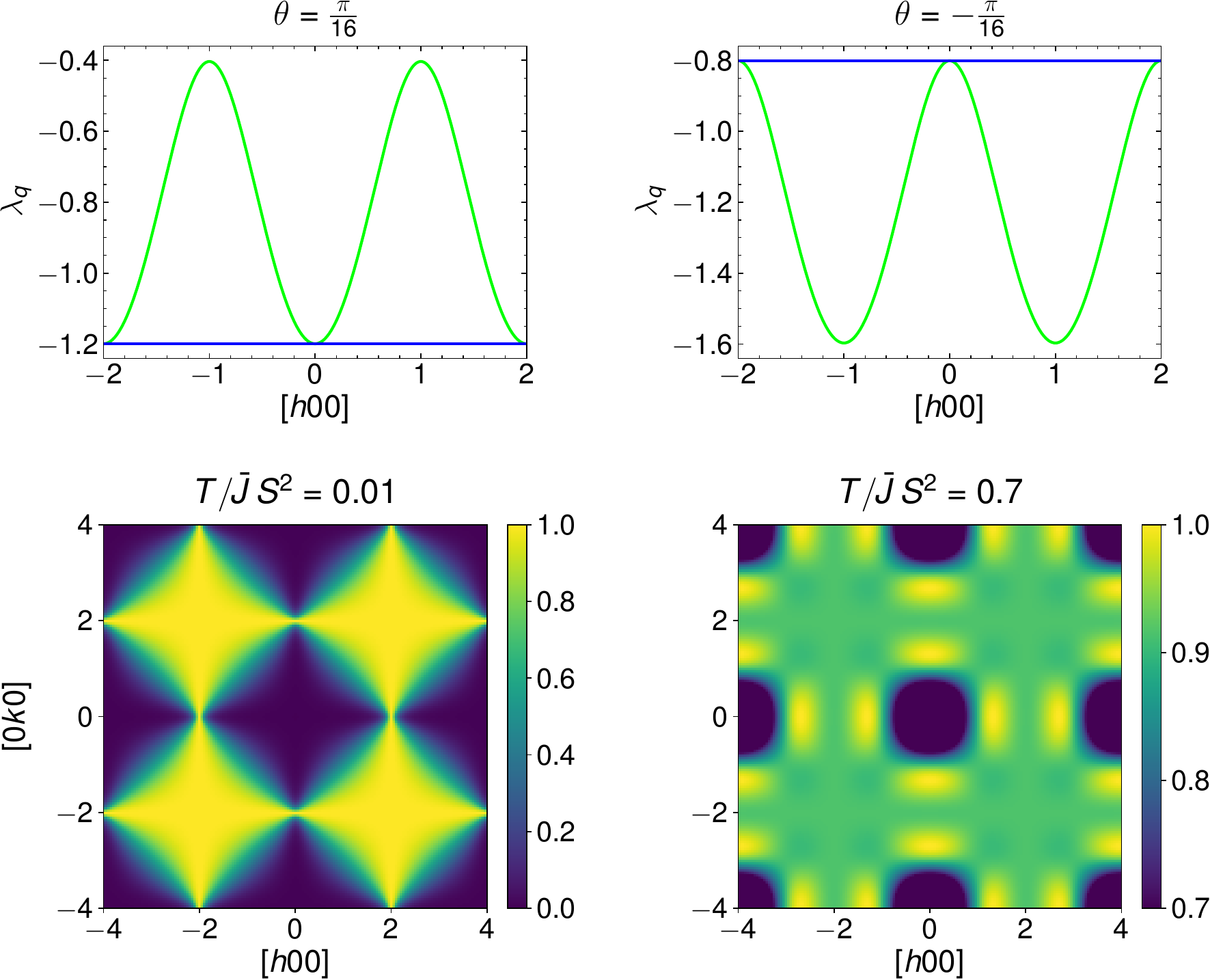}
\caption{
\textbf{Static structure factors of the breathing pyrochlore} obtained from SCGA, showing the pinch points for $\theta=\pi/16$ at very low temperature, $T/(\bar{J}S^2)=0.01$ (bottom left) and the formation of diffuse patterns for $\theta=-\pi/16$ at intermediate temperatures, $T/(\bar{J}S^2)=0.7$ (bottom right). This is because the (green) dispersive band responsible for the diffuse patterns moves from above the (blue) flat band for $\theta > 0$ (top left) to below the (blue) flat band for $\theta < 0$ (top right). Only the low-energy bands along a $[h00]$ cut in Fourier space are shown in the top panels for clarity. The elongation of the diffuse patterns upon cooling is responsible for the square patterns observed in Fig.~\ref{fig:cmc_sq}, in a way reminiscent of the formation of the star patterns for the $\{J_{1},J_{2}=J_{3}\}$ model~\cite{Mizoguchi18a}.
}
\label{fig:scga_halfmoons}
\end{figure}

The change from Coulomb correlations to mixed-sign effective-fcc correlations can be followed more closely within the SCGA.  Figure~\ref{fig:scga_halfmoons} shows representative structure factors together with the corresponding low-energy bands of the Fourier-space interaction matrix along a \([h00]\) cut.  For small positive \(\theta\), the lowest mode is the flat band associated with the Coulomb constraint.  The structure factor therefore develops the usual pinch-point features at low temperature.  For small negative \(\theta\), the dispersive band drops below the flat band near the \(X\) points.  The dominant weight is then displaced away from the Coulomb pinch points and appears as diffuse scattering surrounding the \(\Gamma\) points where the pinch points were.

These patterns are the local precursors of the square-ring pattern in Fig.~\ref{fig:cmc_sq}.  They are reminiscent of the half-moon patterns of frustrated Ising and Heisenberg models~\cite{Udagawa16a,Rau16a,Mizoguchi18a}; both patterns indicate the minimum manifold of the lowest dispersive band, in proximity to a flat-band Coulomb phase.  The two cases are nevertheless distinct.  In the \(J_1\)--\(J_2=J_3\) model the dispersive band sinks below the flat band gradually, so that its minimum manifold is a closed contour that starts as an infinitesimal circle around \(\Gamma\) and grows with \(J_2=J_3\), and the resulting arcs are dispersive images of the pinch points~\cite{Mizoguchi18a,Yan18a}.  Here the dispersive band drops below the flat band as soon as \(J_B\) becomes ferromagnetic, its minimum manifold, the soft lines of the effective-fcc model, does not move with \(J_B\), and the diffuse patterns are straight rather than arc-like.  What the two situations share is that the dispersive band must touch the flat band at \(\Gamma\), where the pinch point of the neighboring Coulomb phase sits, for \(0<|J_B|\le J_A\)~\cite{Essafi17a}: the energy minimum is therefore forced away from \(\Gamma\) onto a closed contour around it, and the minimum is shallow, its depth being set by \(|J_B|\) while the gap to the next dispersive band is set by \(J_A\).  In the present model their origin is tied to the sign change of one breathing-pyrochlore exchange.  The antiferromagnetic quadrant is controlled by the tetrahedral zero-magnetization constraint, whereas the mixed-sign quadrant is controlled by the soft modes of the effective fcc antiferromagnet.

%------------------------------------------------------------------------
%------------------------------------------------------------------------

Figures~\ref{fig:cmc_thermo}--\ref{fig:scga_halfmoons} establish the classical reference point for the remainder of the paper.  In the mixed-sign breathing pyrochlore, the effective-fcc line degeneracy survives as diffuse square-ring correlations in the paramagnetic regime, but it is lifted at finite temperature by order-by-disorder.  The selected state is the collinear Type-I antiferromagnet at \(X=(1,0,0)\).  The quantum calculations below should therefore be compared against this classical outcome: a nonmagnetic regime at finite \(S\) corresponds to a quantum suppression of the thermally selected effective-fcc order, rather than to an ambiguity in the classical ordering tendency.

\section{Quantum phase diagram from pseudofermion FRG}
\label{sec:pffrg_results}

The classical simulations in Sec.~\ref{sec:classical_cmc} show that thermal fluctuations select the Type-I antiferromagnet in the mixed-sign regime.  We now ask how this result is modified by quantum fluctuations.  For this purpose we use the pseudofermion functional renormalization group (pf-FRG), which has been widely applied to frustrated quantum spin models in two and three dimensions~\cite{Reuther2010,Reuther2011,Reuther2011_2,Reuther2011_3,Yasir2016,yasir2016_2,Finn2016,Reuther2017,Finn2018,Finn2018_2,Finn2019,Yasir2019,Dominik2020,Finn2021,Vincent2022,Yasir2023,fukui2023,Lasse2023,Lasse2024,Chern2024}.  Technical details of the implementation are summarized in Appendix~\ref{app:pffrg}.  Here we focus on the physical information obtained from the RG flows and from the momentum-resolved static susceptibility.

%------------------------------------------------------------------------
\begin{figure*}[!t]
\centering
\includegraphics[width=\textwidth]{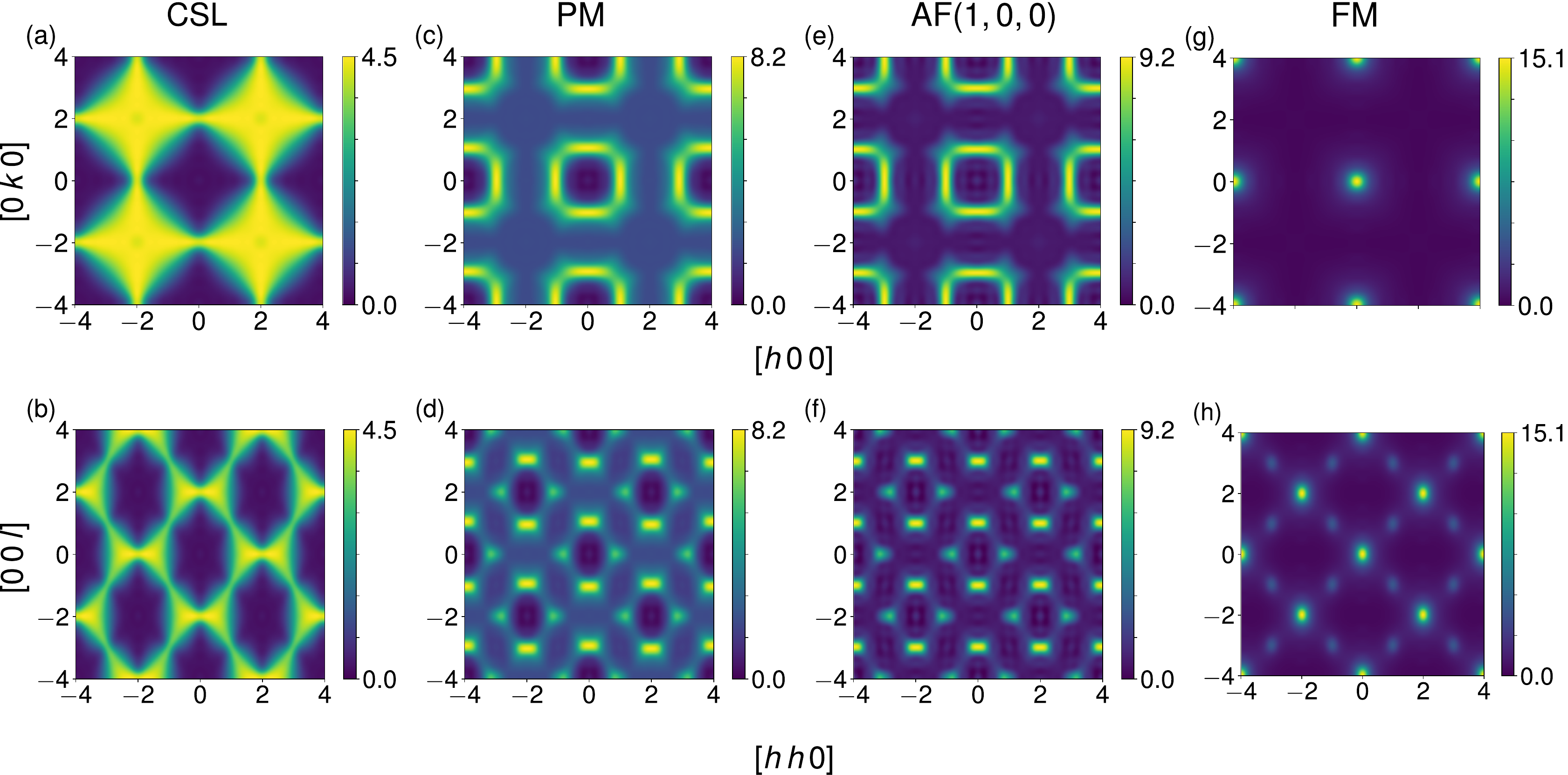}
\caption{
Static susceptibility \(\chi^{zz,\Lambda}(\mathbf{k})\) obtained from pf-FRG at representative points of the phase diagram marked in Fig.~\ref{fig:PD}.  The upper row shows cuts in the \([hk0]\) plane, with horizontal axis \([h00]\) and vertical axis \([0k0]\).  The lower row shows cuts in the \([hhl]\) plane, with horizontal axis \([hh0]\) and vertical axis \([00l]\).  Panels~(a),(b) show the antiferromagnetic quadrant, where the susceptibility has Coulomb-like bow-tie features and no magnetic instability is found.  Panels~(c),(d) show the mixed-sign nonmagnetic regime: the response is diffuse but already concentrated near the \(X\)-point structure of the neighboring ordered phase.  Panels~(e),(f) show the mixed-sign ordered regime, where the susceptibility is evaluated just above the pf-FRG breakdown scale and has dominant maxima at \(X=(1,0,0)\) and symmetry-related wave vectors.  Panels~(g),(h) show the ferromagnetic quadrant, with maxima at \(\mathbf{k}=0\) and equivalent zone centers.
}
\label{fig:pffrg_suscep}
\end{figure*}
%------------------------------------------------------------------------

In pf-FRG, the spin operators are represented by auxiliary fermions, and an infrared cutoff \(\Lambda\) is introduced in Matsubara frequency.  The RG flow generates a cutoff-dependent static spin susceptibility \(\chi^{zz,\Lambda}(\mathbf{k})\), defined in Appendix~\ref{app:pffrg}, whose momentum dependence gives the dominant spin correlations at scale \(\Lambda\).  A smooth flow down to the lowest accessible cutoff is taken as evidence against conventional dipolar magnetic order within the resolution of the calculation.  Conversely, a kink, cusp, or breakdown of the flow at a finite scale \(\Lambda_c\) signals an instability toward magnetic order, with the ordering wave vector inferred from the maximum of \(\chi^{zz,\Lambda}(\mathbf{k})\) close to \(\Lambda_c\).  This criterion is standard in pf-FRG studies of frustrated magnets~\cite{Reuther2010,Reuther2011,Yasir2016,Yasir2019,Tobi2024}.  It should be kept in mind, however, that a smooth flow does not by itself distinguish a featureless paramagnet from a valence-bond, plaquette, tetramerized, nematic, or topologically ordered state. The present pf-FRG analysis therefore gives a phase diagram of dipolar magnetic ordering tendencies and the associated spin-correlation profiles. The internal structure of the nonmagnetic regime will be addressed in Secs.~\ref{sec:dmrg} and~\ref{sec:effective_pseudospin}.

Figure~\ref{fig:PD} summarizes the resulting phase diagrams for \(S=1/2\) and \(S=1\).  The quantum phase diagrams retain the broad structure of the classical problem: the ferromagnetic quadrant orders ferromagnetically, the mixed-sign regime away from the axes orders at \(X=(1,0,0)\), and the antiferromagnetic quadrant remains nonmagnetic with Coulomb-like spin correlations.  The important difference from the classical result is the appearance of a finite nonmagnetic region adjacent to the decoupled antiferromagnetic-tetrahedron limits.  In the classical model, the mixed-sign manifold is selected by order-by-disorder for any nonzero coupling of the opposite sign.  In the quantum model, by contrast, pf-FRG finds that a finite ferromagnetic perturbation is required before the system enters the \(X=(1,0,0)\) ordered phase.  This effect is strongest for \(S=1/2\) and is still visible, though reduced, for \(S=1\).

\subsection{Coulomb-like correlations in the antiferromagnetic quadrant}

We first consider the quadrant \(J_A>0\), \(J_B>0\).  At the classical level this is the breathing deformation of the pyrochlore Coulomb phase.  For finite \(S\), pf-FRG does not find a magnetic instability in this quadrant for either \(S=1/2\) or \(S=1\), in agreement with previous pf-FRG work on the antiferromagnetic breathing pyrochlore~\cite{Yasir2019}.  Representative susceptibility profiles for the isotropic point \(J_A=J_B>0\) are shown in Fig.~\ref{fig:pffrg_suscep}(a) and~\ref{fig:pffrg_suscep}(b).  The response is broad and structured rather than Bragg-like.  In the \([hk0]\) and \([hhl]\) planes it displays the bow-tie features characteristic of pyrochlore Coulomb correlations, although the singular pinch points of the classical zero-temperature limit are rounded by quantum fluctuations and by the finite cutoff at which the susceptibility is evaluated. The profiles in Figs.~\ref{fig:pffrg_suscep}(a) and~\ref{fig:pffrg_suscep}(b) are evaluated at the lowest cutoff reached in the simulation, $\Lambda=10^{-4}$.

The corresponding RG flow is shown in Fig.~\ref{fig:rgflow}.  The flow remains smooth down to the lowest simulated cutoff.  We therefore label this region a Coulomb spin liquid (CSL) in Fig.~\ref{fig:PD}, with the qualification that pf-FRG directly diagnoses the absence of dipolar order and the presence of Coulomb-like two-spin correlations, but does not by itself establish a complete low-energy gauge theory.

\subsection{Quantum paramagnet near decoupled antiferromagnetic tetrahedra}

We now turn to the mixed-sign quadrant \(J_A>0\), \(J_B<0\), starting near the decoupled limit \(J_B=0^-\).  Classically, this is precisely the regime where the ferromagnetic tetrahedra form composite moments and the ground-state manifold maps onto the nearest-neighbor fcc antiferromagnet.  The Monte Carlo results of Sec.~\ref{sec:classical_cmc} show that thermal fluctuations then select the Type-I state at \(X=(1,0,0)\).

The pf-FRG result is different.  For \(S=1/2\), the RG flow remains smooth over a finite interval of ferromagnetic \(J_B\), extending approximately to
\begin{equation}
    \theta_c/\pi \simeq -0.175(13),
    \qquad
    (J_B/J_A)_c=\tan\theta_c \simeq -0.61(5),
    \label{eq:theta_c}
\end{equation}
in the parametrization \(J_A=\bar{J}\cos\theta\), \(J_B=\bar{J}\sin\theta\).  The corresponding region is shown in gray in Fig.~\ref{fig:PD}(b).  A smaller but still visible nonmagnetic region survives for \(S=1\), as shown in Fig.~\ref{fig:PD}(c), extending only to
\begin{equation}
    \theta_c/\pi \simeq -0.025(6),
    \qquad
    (J_B/J_A)_c \simeq -0.08(2) .
    \label{eq:theta_c_s1}
\end{equation}
The nonmagnetic window therefore narrows by roughly a factor of seven between
\(S=1/2\) and \(S=1\), which is the expected direction: the window is a quantum
effect, and increasing the spin length drives the model toward the classical
limit in which any infinitesimal ferromagnetic \(J_B\) already selects the
ordered state.  Both boundaries are obtained from the position of a kink in the
susceptibility flow and should be read as rough estimates; we return to this
point in Sec.~\ref{sec:dmrg_sq}, where the pf-FRG window is compared with
DMRG.

%------------------------------------------------------------------------
\begin{figure}[t]
\centering
\includegraphics[width=1.0\linewidth]{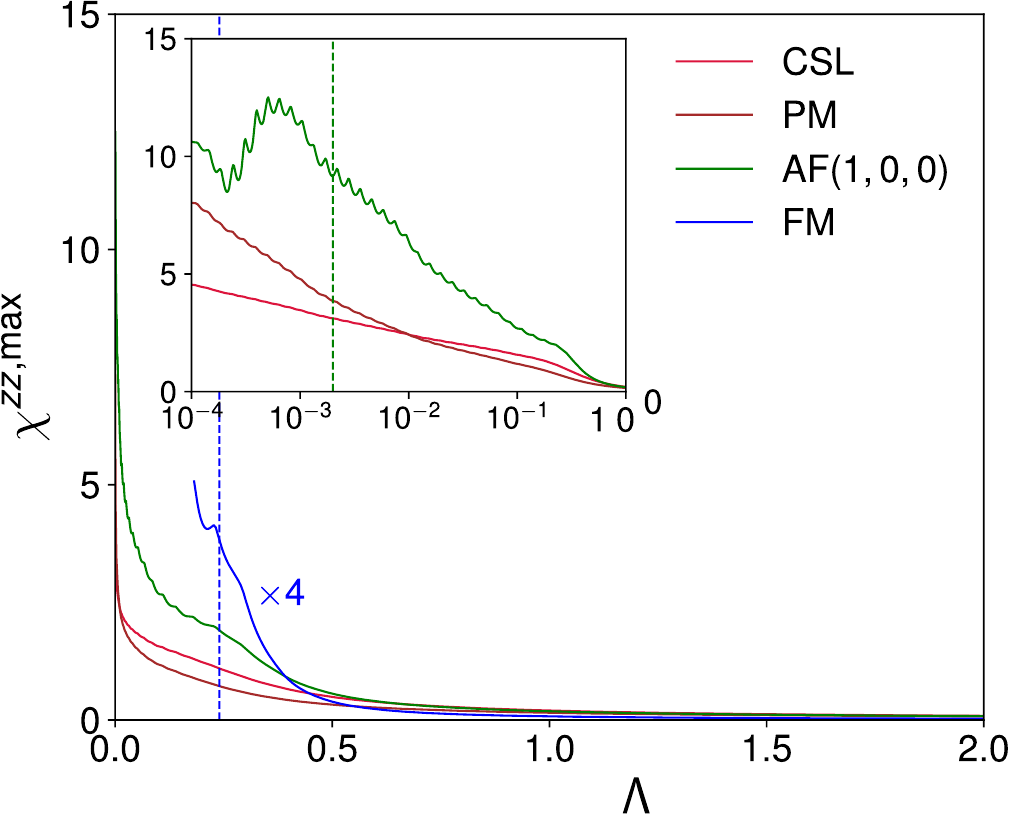}
\caption{
Cutoff dependence of the maximum static susceptibility \(\chi^{zz,\Lambda}_{\rm max}\) for the representative points shown in Fig.~\ref{fig:pffrg_suscep}.  Smooth flows down to the lowest simulated cutoff are found in the Coulomb-like antiferromagnetic regime and in the mixed-sign nonmagnetic regime.  The mixed-sign \(X=(1,0,0)\) phase and the ferromagnetic phase instead show flow instabilities at finite cutoff, signaling the onset of magnetic order.  The inset displays the low-cutoff behavior on a logarithmic scale.  Vertical dashed lines indicate the cutoffs at which the susceptibility maps in Fig.~\ref{fig:pffrg_suscep} are evaluated.  The ferromagnetic curve in the main panel is plotted after rescaling by $1/4$ to fit the vertical scale; the $\times 4$ label denotes the factor required to recover the original values.
}
\label{fig:rgflow}
\end{figure}
%------------------------------------------------------------------------

Representative susceptibility profiles inside this nonmagnetic mixed-sign regime are shown in Figs.~\ref{fig:pffrg_suscep}(c) and~\ref{fig:pffrg_suscep}(d).  The response is no longer Coulomb-like.  Instead, the spectral weight is redistributed toward the same \(X\)-point structure that controls the classical order-by-disorder state.  In the \([hk0]\) plane this appears as a diffuse square-like pattern, while in the \([hhl]\) plane the maxima occur at the corresponding symmetry-related positions.  The correlations therefore retain a clear memory of the effective-fcc manifold.  At the same time, the RG flow in Fig.~\ref{fig:rgflow} remains smooth. The corresponding susceptibility maps in Figs.~\ref{fig:pffrg_suscep}(c) and~\ref{fig:pffrg_suscep}(d) are likewise evaluated at the lowest simulated cutoff, 
$\Lambda=10^{-4}$, for the same reason. Thus, the state is not an \(X\)-ordered antiferromagnet within pf-FRG, even though its strongest spin correlations already occur at the momenta of the neighboring ordered phase.

The pf-FRG result therefore does not indicate that the nonmagnetic regime is unrelated to the classical manifold; it indicates that quantum fluctuations suppress static dipolar order while leaving soft \(X\)-point correlations behind.  Determining what replaces the ordered moment requires probes of real-space and higher-order correlations, which we address using DMRG in Sec.~\ref{sec:dmrg}.

\subsection{\texorpdfstring{\(X=(1,0,0)\)}{X=(1,0,0)} order at stronger mixed-sign coupling}

Upon increasing \(|J_B|\) further in the mixed-sign quadrant, the smooth pf-FRG flow gives way to a magnetic instability.  The susceptibility profiles immediately above the breakdown scale are shown in Figs.~\ref{fig:pffrg_suscep}(e) and~\ref{fig:pffrg_suscep}(f).  The dominant weight is concentrated at the \(X\) points, identifying the ordered phase as the Type-I antiferromagnet with ordering wave vector \(X=(1,0,0)\).  This is the same state selected thermally in the classical Monte Carlo simulations.

The RG flow for this parameter point is shown in Fig.~\ref{fig:rgflow}.  In contrast to the CSL and paramagnetic (PM) flows, the susceptibility develops a clear instability at a finite cutoff.  The susceptibility profiles in Figs.~\ref{fig:pffrg_suscep}(e) and~\ref{fig:pffrg_suscep}(f) are plotted at $\Lambda_c = 0.002$, cutoff just above the breakdown, where the momentum structure of the incipient ordered state can still be read reliably.  The agreement between the ordering vector found by pf-FRG and the one selected by classical order-by-disorder is a useful consistency check: the finite-\(S\) calculation does not introduce a competing dipolar order.  Instead, it shifts the onset of the same \(X\)-ordered state away from the decoupled-tetrahedron point, leaving a quantum nonmagnetic window in between.

\subsection{Ferromagnetic quadrant}

Finally, for \(J_A<0\) and \(J_B<0\), both tetrahedral sublattices favor ferromagnetic alignment.  The pf-FRG flow breaks down at a finite scale, and the maximum of the susceptibility occurs at \(\mathbf{k}=0\), as expected for a ferromagnet.  Representative profiles are shown in Figs.~\ref{fig:pffrg_suscep}(g) and~\ref{fig:pffrg_suscep}(h) for $\Lambda_c = 0.24$.  The strongest response occurs at the Brillouin-zone centers, with symmetry-related periodic copies.  The corresponding flow in Fig.~\ref{fig:rgflow} shows the most direct instability among the four representative cases.

The pf-FRG results provide the quantum counterpart of the classical analysis.  They show that the \(X=(1,0,0)\) state selected by thermal order-by-disorder remains the leading dipolar instability at sufficiently strong mixed-sign coupling.  At the same time, finite-\(S\) fluctuations open a nonmagnetic regime next to the decoupled antiferromagnetic-tetrahedron limit.  This regime is not featureless at the level of two-spin correlations: its susceptibility is already peaked near the \(X\) points of the proximate ordered phase.  The remaining question is therefore not whether the nonmagnetic state is connected to the effective-fcc manifold, but how the local tetrahedral degrees of freedom arrange themselves once dipolar order is suppressed.  We address this question next using DMRG.

\section{DMRG: tetramer correlations and the approach to \texorpdfstring{\(X\)}{X}-point order}
\label{sec:dmrg}

The pf-FRG analysis of Sec.~\ref{sec:pffrg_results} shows that finite-\(S\) fluctuations open a nonmagnetic regime near the decoupled antiferromagnetic-tetrahedron limit.  It also shows that this regime is not disconnected from the classical mixed-sign manifold: the dominant two-spin correlations already occur near the same \(X\) points that characterize the neighboring Type-I antiferromagnet.  What pf-FRG does not determine is the internal structure of the nonmagnetic state.  In particular, a smooth pf-FRG flow rules out conventional dipolar magnetic order within the resolution of the calculation, but it does not distinguish between a featureless paramagnet, a valence-bond solid, a tetramerized state, or a more subtle nonmagnetic phase.

We therefore turn to DMRG calculations on finite three-dimensional clusters.  The aim is less to reproduce the pf-FRG phase boundary than to identify the short-range correlations that develop when the \(X\)-ordered moment is suppressed.  We focus on the mixed-sign regime with antiferromagnetic up-tetrahedra and ferromagnetic down-tetrahedra, denoting the latter coupling by \(J_B<0\) (we set $J_A=1$ in this section).  The calculations are performed in an SU(2)-symmetric implementation, retaining up to 6000 SU(2) states for the structure-factor data shown below. We consider two cubic clusters ($N=64$ and $N=128$ sites) that are commensurate with the \(X=(1,0,0)\) ordered phase. For their precise definitions, see Table II of Ref.~\cite{Hagymasi2026}. Since the clusters are finite and three-dimensional, and since only a limited set of system sizes is presently available, we use the DMRG results primarily as a diagnostic of local and incipient ordering tendencies rather than as a stand-alone determination of a thermodynamic phase boundary.

\subsection{Local correlations: evidence for tetramer formation}
\label{sec:dmrg_local}

\begin{figure}[t]
    \centering
    \includegraphics[width=\columnwidth]{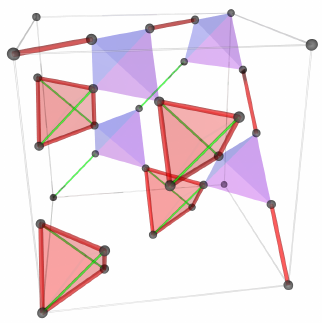}
    \caption{
    Nearest-neighbor spin correlations
    \(\langle \mathbf{S}_i\cdot\mathbf{S}_j\rangle\)
    in the DMRG ground state on the \(N=64\) cluster for $J_B=-0.2$.  Only one cubic unit cell is shown.  Line widths indicate the magnitude of the correlation.  The shaded tetrahedra are guides to the eye and distinguish the two breathing-pyrochlore sublattices.  On the antiferromagnetic tetrahedra, four bonds have correlations close to
    \(\langle \mathbf{S}_i\cdot\mathbf{S}_j\rangle\simeq -0.49\),
    while the two opposite bonds have correlations close to
    \(\langle \mathbf{S}_i\cdot\mathbf{S}_j\rangle\simeq +0.24\).
    This is close to the ideal tetramer-singlet pattern
    \(-1/2\) and \(+1/4\), respectively.
    }
    \label{fig:dmrg_tetramer_correlations}
\end{figure}

Figure~\ref{fig:dmrg_tetramer_correlations} shows the nearest-neighbor spin correlations in the DMRG ground state on the \(N=64\) cluster, with one cubic unit cell displayed.  The most striking feature is the strong differentiation of bonds on the antiferromagnetic tetrahedra.  The same four bonds on each selected tetrahedron carry antiferromagnetic correlations close to
\begin{equation}
    \langle \mathbf{S}_i\cdot\mathbf{S}_j\rangle \simeq -0.49 ,
\end{equation}
while the two remaining opposite bonds carry positive correlations close to
\begin{equation}
    \langle \mathbf{S}_i\cdot\mathbf{S}_j\rangle \simeq +0.24 .
\end{equation}
These numbers are very close to the ideal values
\begin{equation}
    -\frac{1}{2},\qquad +\frac{1}{4},
    \label{eq:ideal_tetramer_correlations}
\end{equation}
expected for a particular tetrahedral singlet in which four of the six bonds form a correlated antiferromagnetic loop and the remaining two opposite bonds are ferromagnetic.  This pattern is distinct from both a product of two isolated singlet dimers and from the fully symmetric tetrahedral singlet.  A product of two dimers would have two bonds with correlations \(-3/4\) and four nearly inactive bonds, whereas a symmetric tetrahedral singlet would give the same value, \(-1/4\), on all six bonds.  The DMRG pattern instead corresponds to a four-spin tetramer, or quadrumer, singlet on the antiferromagnetic tetrahedron.

Two points follow.  The nonmagnetic regime is not structureless at the level of local spin correlations: the antiferromagnetic tetrahedra do not retain an undifferentiated singlet character, but select one of the three inequivalent ways of singling out a pair of opposite bonds, that is, one of three tetramer orientations.  In addition, the near-ideal values in Eq.~\eqref{eq:ideal_tetramer_correlations} indicate that the relevant low-energy Hilbert space is the two-dimensional singlet manifold of an isolated spin-\(1/2\) tetrahedron, which is the setting in which the strong-breathing pseudospin description applies.

The finite cluster does not by itself prove long-range tetramer order in the thermodynamic limit.  Boundary conditions, cluster geometry, and the DMRG sweeping order can favor a particular member of a nearly degenerate manifold.  The local correlation pattern is nevertheless stable across the parameters we examined, and the microscopic spin model develops tetrahedral singlet correlations of the same form as those obtained from the effective pseudospin theory of Sec.~\ref{sec:effective_pseudospin}.

%------------------------------------------------------------------------
%------------------------------------------------------------------------

\subsection{Momentum-space correlations and proximity to Type-I order}
\label{sec:dmrg_sq}

The local tetramer pattern does not exhaust the physics of the nonmagnetic regime.  The pf-FRG susceptibility already indicated that the strongest spin correlations remain tied to the \(X\) points of the effective-fcc manifold.  To examine this in the microscopic wave function, we compute the equal-time structure factor of Eq.~\eqref{eq:sq}, now evaluated in the DMRG ground state.

\begin{figure}[t]
    \centering
    \includegraphics[width=\columnwidth]{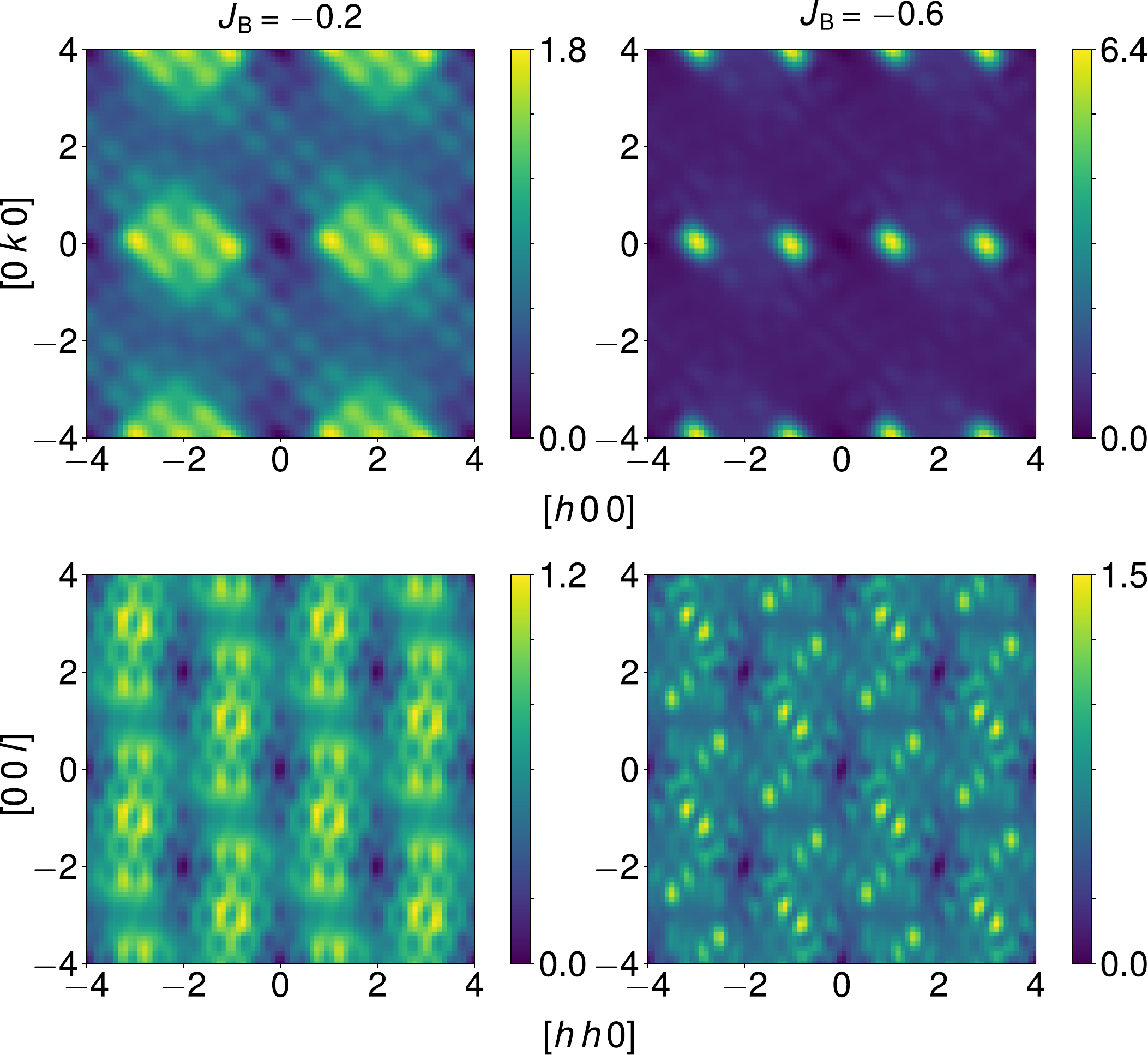}
    \caption{
    Equal-time spin structure factor \(S(\mathbf{q})\), Eq.~\eqref{eq:sq}, obtained from DMRG on the \(N=64\) cluster.  The upper row shows the \([hk0]\) plane, while the lower row shows the \([hhl]\) plane.  The left column corresponds to weak ferromagnetic coupling, \(J_B=-0.2\), where the response remains broad.  The right column corresponds to stronger ferromagnetic coupling, \(J_B=-0.6\), where pronounced maxima develop at \(X\)-type wave vectors, consistent with the onset of Type-I antiferromagnetic correlations.  The data were obtained using 6000 SU(2) states, and correspond to a single symmetry-broken member of the threefold tetramer manifold; they are not symmetrized over the cubic point group (see text).
    }
    \label{fig:dmrg_structure_factor}
\end{figure}

Figure~\ref{fig:dmrg_structure_factor} shows \(S(\mathbf{q})\) for the \(N=64\) cluster in two reciprocal-space cuts.  For weak ferromagnetic coupling, represented by \(J_B=-0.2\), the response is broad.  It does not show sharp Bragg peaks, in agreement with the pf-FRG identification of a nonmagnetic regime near the decoupled antiferromagnetic-tetrahedron limit.  At the same time, the weight is not completely featureless.  It is distributed in a way that already reflects the soft \(X\)-point structure of the nearby mixed-sign manifold.

For stronger ferromagnetic coupling, represented by \(J_B=-0.6\), the same structure factor develops much sharper maxima at \(X\)-type wave vectors.  This is the momentum-space signature expected for the Type-I antiferromagnet selected in the classical Monte Carlo simulations and found as the leading pf-FRG instability at larger \(|J_B|\).  The DMRG data therefore support a continuous physical interpretation across methods: the nonmagnetic state at small \(|J_B|\) retains short-range \(X\)-point correlations, while increasing the ferromagnetic coupling strengthens these correlations and drives the system toward the same Type-I order selected classically.

One point deserves comment.  The pf-FRG boundary of Eq.~\eqref{eq:theta_c} places
the onset of \(X\) order at \((J_B/J_A)_c\simeq-0.61(5)\), so the DMRG
representative of the ordered side, \(J_B=-0.6\), lies within one standard error
of that boundary rather than well inside the ordered phase.  The two results are
nonetheless consistent, because a pf-FRG phase boundary obtained from the
position of a kink in the susceptibility flow is only a rough estimate.  As
discussed in Ref.~\cite{Hagymasi2024} for the \(J_1\)--\(J_2\) spin-1 pyrochlore
antiferromagnet, the cutoff at which a weak kink appears depends sensitively on
implementation details such as the frequency grid, so that the extent of the
ordered region obtained in this way is best read as a lower bound, and the
nonmagnetic window correspondingly as an upper bound.  In that model the
comparison is stark: pf-FRG places the boundary at \(J_2/J_1=0.09(2)\), whereas
DMRG finds the nonmagnetic phase ending near \(J_2/J_1\sim0.02\), with a
pseudo-Majorana FRG treatment that extrapolates a finite-temperature transition
to \(T=0\) giving \(0.035(8)\), close to the DMRG value.  The caveat applies
directly here, since our \(\theta_c\) is likewise obtained by locating a kink by
eye (Appendix~\ref{app:pffrg_numerics}).  The sharpening of the \(X\)-point
maxima at \(J_B=-0.6\) is therefore best read as the onset of Type-I order in a
regime where the true phase boundary plausibly lies at smaller \(|J_B|\) than the
pf-FRG estimate.

A remark on the comparison with the pf-FRG susceptibility profiles of Fig.~\ref{fig:pffrg_suscep} is in order.  The DMRG structure factor of Fig.~\ref{fig:dmrg_structure_factor} is not symmetrized over the cubic point group.  As discussed in Sec.~\ref{sec:dmrg_local}, the ground state on the finite cluster selects one of the three tetramer orientations.  The three states so obtained are related by the point-group symmetry of the cubic cluster and are therefore expected to be degenerate, in line with the exact degeneracy found in the 32-site pseudospin model of Sec.~\ref{sec:effective_pseudospin}; the state reached by DMRG is one member of this manifold, singled out by the random initial state and by the mapping of the three-dimensional cluster onto the one-dimensional DMRG chain.  Its structure factor therefore carries the reduced symmetry of that member.  The pf-FRG susceptibility, by contrast, cannot break the point-group symmetry and by construction yields the symmetric response, i.e.\ the average over the three orientations.  Restoring the symmetry on the DMRG side would require the orthogonal partner states, which can be obtained by successive simulations with an orthogonality constraint, as was done for the \(J_1\)--\(J_2\) spin-1 pyrochlore antiferromagnet in Ref.~\cite{Hagymasi2024}; for the present cluster sizes this is a considerably more demanding calculation, and it has not been attempted here.  The comparison between the two methods should therefore be made at the level of the dominant wave vectors, which agree, rather than of the detailed intensity distribution.

Enhanced \(X\)-point weight in the nonmagnetic regime does not by itself mean that the state is magnetically ordered.  On a finite cluster, broad maxima can arise from short-range correlations whose characteristic wave vector is fixed by the proximate ordered phase.  The distinction between short-range \(X\)-centered correlations and true long-range Type-I order must be made through finite-size growth, to which we now turn.

%------------------------------------------------------------------------
%------------------------------------------------------------------------

\subsection{Growth of the \texorpdfstring{\(X\)}{X}-point structure factor}

\begin{figure}[t]
    \centering
    \includegraphics[width=\columnwidth]{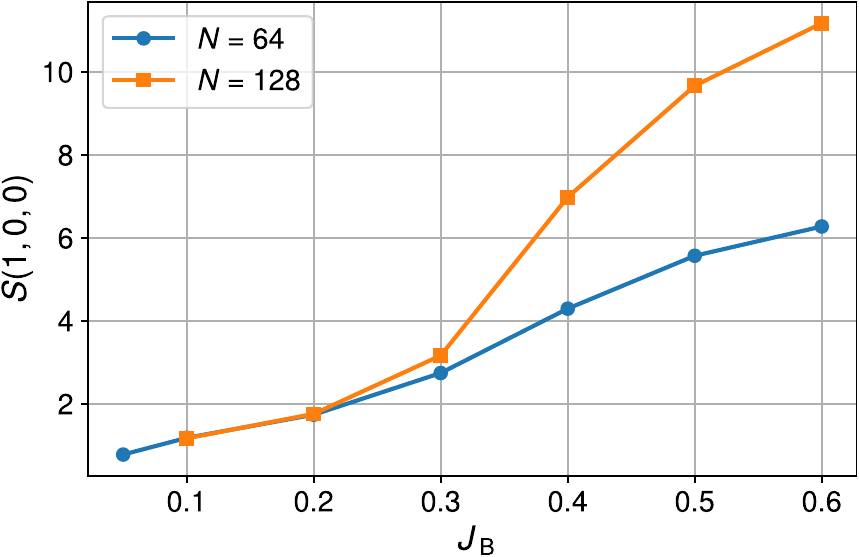}
    \caption{
    \(X\)-point structure factor \(S(\mathbf{q}=X)\) as a function of the magnitude of the ferromagnetic down-tetrahedron coupling \(|J_B|\), shown for \(N=64\) and \(N=128\).  At weak coupling the two sizes remain close, consistent with short-range \(X\)-centered correlations.  At stronger coupling the \(N=128\) result grows much more rapidly, indicating the development of Type-I antiferromagnetic order.  The finite-size trend is consistent with the pf-FRG phase diagram, although the available sizes do not by themselves fix a sharp transition point.
    }
    \label{fig:dmrg_x_scaling}
\end{figure}

Figure~\ref{fig:dmrg_x_scaling} tracks the structure factor at the representative \(X\)-point wave vector \(\mathbf{q}=X=(1,0,0)\) as a function of \(|J_B|\) for \(N=64\) and \(N=128\).  For small \(|J_B|\), the two sizes give comparable values and the structure factor remains modest.  This behavior is consistent with a nonmagnetic regime in which \(X\)-centered correlations are present but short ranged.  Beyond intermediate coupling, the \(N=128\) data grow much more rapidly than the \(N=64\) data.  Such finite-size enhancement is the expected trend if the system is approaching or entering a phase with Type-I antiferromagnetic order.

The available sizes do not allow a precise extrapolation of the ordering threshold.  In particular, the apparent onset region should not be interpreted as a sharply determined critical coupling from DMRG alone.  Nevertheless, the trend in Fig.~\ref{fig:dmrg_x_scaling} is consistent with the broader phase diagram obtained from pf-FRG: a finite nonmagnetic interval is present near \(J_B=0^{-}\), and stronger ferromagnetic coupling eventually produces the \(X=(1,0,0)\) ordered phase.  The important point is that DMRG links these two regimes microscopically.  The nonmagnetic state already contains the local tetrahedral singlet texture and the soft \(X\)-point correlations out of which the ordered phase develops.

%------------------------------------------------------------------------
%------------------------------------------------------------------------

\subsection{Physical interpretation and connection to the strong-breathing theory}

The DMRG results give a more detailed picture of the nonmagnetic regime than is accessible from pf-FRG alone.  There are two intertwined tendencies.  On the one hand, the spin structure factor retains the momentum-space memory of the effective-fcc manifold: the dominant correlations are centered near the \(X\) points and evolve into Type-I antiferromagnetic order as \(|J_B|\) is increased.  On the other hand, the real-space correlations show that, before dipolar order appears, the antiferromagnetic tetrahedra group into nearly ideal tetramer singlets.

This combination is natural in the strong-breathing limit.  At \(J_B=0\), the system is a product of isolated antiferromagnetic tetrahedra.  For spin \(1/2\), each tetrahedron has a two-dimensional singlet ground-state manifold.  The low-energy degree of freedom is therefore not a magnetic moment, but a pseudospin labeling different singlet combinations on a tetrahedron.  A finite inter-tetrahedron coupling then generates an effective Hamiltonian acting on these pseudospins.  For the usual antiferromagnetic perturbation, this problem was analyzed by Tsunetsugu and in related work~\cite{Tsunetsugu-2001a,Tsunetsugu-2001b,Tsunetsugu-2002,Harris-1991,Isoda-1998,Koga-2001,Canals-1998,Canals-2000}.  In the present mixed-sign problem, the sign of the perturbation is reversed.  The DMRG tetramer pattern suggests that this sign reversal favors pseudospin configurations corresponding to the tetramer singlets of Eq.~\eqref{eq:ideal_tetramer_correlations}.

This observation sets the stage for the next section.  We now derive the effective pseudospin model in the tetrahedral singlet manifold and study it by exact diagonalization.  The purpose of that analysis is not only to reproduce the local DMRG bond pattern, but also to clarify how the three possible tetramer orientations on each tetrahedron are selected and correlated across the effective fcc lattice.

\section{Strong-breathing pseudospin theory and exact diagonalization}
\label{sec:effective_pseudospin}

The DMRG results of Sec.~\ref{sec:dmrg} point to nearly ideal tetramer correlations on the antiferromagnetic tetrahedra.  We now show that this tendency follows naturally from the strong-breathing limit.  The analysis closely parallels Tsunetsugu's treatment of the singlet sector of the spin-$1/2$ pyrochlore antiferromagnet~\cite{Tsunetsugu-2001a,Tsunetsugu-2001b,Tsunetsugu-2002}, with one important change: in the present problem, the inter-tetrahedron perturbation is ferromagnetic.  Since the first term that lifts the degeneracy of the tetrahedral singlet manifold is third order in this perturbation, reversing the sign of \(J_B\) reverses the ordering tendency within the effective pseudospin model.

\subsection{Tetrahedral singlet doublet}

We consider the strong-breathing limit
\begin{equation}
    J_A>0,\qquad |J_B|\ll J_A ,
\end{equation}
where the \(A\)-tetrahedra are antiferromagnetic and weakly coupled.  At \(J_B=0\), the ground state is a product of isolated tetrahedron ground states.  A single spin-$1/2$ tetrahedron has two singlet ground states, three triplet multiplets, and one quintet multiplet.  The triplet sector is separated from the singlet doublet by an energy of order \(J_A\).  The low-energy Hilbert space is therefore the tensor product of two-dimensional singlet spaces, one for each \(A\)-tetrahedron.

We use the chiral basis for this singlet doublet.  With
\begin{equation}
    \omega=e^{2\pi i/3},
\end{equation}
we choose
\begin{subequations}
\label{eq:chiral_singlet_basis}
\begin{align}
    |+\rangle
    &=
    \frac{1}{\sqrt{6}}
    \Big[
    |\downarrow\downarrow\uparrow\uparrow\rangle
    +|\uparrow\uparrow\downarrow\downarrow\rangle
    +\omega
    \big(
    |\downarrow\uparrow\downarrow\uparrow\rangle
    +|\uparrow\downarrow\uparrow\downarrow\rangle
    \big)
    \nonumber\\
    &\hspace{3.4cm}
    +\omega^{2}
    \big(
    |\downarrow\uparrow\uparrow\downarrow\rangle
    +|\uparrow\downarrow\downarrow\uparrow\rangle
    \big)
    \Big],
    \\
    |-\rangle
    &=
    \frac{1}{\sqrt{6}}
    \Big[
    |\downarrow\downarrow\uparrow\uparrow\rangle
    +|\uparrow\uparrow\downarrow\downarrow\rangle
    +\omega^{2}
    \big(
    |\downarrow\uparrow\downarrow\uparrow\rangle
    +|\uparrow\downarrow\uparrow\downarrow\rangle
    \big)
    \nonumber\\
    &\hspace{3.4cm}
    +\omega
    \big(
    |\downarrow\uparrow\uparrow\downarrow\rangle
    +|\uparrow\downarrow\downarrow\uparrow\rangle
    \big)
    \Big].
\end{align}
\end{subequations}
The two states have opposite scalar chirality.  We introduce a pseudospin-$1/2$ operator \(\bm{\tau}\) acting in this doublet, with
\begin{equation}
    \tau^z|\pm\rangle=\pm\frac{1}{2}|\pm\rangle .
\end{equation}
The centers of the \(A\)-tetrahedra form an fcc lattice, so the strong-breathing problem becomes an effective pseudospin model on the fcc lattice.

The pseudospin should not be mistaken for a magnetic moment.  Its \(z\)-component distinguishes the two chiral singlets and is odd under time reversal.  The transverse components encode the distribution of bond energies within a tetrahedron.  A general time-reversal-invariant state in the equatorial plane can be written as
\begin{equation}
    |\varphi\rangle
    =
    \frac{|+\rangle+e^{i\varphi}|-\rangle}{\sqrt{2}} .
    \label{eq:pseudospin_theta}
\end{equation}
The three angles
\begin{equation}
    \varphi=\pi,\quad \frac{\pi}{3},\quad \frac{5\pi}{3}
\end{equation}
represent the three possible dimer-pair singlets on a tetrahedron, while
\begin{equation}
    \varphi=0,\quad \frac{2\pi}{3},\quad \frac{4\pi}{3}
    \label{eq:tetramer_angles}
\end{equation}
represent the three tetramer singlets.  For example, the \(\varphi=0\) tetramer has
\begin{subequations}
\label{eq:tetramer_bond_correlations}
\begin{align}
    \langle\mathbf{S}_i\cdot\mathbf{S}_j\rangle
    &=-\frac{1}{2},
    &&
    (i,j)\in\{(1,3),(3,2),(2,4),(4,1)\},
    \\
    \langle\mathbf{S}_i\cdot\mathbf{S}_j\rangle
    &=+\frac{1}{4},
    &&
    (i,j)\in\{(1,2),(3,4)\}.
\end{align}
\end{subequations}
These are the ideal values approached by the microscopic DMRG bond correlations in Fig.~\ref{fig:dmrg_tetramer_correlations}.  The strong-breathing description therefore gives a microscopic interpretation of the DMRG pattern: the nonmagnetic state chooses a direction in the transverse pseudospin plane, corresponding to a definite tetrahedral energy-density pattern, without developing a magnetic moment.

%------------------------------------------------------------------------
\begin{figure}[t]
    \centering
    \includegraphics[width=\columnwidth]{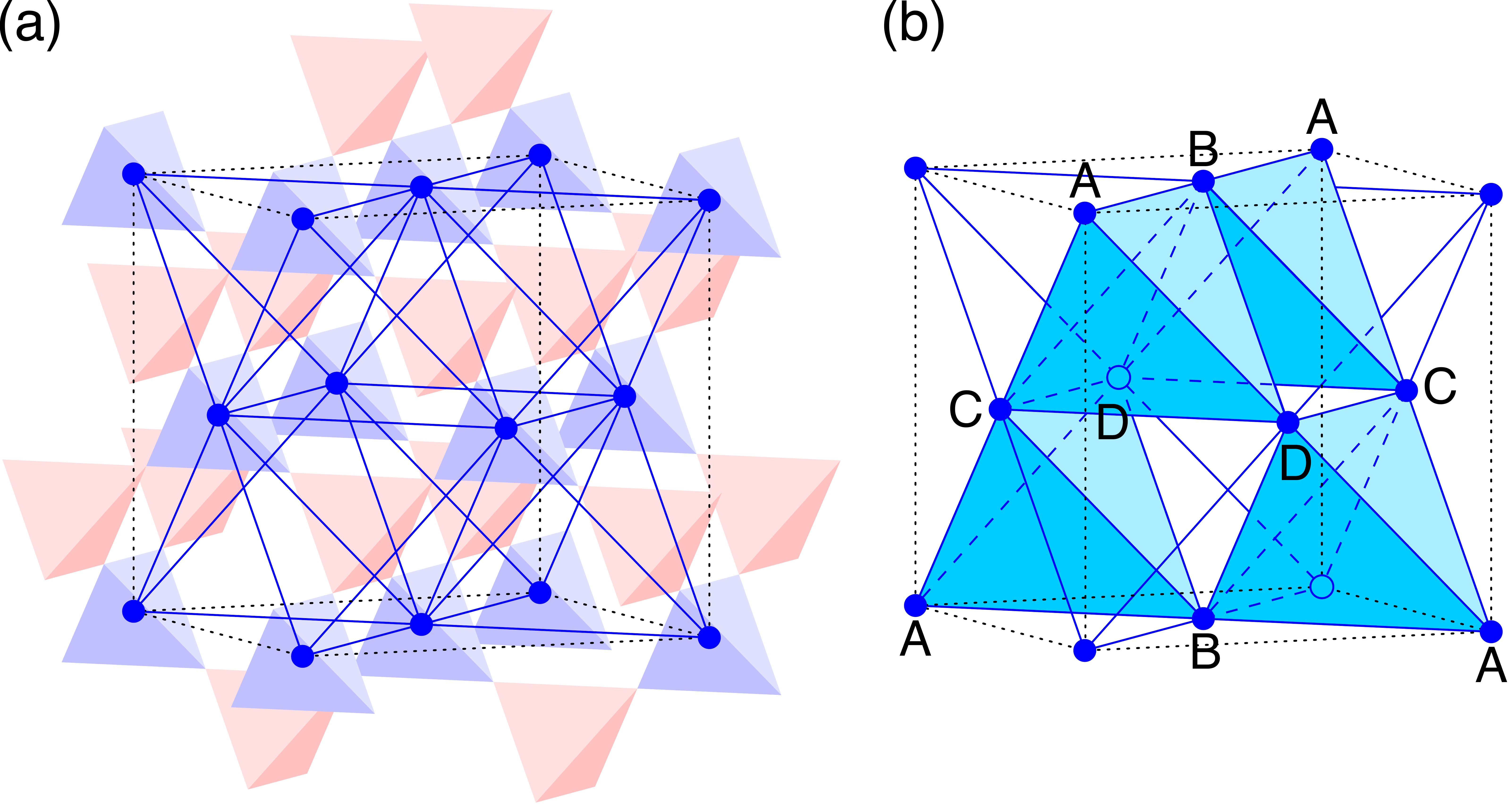}
    \caption{
    Geometry of the strong-breathing pseudospin model.  (a) The centers of the antiferromagnetic \(A\)-tetrahedra form an fcc lattice.  Each tetrahedron contributes one pseudospin-$1/2$ degree of freedom representing its two singlet states.  (b) The third-order processes act on the oriented triangular faces of the fcc lattice whose three inter-tetrahedron bonds belong to three different \(B\) tetrahedra.  The highlighted triangles indicate the elementary triples entering Eq.~\eqref{eq:model_hamiltonian}.
    The labels \(A,B,C,D\) denote the four cubic-cell sublattices of this fcc lattice, used in Eq.~\eqref{eq:oriented_faces} to orient the 3-body terms in Eq.~\eqref{eq:model_hamiltonian}.}  
    \label{fig:fcc_lattice}
\end{figure}
%------------------------------------------------------------------------

\subsection{Third-order effective Hamiltonian}
\label{sec:third_order}

We now turn on \(J_B\).  A single spin operator takes a tetrahedral singlet into
the triplet sector, so the first-order correction has no matrix elements within
the singlet manifold.  The second-order term gives only a state-independent
energy shift, equal to \(-\tfrac{9}{64}J_B^2/J_A\) per spin. 
The first term that splits the \(2^{N_A}\)-fold singlet manifold, where \(N_A\)
is the number of \(A\) tetrahedra, therefore appears at third order in
\(J_B/J_A\)~\cite{Harris-1991,Tsunetsugu-2001a,Tsunetsugu-2001b}.  It arises from
virtual processes involving three neighboring \(A\)-tetrahedra connected by three
inter-tetrahedron bonds, which live on triangular faces of the fcc lattice of
\(A\)-tetrahedron centers, as shown in Fig.~\ref{fig:fcc_lattice}.

Not all such triangles contribute.  Each site of the fcc lattice of
\(A\)-tetrahedron centers
belongs to \(24\) elementary triangles, which split into two families of \(12\).
In the first family, the three $A$-tetrahedra are vertices of a common $B$-tetrahedron, the two bonds acting on each of the $A$-tetrahedra enter and leave on the same site. The virtual process is effectively proportional to $\mathbf{S}_i\cdot\mathbf{S}_i = S(S+1)$ on each tetrahedron, so the induced operator is proportional to the identity. This family contributes a further energy shift but does not lift the degeneracy.
Only the second family contributes: there the
three inter-tetrahedron bonds belong to three different \(B\) tetrahedra, and the two bonds that act on each $A$ tetrahedron enter and leave through different sites $i\neq j$. The induced operator contains $\mathbf{S}_i\cdot\mathbf{S}_j$ terms which act nontrivially on the $A$ tetrahedra and split the degeneracy. These triangular terms form the faces of a supertetrahedron in the fcc lattice in the same orientation as the strong tetrahedra in the breathing pyrochlore lattice, as shown in Fig.~\ref{fig:fcc_lattice}. 

Carrying out degenerate perturbation theory for an fcc supertetrahedron containing four such triangles, we obtain, in close analogy with Tsunetsugu's third-order Hamiltonian for the antiferromagnetic case~\cite{Tsunetsugu-2001a,Tsunetsugu-2001b},
\begin{equation}
    \mathcal{H}_{\rm eff,tet}^{(3)}
    =
    -K_3
    \sum_{\Delta\in\mathcal{C}}\left[-1+
    \prod_{\mu=0}^{2}
    \left(
    1-4\,\mathbf{w}_{\mu}\cdot\bm{\tau}_{\Delta_\mu}
    \right)\right],
    \label{eq:model_hamiltonian}
\end{equation}
where
\begin{equation}
    K_3=\frac{1}{384}\,\frac{J_B^3}{J_A^2},
    \label{eq:K3}
\end{equation}
\(\bm{\tau}=\bm{\sigma}/2\) denotes the pseudospin-1/2 operator acting in the
singlet doublet, and \(\Delta_0,\Delta_1,\Delta_2\) are the three pseudospins of
the oriented triangle \(\Delta\), taken in the order fixed by
Eq.~\eqref{eq:oriented_faces}.  
The additive constant -1 is chosen such that the operator is traceless.
Because
\(|\mathbf{w}_\mu|=1\), each factor in Eq.~\eqref{eq:model_hamiltonian} has
eigenvalues \(-1\) and \(+3\).  The three unit vectors in the pseudospin
\(xy\)-plane are
\begin{subequations}
\label{eq:w_vectors}
\begin{align}
    \mathbf{w}_0 &= (1,0,0),\\
    \mathbf{w}_1 &= \left(-\frac{1}{2},\frac{\sqrt{3}}{2},0\right),\\
    \mathbf{w}_2 &= \left(-\frac{1}{2},-\frac{\sqrt{3}}{2},0\right),
\end{align}
\end{subequations}
and the set \(\mathcal{C}\) contains the four oriented triangular faces of an
elementary fcc supertetrahedron of the contributing family,
\begin{equation}
    \mathcal{C}
    =
    \{(C,A,B),(D,B,A),(A,C,D),(B,D,C)\},
    \label{eq:oriented_faces}
\end{equation}
where the global effective operator $\mathcal{H}_{\rm eff}^{(3)}$ sums $\mathcal{H}_{\rm eff,tet}^{(3)}$ over all fcc translations.
The labels \(A,B,C,D\) denote the four
sublattices of the cubic cell of the fcc lattice of \(A\)-tetrahedron centers,
i.e.\ the sites at \(\mathbf{0}\), \(\tfrac{a}{2}(1,1,0)\),
\(\tfrac{a}{2}(1,0,1)\) and \(\tfrac{a}{2}(0,1,1)\).  The orientation of the
triangles matters, because the vector \(\mathbf{w}_{\mu}\) attached to a given
pseudospin is fixed by the pair of sites through which the corresponding physical
tetrahedron participates in the virtual process.  This geometrical dependence is
the characteristic feature of Tsunetsugu's effective
Hamiltonian~\cite{Tsunetsugu-2001a,Tsunetsugu-2001b,Tsunetsugu-2002}.
This Hamiltonian has been checked against an exact treatment on a small spin cluster.

Equation~\eqref{eq:model_hamiltonian} is easiest to read for equatorial product
states.  Writing \(\bm{\tau}\) of every tetrahedron at angle \(\varphi\) in the
\(xy\)-plane, the product in
Eq.~\eqref{eq:model_hamiltonian} evaluates, for a uniform state, to
\begin{equation}
    \prod_{\mu=0}^{2}\left(1-2\cos(\varphi-\varphi_\mu)\right)
    = -2-2\cos 3\varphi ,
    \label{eq:uniform_product}
\end{equation}
with \(\varphi_\mu=0,2\pi/3,4\pi/3\).  This vanishes at the dimer angles
\(\varphi=\pi/3,\pi,5\pi/3\) and reaches its minimum, \(-4\), at the tetramer
angles \(\varphi=0,2\pi/3,4\pi/3\) of Eq.~\eqref{eq:tetramer_angles}.  The sign
of \(K_3\) therefore decides between the two families.  For \(J_B>0\) one has
\(K_3>0\), the energy is lowered by making the product as large as possible, and
one recovers the partially dimerized four-sublattice state of
Tsunetsugu~\cite{Tsunetsugu-2001a,Tsunetsugu-2001b,Tsunetsugu-2002}.  For the
mixed-sign case studied here, \(J_B<0\) gives \(K_3<0\), the product is instead
minimized, and uniform tetramer configurations are selected.  This is the pattern
found in the DMRG correlations of Fig.~\ref{fig:dmrg_tetramer_correlations}.  

The perturbative model cannot be used to locate the phase boundary of the original
spin Hamiltonian.  It serves the more limited purpose of identifying the singlet
pattern selected close to the decoupled-tetrahedron limit.

\subsection{Exact diagonalization of the effective model}

We diagonalize Eq.~\eqref{eq:model_hamiltonian}, with \(K_3<0\) and its magnitude set to unity, on a 32-site pseudospin cluster, corresponding to 128 microscopic spin-\(1/2\) moments in the original breathing-pyrochlore lattice.  Translation symmetries are implemented using the XDiag package~\cite{Wietek,Wietek2}.  The ground state is found in the \(\Gamma\) sector.  A calculation without imposing momentum quantum numbers gives the same ground-state energy, which checks that no lower state is missed in another momentum sector.

The observables in this effective model are pseudospin correlations.  We define
\begin{subequations}
\begin{align}
    C^{ab}
    &=
    \frac{1}{N-1}
    \sum_{i>0}
    \langle
    \tau_0^a\tau_i^b
    \rangle , 
    \quad a,b\in\{x,y,z\} 
    \label{eq:pseudospin_two_point}
\end{align}
\end{subequations}
where $N$ is the number of pseudospins in the lattice and $\tau_0$ is an arbitrary reference pseudospin, and similarly for three-pseudospin correlators.  Since the effective Hamiltonian contains only \(\tau^x\) and \(\tau^y\), any ordered state is expected to lie in the equatorial plane.  For a pseudospin-ferromagnetic product state with every tetrahedron in the same state \(|\varphi\rangle\), one obtains, for \(i\neq j\), $\langle \tau_i^a\tau_j^b\rangle=C^{ab}$ and 
\begin{subequations}
\label{eq:single_theta_correlators}
\begin{align}
    \langle \tau_i^x\tau_j^x\rangle
    &=\frac{1}{4}\cos^2\varphi,\\
    \langle \tau_i^y\tau_j^y\rangle
    &=\frac{1}{4}\sin^2\varphi,\\
    \langle \tau_i^x\tau_j^y\rangle
    &=\frac{1}{8}\sin 2\varphi,\\
    \langle \bm{\tau}_i\cdot\bm{\tau}_j\rangle
    &=\frac{1}{4}.
\end{align}
\end{subequations}
Long-range saturation of \(\langle\bm{\tau}_i\cdot\bm{\tau}_j\rangle\) therefore signals pseudospin ferromagnetism.  In the original spin model, this does not mean magnetic ferromagnetism.  It means that the same tetrahedral singlet pattern is chosen on all \(A\)-tetrahedra.

%------------------------------------------------------------------------
\begin{figure}[t]
    \centering
    \includegraphics[width=\columnwidth]{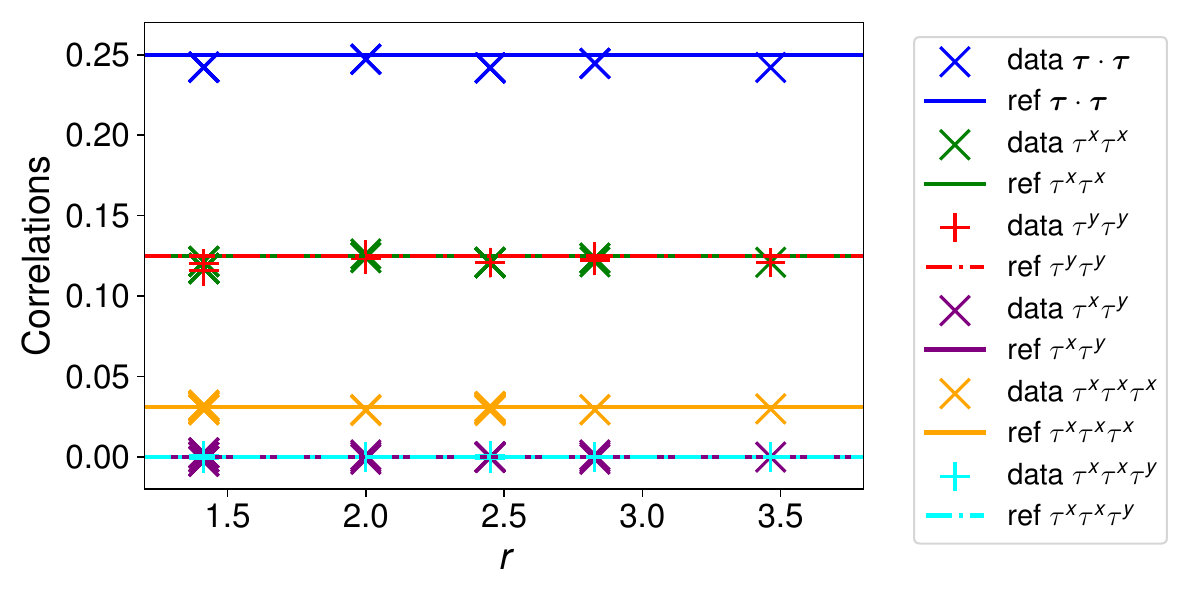}
    \caption{
    Two- and three-pseudospin correlations in the ED ground state of the 32-site effective fcc model.  The horizontal lines show the values expected for the equal-weight superposition of the three pseudospin-ferromagnetic tetramer states with \(\varphi=0,2\pi/3,4\pi/3\).  The near saturation of \(\langle\bm{\tau}_i\cdot\bm{\tau}_j\rangle\) indicates pseudospin-ferromagnetic order.  The values
    \(\langle \tau^x\tau^x\rangle\simeq
    \langle \tau^y\tau^y\rangle\simeq1/8\),
    \(\langle \tau^x\tau^y\rangle\simeq0\), and
    \(\langle \tau^x\tau^x\tau^x\rangle\simeq1/32\)
    identify the finite-cluster ground state as the symmetric superposition of the three tetramer orientations.
    }
    \label{fig:all_correlations_plot_cross}
\end{figure}
%------------------------------------------------------------------------

Fig.~\ref{fig:all_correlations_plot_cross} shows the ED results.  The pseudospin dot product is nearly saturated,
\begin{equation}
    \frac{\langle\bm{\tau}_i\cdot\bm{\tau}_j\rangle}
    {(\langle\bm{\tau}_i\cdot\bm{\tau}_j\rangle)_{\rm sat}}
    \simeq 0.97, 
\end{equation}
and shows no visible decay over the distances available on the cluster.  This indicates pseudospin-ferromagnetic order.  The individual components are
\begin{equation}
    C^{xx}\simeq 0.121,
    \qquad
    C^{yy}\simeq 0.121,
    \qquad
    C^{xy}\simeq 0 .
    \label{eq:ed_two_body_values}
\end{equation}
The first two values are close to \(1/8\), not to the \(\varphi\)-dependent values of a single symmetry-broken product state in Eq.~\eqref{eq:single_theta_correlators}.  This is what one expects on a finite periodic cluster when the exact eigenstate preserves the discrete symmetry: instead of selecting one orientation, it forms a symmetric superposition of symmetry-related pseudospin-ferromagnetic states.

To make this explicit, consider
\begin{equation}
    |\Psi_{\varphi,n}\rangle
    =
    \frac{1}{\sqrt{n}}
    \sum_{m=0}^{n-1}
    \prod_j
    \left|
    \varphi+\frac{2\pi m}{n}
    \right\rangle_j .
    \label{eq:n_cat_state}
\end{equation}
For any \(n>2\), the two-body correlations in the \(xy\)-plane average to
\begin{equation}
    \langle \tau_i^x\tau_j^x\rangle
    =
    \langle \tau_i^y\tau_j^y\rangle
    =
    \frac{1}{8},
    \qquad
    \langle \tau_i^x\tau_j^y\rangle=0 .
\end{equation}
Thus, the two-body correlations show that the ED ground state is not a single equatorial ferromagnet, but they do not by themselves identify how many symmetry-related components are involved.

The three-pseudospin correlations do.  For the state in Eq.~\eqref{eq:n_cat_state},
\begin{subequations}
\label{eq:three_body_cat_values}
\begin{align}
    \langle \tau_i^x\tau_j^x\tau_k^x\rangle
    &=
    \begin{cases}
    \dfrac{\cos^{3}\varphi}{8}, & n=1,\\[0.4em]
    \dfrac{\cos 3\varphi}{32}, & n=3,\\[0.4em]
    0, & \text{otherwise},
    \end{cases}
    \\
    \langle \tau_i^x\tau_j^x\tau_k^y\rangle
    &=
    \begin{cases}
    \dfrac{\cos^{2}\varphi\,\sin\varphi}{8}, & n=1,\\[0.4em]
    \dfrac{\sin 3\varphi}{32}, & n=3,\\[0.4em]
    0, & \text{otherwise}.
    \end{cases}
\end{align}
\end{subequations}
The ED ground state gives
\begin{equation}
    C^{xxx}
    \simeq 3.0\times10^{-2},
    \qquad
    C^{xxy}
    \simeq 0 ,
\end{equation}
which is very close to
\begin{equation}
    \frac{1}{32}=3.125\times 10^{-2},
\end{equation}
as expected for \(n=3\) with \(\varphi\equiv0 \pmod{2\pi/3}\).
The finite-cluster ground state is therefore well described by
\begin{equation}
    |\Psi_{\rm ED}\rangle
    \approx
    \frac{
    |\Theta_0\rangle
    +
    |\Theta_{2\pi/3}\rangle
    +
    |\Theta_{4\pi/3}\rangle
    }{\sqrt{3}},
    \label{eq:three_tetramer_cat}
\end{equation}
where \(|\Theta_{\varphi}\rangle=\prod_j|\varphi\rangle_j\).  In the thermodynamic limit, one of the three components of Eq.~\eqref{eq:three_tetramer_cat} may be selected, corresponding to one of the three tetramer orientations in Eq.~\eqref{eq:tetramer_angles}.  On the finite periodic cluster, the exact eigenstate preserves the symmetry and appears as their equal-weight superposition.

The low-lying spectrum supports this interpretation.  The first excited state in the pseudospin-ferromagnetic manifold lies extremely close to the ground state, with a relative splitting of order \(10^{-6}\).  The separation from the next non-ferromagnetic level is larger by a factor of about \(5\times10^{2}\).  This small splitting should be viewed as a finite-size remnant of the discrete symmetry-breaking manifold, not as a separate scale of the microscopic spin model.  The main conclusion is that \(\mathcal{H}_{\rm eff}^{(3)}\) with \(K_3<0\), appropriate to \(J_B<0\), selects a pseudospin ferromagnet built from the three tetramer states, rather than the dimer pattern favored for the antiferromagnetic perturbation.

\subsection{Interpretation for the microscopic breathing pyrochlore}

The strong-breathing ED gives a direct interpretation of the DMRG correlations.  In the original spin language, pseudospin ferromagnetism means that all antiferromagnetic tetrahedra choose the same singlet texture.  For the mixed-sign perturbation, this texture is one of the three tetramers.  The corresponding microscopic bond correlations are those of Eq.~\eqref{eq:tetramer_bond_correlations}: four antiferromagnetic bonds near \(-1/2\) and two opposite ferromagnetic bonds near \(+1/4\).  This is the same pattern seen in the DMRG ground-state correlations.

This also clarifies how the quantum nonmagnetic regime is related to the classical effective-fcc picture.  In the classical mixed-sign problem, the ferromagnetic tetrahedra behave as composite magnetic moments, and thermal fluctuations select Type-I fcc antiferromagnetic order.  In the strong-breathing quantum problem, the antiferromagnetic tetrahedra remain in their singlet manifold.  The fcc lattice is still present, but the active variables are not magnetic moments.  They are tetrahedral singlet pseudospins.  The ferromagnetic inter-tetrahedron coupling then selects a uniform tetramer orientation in this singlet space.

Thus the nonmagnetic regime found by pf-FRG and DMRG can be understood as a quantum counterpart of the same fcc geometry.  The nearby \(X\)-point spin correlations remain visible in the structure factor, but the local order parameter close to the decoupled-tetrahedron limit is the tetramer energy-density pattern rather than a dipolar moment.

The perturbative treatment is quantitatively controlled only for \(|J_B|/J_A\ll1\), where triplet excitations on the \(A\)-tetrahedra can be projected out.  It should therefore not be used to locate the transition into the \(X=(1,0,0)\) ordered phase.  Its role here is to explain why the nonmagnetic regime next to the decoupled-tetrahedron point naturally develops tetramer correlations, and why those correlations have the bond pattern observed in DMRG.

The same conclusion is obtained at the product-state level.  Minimizing
Eq.~\eqref{eq:model_hamiltonian} over equatorial pseudospin coherent
states on a single supertetrahedron reproduces Tsunetsugu's partially
dimerized four-sublattice solution for \(J_B>0\), whereas for \(J_B<0\)
the minima occur at the tetramer angles \(\varphi=0,2\pi/3,4\pi/3\).  The uniform tetramer
states are among these minima; the exact-diagonalization results of
Sec.~\ref{sec:effective_pseudospin} show that the finite periodic cluster selects the symmetric superposition of the
three uniform tetramer orientations.  Details of this mean-field
comparison and of the correlator identities used to diagnose the ED
state are given in Appendix~\ref{sec:appendix_MF}.

\section{Finite-temperature spin dynamics from high-temperature expansion}
\label{sec:hte_dynamics}

The preceding sections described the mixed-sign breathing pyrochlore from
three complementary viewpoints.  Classical Monte Carlo and SCGA showed
that the dimensionally reduced mixed-sign manifold is selected, by thermal
order-by-disorder, into the Type-I state with ordering wave vector
\(X=(1,0,0)\).  The pf-FRG calculation then showed that quantum
fluctuations suppress this dipolar order over a finite interval near the
decoupled antiferromagnetic-tetrahedron limit, while retaining diffuse
\(X\)-centered spin correlations.  Finally, DMRG and the strong-breathing
pseudospin theory showed that the nonmagnetic regime has a local singlet
texture: the antiferromagnetic tetrahedra develop nearly ideal tetramer
correlations.

We now ask how this physics appears in finite-temperature spin dynamics.
This is not a redundant question, because the dynamical spin structure
factor probes a different part of the problem.  The tetramer order is most
naturally described in the singlet sector, or equivalently in terms of
bond-energy pseudospins on the antiferromagnetic tetrahedra.  By contrast,
neutron scattering measures the dipolar spin correlator.  A single spin
operator takes an isolated tetrahedral singlet into the triplet sector; it
does not directly measure the tetramer pseudospin.  Thus, one should not
expect a sharp ``tetramer mode'' in \(S(\mathbf{k},\omega)\).  The relevant
question is subtler: how do the local magnetic excitations of
antiferromagnetic tetrahedra evolve into the collective \(X\)-centered
fluctuations associated with the mixed-sign effective-fcc manifold?

To address this question, we use the dynamic high-temperature expansion
(Dyn-HTE) of Refs.~\cite{Burkard-2026-PRL,Burkard-2026-PRB}.  The method
starts from the Matsubara spin correlator and expands it in powers of the exchange
couplings divided by temperature.  The high-frequency coefficients of this
expansion are identified with frequency moments of the real-frequency
spectral function, from which the dynamical structure factor is
reconstructed using a continued-fraction representation.  This gives
direct access to finite-temperature real-frequency spectra in the
thermodynamic limit, without analytic continuation of noisy numerical
Matsubara data. When summing over all Matsubara frequencies, the Dyn-HTE result reproduces the standard high-temperature expansion (HTE) \cite{oitmaaSeriesExpansion2006} for the equal-time structure factor.

%------------------------------------------------------------------------
\begin{figure*}[!t]
    \centering
    \includegraphics[width=\textwidth]{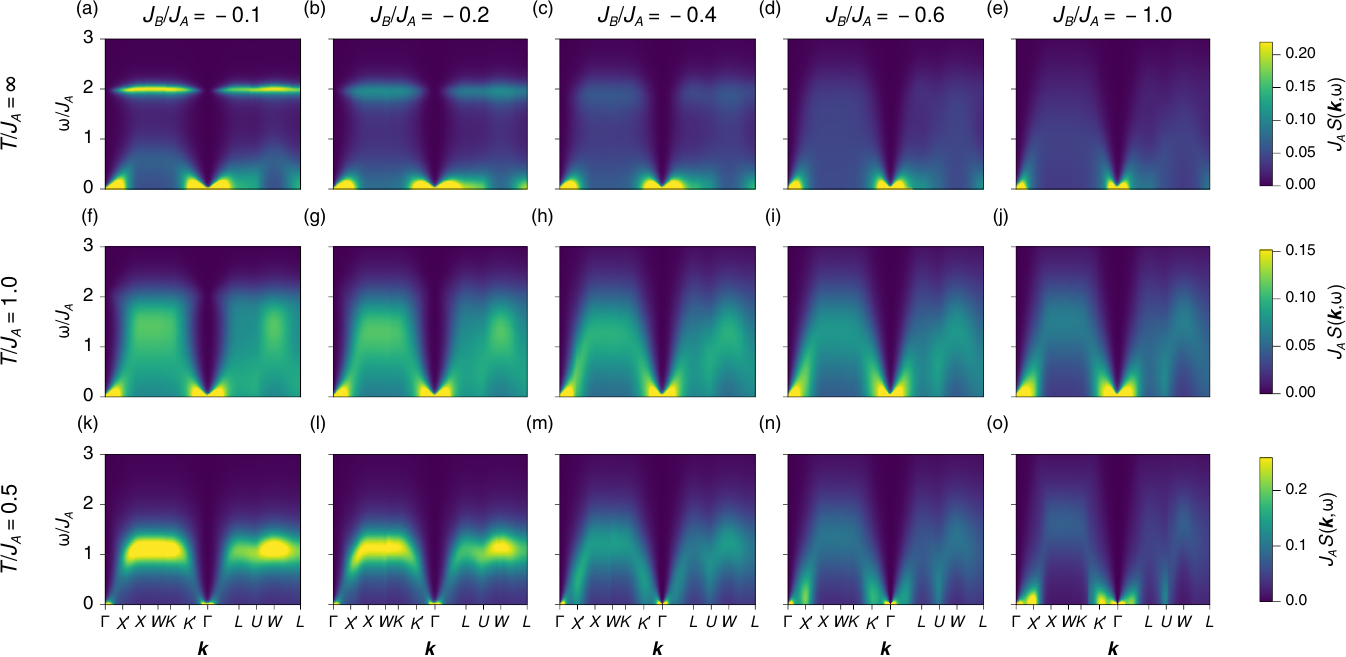}
    \caption{
    Dynamical structure factor \(J_A\,S(\mathbf{k},\omega)\) obtained from
    dynamic high-temperature expansion along the high-symmetry path
    \(\Gamma\)--\(X'\)--\(X\)--\(W\)--\(K\)--\(K'\)--\(\Gamma\)--\(L\)--\(U\)--\(W\)--\(L\).  The columns correspond to
    \(J_B/J_A=-0.1,-0.2,-0.4,-0.6\), and \(-1\), with \(J_A>0\) on the
    antiferromagnetic \(A\)-tetrahedra and \(J_B<0\) on the
    ferromagnetic \(B\)-tetrahedra.  The rows show \(T/J_A=\infty\),
    \(1\), and \(0.5\).  Each row has its own color scale.  At weak
    mixed-sign coupling, the response retains nearly local tetrahedral
    features associated with the multiplet structure of an
    antiferromagnetic tetrahedron.  With increasing \(|J_B|/J_A\),
    these local features broaden and lose their isolation, while
    low-energy weight develops near the \(X\)-type wave vectors of the
    effective-fcc manifold.
}
    \label{fig:hta_suscep_dynamic}
\end{figure*}
%------------------------------------------------------------------------

Throughout this section we take \(J_A>0\) and \(J_B<0\), and quote temperatures in units of \(J_A\).  The results below should be
viewed as finite-temperature precursors of the low-temperature phases
discussed above.  They are not intended to determine the zero-temperature
phase boundary, nor to resolve singlet-sector excitations inside the
tetramer manifold.

\subsection{Local tetrahedron scales in the dynamical response}

A useful reference point is the isolated antiferromagnetic tetrahedron.
For four spin-\(1/2\) moments with exchange \(J_A>0\),
\begin{equation}
    H_{\rm tet}=J_A\sum_{\langle ij\rangle\in {\rm tet}}
    \mathbf{S}_i\cdot\mathbf{S}_j
    =
    \frac{J_A}{2}
    \left[
    \mathbf{S}_{\rm tot}^2-3
    \right] .
\end{equation}
The two singlets have energy \(-3J_A/2\), the triplets have energy
\(-J_A/2\), and the quintet has energy \(3J_A/2\).  The singlet--triplet
gap is therefore \(J_A\), while the triplet--quintet separation is
\(2J_A\).  These local scales are visible in the high-temperature
dynamics when \(|J_B|/J_A\) is small.

Figure~\ref{fig:hta_suscep_dynamic} shows \(J_A\,S(\mathbf{k},\omega)\) along
the path
\(\Gamma\)--\(X'\)--\(X\)--\(W\)--\(K\)--\(K'\)--\(\Gamma\)--\(L\)--\(U\)--\(W\)--\(L\), for five values of
\(J_B/J_A\) and three temperatures.  At \(J_B/J_A=-0.1\), the
infinite-temperature spectrum has a broad, weakly dispersive feature
near \(\omega/J_A\simeq 2\).  This is naturally understood as a remnant
of local tetrahedral transitions involving thermally populated triplet
and quintet states.  It is almost horizontal along the momentum path
because, at this weak inter-tetrahedron coupling, the excitation is still
largely local.

Upon lowering the temperature to \(T/J_A=1\) and \(0.5\), spectral weight
moves toward lower frequencies.  This is also consistent with the
isolated-tetrahedron picture: the lower-temperature ensemble gives more
weight to transitions out of the singlet sector into the triplets, whose
characteristic energy is \(J_A\).  The spectrum is not a set of sharp
delta functions, because finite \(J_B\) hybridizes the tetrahedral
multiplets between neighboring tetrahedra and because the
continued-fraction reconstruction produces a smooth spectral function
from a finite number of moments.  The important feature is not the
precise linewidth, but the persistence of a local tetrahedral magnetic
scale at weak mixed-sign coupling.

This local feature weakens steadily as \(|J_B|/J_A\) grows.  By
\(J_B/J_A=-0.4\), it is no longer a nearly isolated horizontal band.  By
\(J_B/J_A=-0.6\) and \(-1\), the high-energy response has merged into a
broader continuum.  This trend is physically significant.  It shows that
the system is leaving the regime of weakly communicating
antiferromagnetic tetrahedra and entering a regime in which the
inter-tetrahedron mixed-sign correlations govern the magnetic response.
This is the finite-temperature analogue of the crossover seen earlier:
near \(J_B=0^{-}\), the strong-breathing singlet description is natural;
at larger \(|J_B|\), the effective-fcc correlations increasingly control
the spin sector.

\subsection{Growth of low-energy \texorpdfstring{\(X\)}{X}-centered fluctuations}

The second, and more directly collective, feature in
Fig.~\ref{fig:hta_suscep_dynamic} is the growth of low-energy spectral
weight near \(X\)-type momenta.  Already for \(J_B/J_A=-0.2\), the
low-energy response is no longer momentum-independent.  For
\(J_B/J_A=-0.4\) and \(-0.6\), a V-shaped or cusp-like envelope becomes
visible along the high-symmetry path, with enhanced intensity near the
same wave vectors that organize the Type-I state.

This is the dynamical counterpart of the static results obtained earlier
in the paper.  In the classical model, the mixed-sign manifold maps onto
the nearest-neighbor fcc antiferromagnet and has lines of soft modes;
thermal fluctuations select the intersections of these lines, namely the
\(X=(1,0,0)\) points.  In pf-FRG, the nonmagnetic quantum regime still
has diffuse maxima near these same \(X\) points.  In DMRG, the
structure factor remains broad at weak \(|J_B|\), but its weight is
already biased toward \(X\), and the \(X\)-point response grows strongly
upon approaching the ordered phase.  The Dyn-HTE shows that this
same tendency survives at finite temperature as low-energy spectral
weight centered around the Type-I wave vectors.

The temperature dependence reinforces this interpretation.  At
\(T/J_A=\infty\), the response is governed mainly by short-time local
processes.  The \(X\)-centered structure is present only as a broad
modulation.  At \(T/J_A=1\), the low-energy weight becomes more clearly
organized along the path.  At \(T/J_A=0.5\), the same features sharpen
further, especially at intermediate and strong \(|J_B|/J_A\).  This is
what one expects for a finite-temperature approach toward the
Type-I manifold.

These features should not be described as spin waves.  The spectra are
broad, and the temperatures are still in the regime where Dyn-HTE is probing
short- and intermediate-time correlations rather than the asymptotic
ordered-state dynamics.  The calculation therefore shows fluctuating
\(X\)-centered correlations, not the sharp magnon spectrum of a fully
ordered \(\mathrm{AF}(1,0,0)\) phase.

%------------------------------------------------------------------------
\begin{figure*}[!t]
    \centering
    \includegraphics[width=\textwidth]{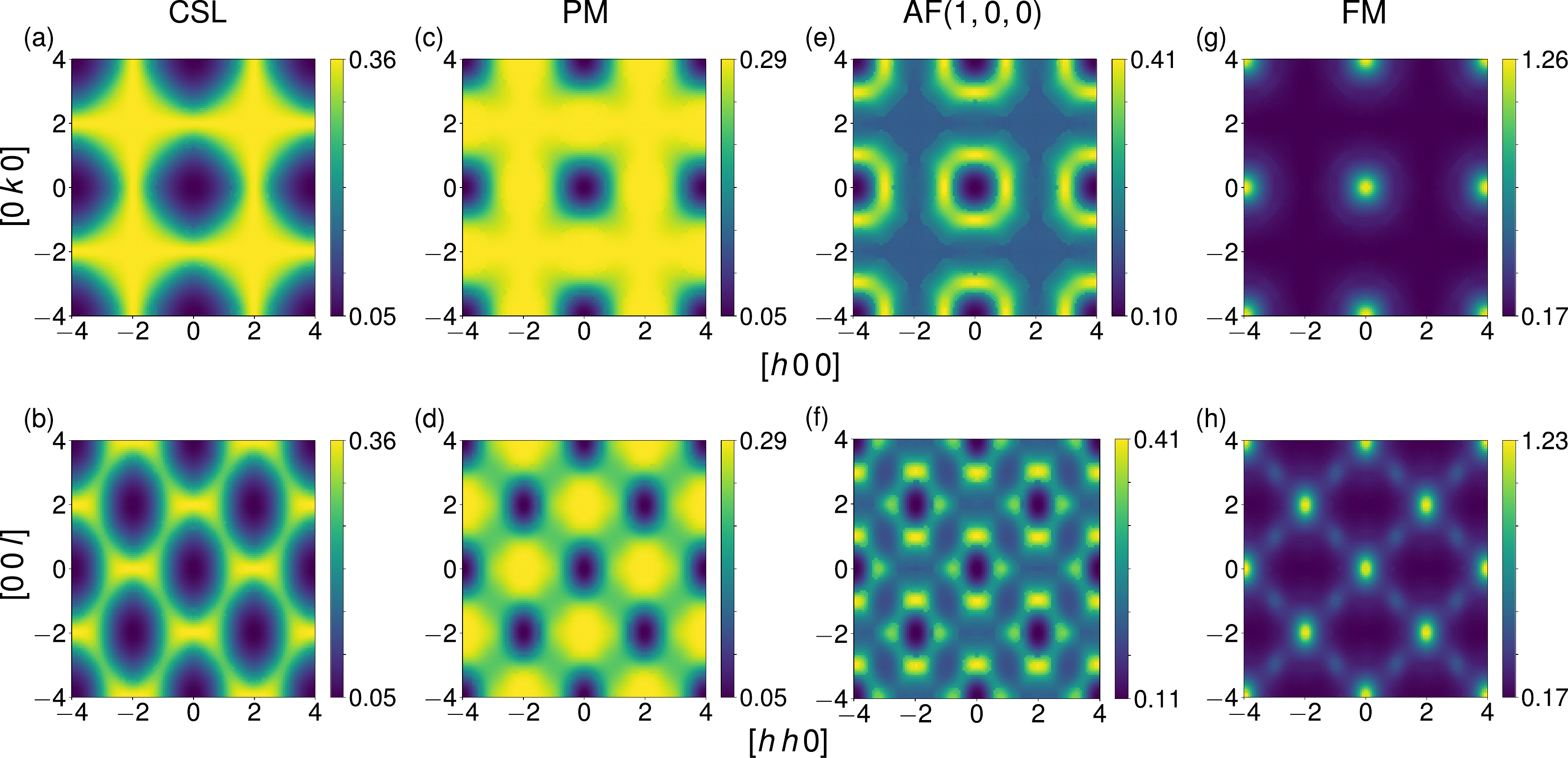}
    \caption{
    Equal-time structure factor \(S(\mathbf{k})\) obtained from
    high-temperature expansion at \(T/J_A=0.4\).  The upper row shows
    the \([hk0]\) plane and the lower row the \([hhl]\) plane.  The
    four columns correspond to representative points in the
    Coulomb-like antiferromagnetic regime (CSL), the mixed-sign
    nonmagnetic regime (PM), the \(\mathrm{AF}(1,0,0)\) regime, and the
    ferromagnetic regime (FM).  The CSL pattern retains rounded
    pinch-point-like structure.  The PM pattern is broad and
    non-Bragg-like but already organized by \(X\)-centered
    effective-fcc correlations.  The \(\mathrm{AF}(1,0,0)\) pattern sharpens the
    same structure toward the Type-I ordering wave vectors.  The FM
    regime is distinguished by strong zone-center intensity.
    The maps show the \((5,5)\) Pad\'e approximant of the \(10\)th-order
    series; at the \(\mathbf{k}\) points where that approximant is singular, the
    average of the neighboring points is displayed instead.
    }
    \label{fig:hte_suscep}
\end{figure*}
%------------------------------------------------------------------------

\subsection{Signatures of tetramer correlations in the spin dynamics}

The DMRG and pseudospin results show that the nonmagnetic regime has
strong tetramer correlations on the antiferromagnetic tetrahedra.  It is
therefore natural to ask whether this tetramer physics appears in
Fig.~\ref{fig:hta_suscep_dynamic}.  The answer is indirect.

In the strong-breathing theory, the tetramer degree of freedom lies
inside the two-dimensional singlet manifold of an antiferromagnetic
tetrahedron.  It is described by a transverse pseudospin, whose direction
specifies a pattern of bond energies.  The dipolar spin operator, by
contrast, connects the singlet manifold to triplet states.  The most
direct tetramer excitations are therefore singlet-sector, or bond-energy,
excitations.  They are not expected to appear as sharp modes in the
two-spin dynamical structure factor.

The absence of a distinct tetramer branch in
Fig.~\ref{fig:hta_suscep_dynamic} therefore does not argue against the
tetramer interpretation; it follows from the observable being computed.  Tetramer order would be more directly probed
by bond-bond correlations, dimer/tetramer structure factors, or
Raman-like singlet-sector response functions.  In the dipolar response,
tetramer physics appears through its consequences: the persistence of
local tetrahedral magnetic scales near the singlet--triplet sector, the
absence of a sharp conventional magnon branch in the nonmagnetic regime,
and the gradual transfer of spectral weight toward the \(X\)-centered
collective fluctuations that become dominant at larger \(|J_B|\).

This is also why the weak-\(|J_B|\) spectra are important.  They retain
a memory of the isolated antiferromagnetic tetrahedra, which are the
building blocks of the tetramerized state.  At the same time, the
low-energy part of the spectrum already knows about the effective-fcc
geometry.  The nonmagnetic regime is therefore not a trivial thermal
paramagnet.  It contains both local tetrahedral singlet physics and
incipient collective correlations at the Type-I wave vectors.

\subsection{Equal-time structure factor and finite-temperature memory of the phase diagram}

The equal-time structure factor in Fig.~\ref{fig:hte_suscep} gives a
complementary view of the same physics.  It integrates over frequency,
\begin{equation}
    S(\mathbf{k})
    =
    \int_{-\infty}^{\infty} d\omega\,S(\mathbf{k},\omega),
\end{equation}
and therefore combines local tetrahedral spectral weight with collective
low-energy correlations.  The result is a finite-temperature map of the
phase diagram in momentum space.

In the CSL column, corresponding to the antiferromagnetic quadrant, the
structure factor has broad bow-tie features.  The pinch points are
rounded, as they must be at finite temperature and within a finite-order
series expansion, but the pattern retains the anisotropic form associated
with the local constraint of vanishing magnetization on each tetrahedron.
This is the finite-temperature counterpart of the Coulomb-like
correlations found in the classical SCGA and in the pf-FRG susceptibility.

The PM column is the most important for the quantum nonmagnetic regime.
There are no Bragg-like singularities, consistent with the absence of
dipolar order in pf-FRG and with the finite-cluster DMRG results.  At
the same time, the response is clearly not momentum independent.  In the
\([hk0]\) plane it has a square-like organization of intensity, while in
the \([hhl]\) plane it forms a broad pattern of maxima and minima tied to
the same \(X\)-centered structure.  This is the equal-time version of the
diffuse \(X\)-point susceptibility found by pf-FRG.  It shows that the
PM regime retains the fingerprint of the effective-fcc manifold, even
though it does not develop a static dipolar moment.

In the \(\mathrm{AF}(1,0,0)\) column, the same structure is sharpened.  The
weight is more strongly concentrated near the Type-I wave vectors, and
the \([hhl]\) cut develops more pronounced bright spots connected by
ridges of intensity.  This should be read as the finite-temperature
precursor of Type-I order.  HTE does not produce the Bragg delta
function of the ordered phase, but it does show the same momentum-space
selection that appears in the classical Monte Carlo, pf-FRG, and DMRG
calculations.

The FM column provides a useful control.  Here the response is dominated
by the zone centers, and its overall intensity is much larger.  This
confirms that the ferromagnetic exchange on the \(B\)-tetrahedra does
not by itself imply ferromagnetic correlations in the spin structure
factor.  In the mixed-sign PM and \(\mathrm{AF}(1,0,0)\) regimes, the
ferromagnetic \(B\)-tetrahedra act instead as part of an effective
frustrated fcc network.  Only when both tetrahedral sublattices favor
parallel alignment does the response become truly ferromagnetic.

\subsection{Range of validity of the expansion}

The interpretation of Figs.~\ref{fig:hta_suscep_dynamic} and
\ref{fig:hte_suscep} rests on the range of validity of the expansion.  The order
of the series, the moments retained, and the parameters of the continued-fraction
reconstruction are collected in Appendix~\ref{app:dynhte}.
The series for the equal-time structure factor is exact to $10$th order and
can therefore reach slightly lower temperatures than the dynamic data.  The range
of validity is defined as the region in which the results obtained from different
nonsingular Pad\'e approximants agree with one another.

The broad redistribution of spectral weight, the transfer from local
tetrahedral scales to low-energy \(X\)-centered fluctuations, and the
equal-time momentum patterns are robust qualitative outputs.  These are
set mainly by low frequency moments and short-range correlations, which
are precisely the quantities controlled by the high-temperature
expansion.

More delicate features should be treated cautiously.  The finite order
of the series and the resummation procedure limit the reach to low
temperature.  Close to an ordering transition, HTE cannot generate a
true Bragg singularity or an infinite correlation length.  Thus the
\(\mathrm{AF}(1,0,0)\) column of Fig.~\ref{fig:hte_suscep} should be interpreted
as a finite-temperature precursor of order, not as an order parameter.

The continued-fraction reconstruction also uses only a finite number of
frequency moments.  It can broaden spectral features and smooth
thresholds.  Consequently, the precise line shape and linewidth of a
feature in \(S(\mathbf{k},\omega)\) are less significant than the energy
scale at which weight appears and the momentum dependence of that
weight.  In particular, weak narrow branches, small gaps, or singlet
bound states should not be inferred from these spectra.

Finally, the observable itself is a dipolar spin correlator.  It is not
the natural correlator for the tetramer order parameter.  The absence of
a sharp tetramer feature in \(S(\mathbf{k},\omega)\) is therefore expected.
A direct dynamical probe of tetramer physics would require bond-bond,
dimer, tetramer, or Raman-like correlation functions.  The spin
structure factor instead shows the magnetic environment in which the
tetramerized nonmagnetic regime lives: local tetrahedral magnetic
excitations at weak mixed-sign coupling and diffuse \(X\)-centered
fluctuations as the effective-fcc correlations grow.

The HTE results therefore place the finite-temperature response in the
same physical picture as the earlier sections.  The local
antiferromagnetic tetrahedra set the high-energy magnetic scale.  The
mixed-sign effective-fcc manifold controls the low-energy momentum
dependence.  The tetramer tendency explains why the weak-coupling
nonmagnetic regime is not a conventional magnet with a small ordered
moment, even though its spin correlations already point toward the
neighboring \(X=(1,0,0)\) phase.

The discussion so far has been deliberately model-centered.  We have
shown how the mixed-sign breathing-pyrochlore Hamiltonian generates, at
successive levels, an effective-fcc classical manifold, an \(X\)-centered
quantum spin response, and tetramer correlations in the strong-breathing
singlet sector.  We now turn to the materials motivation for this
description.  In chromium thiospinels the microscopic exchange network is
more complicated than the nearest-neighbor model of Eq.~\eqref{eq:breathing_hamiltonian},
but a coarse graining over Cr$_4$ tetrahedra leads naturally to effective
fcc models.  This provides a concrete setting in which the finite-temperature
signatures discussed above can be sought.

\section{Effective-fcc description of the breathing chromium thiospinels}
\label{sec:materials}

We now connect the model results to the breathing chromium thiospinels.  These
materials are not described, in full microscopic detail, by the
nearest-neighbor breathing-pyrochlore Hamiltonian alone.  The Cr--Cr exchange
network contains further-neighbor couplings, and the balance between direct
antiferromagnetic exchange and indirect ferromagnetic exchange depends
sensitively on bond length, bond angle, ligand chemistry, and structural
distortion.  Nevertheless, the calculations below show that several breathing
chromium thiospinels admit a much simpler description after coarse graining over
Cr\(_4\) tetrahedra: the tetrahedral units form an effective fcc lattice with a
dominant antiferromagnetic nearest-neighbor coupling.

This observation gives the materials part of the paper a direct connection to the
rest.  The mixed-sign breathing-pyrochlore model led us, first classically and
then in the finite-temperature response, to an effective-fcc manifold with strong
\(X\)-centered correlations.  The chromium thiospinels show that this fcc
description is not only a formal feature of the ideal model.  It can also emerge
from realistic exchange parameters in materials whose microscopic pyrochlore
network is considerably more complicated.  We then ask whether the effective-fcc
placement of each compound agrees with what is known experimentally about its
magnetic ground state.  For one compound the comparison can be made directly, and
the mapping is borne out; for the others it either accounts for the observed
behavior, delimits its own range of validity, or leads to a prediction that has
not yet been tested.

\subsection{Microscopic exchange parameters}
\label{sec:materials_micro}

\begin{figure*}[!t]
\centering
\includegraphics[width=\textwidth]{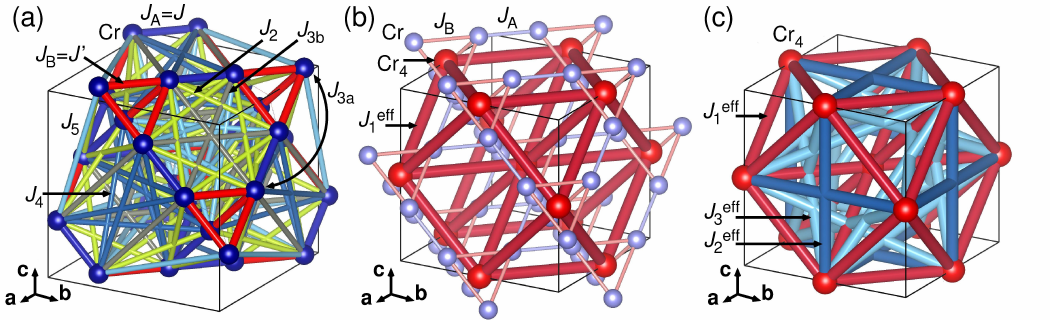}
\caption{
From the microscopic Cr exchange network to the effective fcc model.
(a) The Cr sublattice of a breathing chromium spinel, showing the seven exchange
paths of Table~\ref{tab:materials_couplings}: the two nearest-neighbor breathing
couplings \(J\equiv J_A\) on the small tetrahedra and \(J'\equiv J_B\) on the
large tetrahedra, and the further-neighbor couplings \(J_2\), \(J_{3a}\),
\(J_{3b}\), \(J_4\), and \(J_5\).  The third-neighbor bonds \(J_{3a}\) and
\(J_{3b}\) lie at the same Cr--Cr distance and are distinguished by their local
environment: the two sites of a \(J_{3a}\) bond have a common nearest neighbor
lying between them, so that the bond is spanned by two nearest-neighbor steps,
whereas the two sites of a \(J_{3b}\) bond have none.  Each type has coordination
number six.
(b) Coarse graining over Cr\(_4\) units: each ferromagnetically coupled \(J_B\)
tetrahedron is replaced by a single site carrying the composite moment
\(\mathbf{T}_R=\sum_{i\in R}\mathbf{S}_i\) of Sec.~\ref{sec:materials_fcc}.  The centers of these
units form an fcc lattice whose nearest-neighbor coupling is \(J_1^{\rm eff}\).
(c) The resulting effective fcc model, showing its first three neighbor shells
\(J_1^{\rm eff}\), \(J_2^{\rm eff}\), and \(J_3^{\rm eff}\), with \(12\), \(6\),
and \(24\) neighbors respectively.  The colors of \(J_4\) and \(J_5\) in (a) are
carried over to \(J_2^{\rm eff}\) and \(J_3^{\rm eff}\) in (c) to indicate the
correspondence established by Eq.~\eqref{eq:projection}: \(J_2^{\rm eff}\) is
generated entirely by \(J_4\) and \(J_3^{\rm eff}\) entirely by \(J_5\), while no
microscopic coupling up to \(J_5\) generates a fourth-neighbor effective
coupling.
}
\label{fig:fcc_mapping}
\end{figure*}

We first summarize the microscopic exchange couplings obtained for the breathing
chromium thiospinels.  The calculations were performed within the generalized
gradient approximation plus Hubbard \(U\) (GGA\(+U\)), with a Hund's coupling
\(J_{\rm H}=0.72\)~eV, using the energy-mapping procedure of
Ref.~\cite{ghosh19a}.  Every calculation reported here was performed on a cubic
\(F\bar{4}3m\) structure.  This applies to the low-temperature entries of
Table~\ref{tab:materials_couplings} as well: they use cubic refinements at the
stated temperature, not the orthorhombic ground-state structures of those
compounds that distort.  No \(Imm2\) structure was treated, which sets a limit on
the analysis that we return to in Sec.~\ref{sec:materials_experiment}.  The
exchange constants in
Table~\ref{tab:materials_couplings} are quoted for classical Cr\(^{3+}\) spins of
length \(S=3/2\), without double counting of bonds, in the convention that
positive \(J\) is antiferromagnetic.  Following the materials literature, we
denote the two inequivalent nearest-neighbor breathing-pyrochlore exchanges by
\(J\) and \(J'\), with \(J\equiv J_A\) on the small (\(A\)) tetrahedra and
\(J'\equiv J_B\) on the large (\(B\)) tetrahedra, in the notation of
Eq.~\eqref{eq:breathing_hamiltonian}; \(J_2\), \(J_{3a}\), \(J_{3b}\), \(J_4\),
and \(J_5\) denote the further-neighbor exchanges on the Cr pyrochlore network,
with \(J_{3a}\) and \(J_{3b}\) the two symmetry-inequivalent third-neighbor bonds
at the same Cr--Cr distance.  All seven exchange paths are drawn in
Fig.~\ref{fig:fcc_mapping}(a).

The table illustrates why a purely nearest-neighbor description is not sufficient
as a materials model.  In four of the five compounds the two nearest-neighbor
exchanges have opposite signs, as expected for a competition between
antiferromagnetic direct exchange and ferromagnetic \(90^\circ\) superexchange
channels, and in every case \(J'<0\) is ferromagnetic.  At the same time, the
further-neighbor terms, especially the third-neighbor couplings \(J_{3a}\) and
\(J_{3b}\), are not negligible; in \ce{LiInCr4S8} and \ce{CuAlCr4S8} they are in
fact larger than \(J\) itself.  The microscopic Hamiltonian therefore contains more
structure than the ideal model studied in
Secs.~\ref{sec:model_classical}--\ref{sec:hte_dynamics}.  The useful question is
not whether the materials realize Eq.~\eqref{eq:breathing_hamiltonian} term by
term, but whether the dominant correlations can be reorganized in terms of
tetrahedral units.

\begin{table*}[!t]
\caption{
Microscopic exchange couplings of the breathing chromium thiospinels, calculated
within GGA\(+U\) at \(J_{\rm H}=0.72\)~eV, quoted in Kelvin for classical
Cr\(^{3+}\) spins of length \(S=3/2\), without double counting of bonds.  Positive
values are antiferromagnetic.  The two nearest-neighbor breathing-pyrochlore
exchanges are \(J\equiv J_A\) (small tetrahedra) and \(J'\equiv J_B\) (large
tetrahedra); \(J_2,J_{3a},J_{3b},J_4,J_5\) are the further-neighbor exchanges of
the Cr pyrochlore network.  ``RT'' denotes room temperature; the numerical entries
in the \(T\) column are in Kelvin.  Each entry is obtained by interpolating the
calculated \(U\) dependence of the couplings to the value of \(U\) at which the
computed Curie--Weiss temperature matches the measured one.  Values are quoted to
three significant figures, which is commensurate with the systematic uncertainty
of the energy mapping.
}
\label{tab:materials_couplings}
\begin{ruledtabular}
\begin{tabular}{lccccccccc}
Material & \(T\) & \(U\) & \(J\) & \(J'\) & \(J_2\) & \(J_{3a}\) & \(J_{3b}\) & \(J_4\) & \(J_5\) \\
 & (K) & (eV) & (K) & (K) & (K) & (K) & (K) & (K) & (K) \\
\hline
\ce{LiGaCr4S8} & RT & 1.769 & \(-4.69\)  & \(-13.0\) & \(1.29\)  & \(5.97\) & \(2.80\) & \(-0.452\) & \(0.539\) \\
\ce{LiGaCr4S8} & 10 & 1.729 & \(-7.96\)  & \(-10.3\) & \(1.35\)  & \(6.07\) & \(2.85\) & \(-0.453\) & \(0.532\) \\
\ce{LiInCr4S8} & RT & 1.677 & \(2.94\)   & \(-28.9\) & \(0.660\) & \(5.22\) & \(2.24\) & \(-0.388\) & \(0.533\) \\
\ce{CuInCr4S8} & RT & 1.126 &  \(9.6\)   & \(-22.4\) & \(2.27\)  & \(6.80\) & \(4.53\) & \(-0.907\) & \(0.620\)\\
\ce{CuInCr4S8} & 30 & 1.850 & \(27.2\)   & \(-30.0\) & \(1.34\)  & \(5.02\) & \(3.47\) & \(-0.874\) & \(0.375\) \\
\ce{CuAlCr4S8} & RT & 1.633 & \(1.92\)   & \(-4.21\) & \(1.05\)  & \(5.86\) & \(3.30\) & \(-0.990\) & \(-0.0752\) \\
\ce{CuAlCr4S8} & 12 & 1.684 & \(1.96\)   & \(-4.56\) & \(1.08\)  & \(5.89\) & \(3.32\) & \(-0.972\) & \(-0.0608\) \\
\ce{CuGaCr4S8} & RT & 1.712 & \(11.6\)   & \(-11.0\) & \(2.04\)  & \(5.74\) & \(4.27\) & \(-1.08\)  & \(0.272\)
\end{tabular}
\end{ruledtabular}
\end{table*}

\begin{figure}[t]
\centering
\includegraphics[width=\columnwidth]{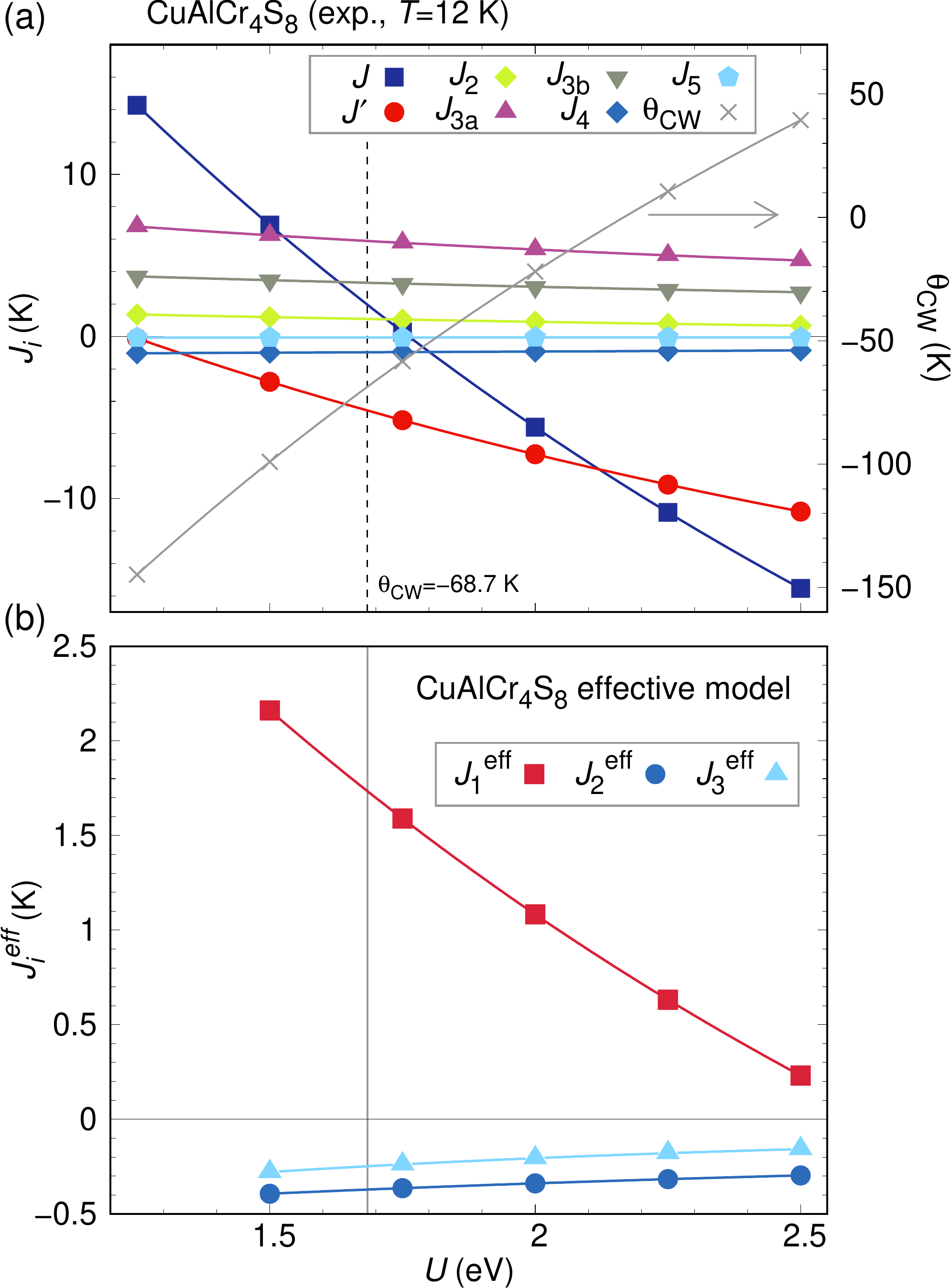}
\caption{
Exchange couplings of \ce{CuAlCr4S8}, obtained for the experimental
low-temperature structure at \(T=12\)~K, calculated within GGA\(+U\) at
\(J_{\rm H}=0.72\)~eV using a \(6\times6\times6\) \(k\)-point mesh in a supercell
containing 16 Cr\(^{3+}\) sites.  (a) Microscopic Cr--Cr exchange couplings of
Table~\ref{tab:materials_couplings} as a function of \(U\), together with the
calculated Curie--Weiss temperature (gray crosses, right axis).  The vertical
line marks the value of \(U\) at which \(\theta_{\rm CW}\) matches the measured
value.  (b) The corresponding effective-fcc couplings of
Table~\ref{tab:materials_fcc_couplings}, obtained from a tenfold supercell
containing ten Cr\(_4\) tetrahedra.  The colors of \(J_2^{\rm eff}\) and
\(J_3^{\rm eff}\) in (b) match those of \(J_4\) and \(J_5\) in (a), the
microscopic couplings that generate them through
Eq.~\eqref{eq:projection}.
}
\label{fig:cas_exchange}
\end{figure}

\begin{figure}[t]
\centering
\includegraphics[width=\columnwidth]{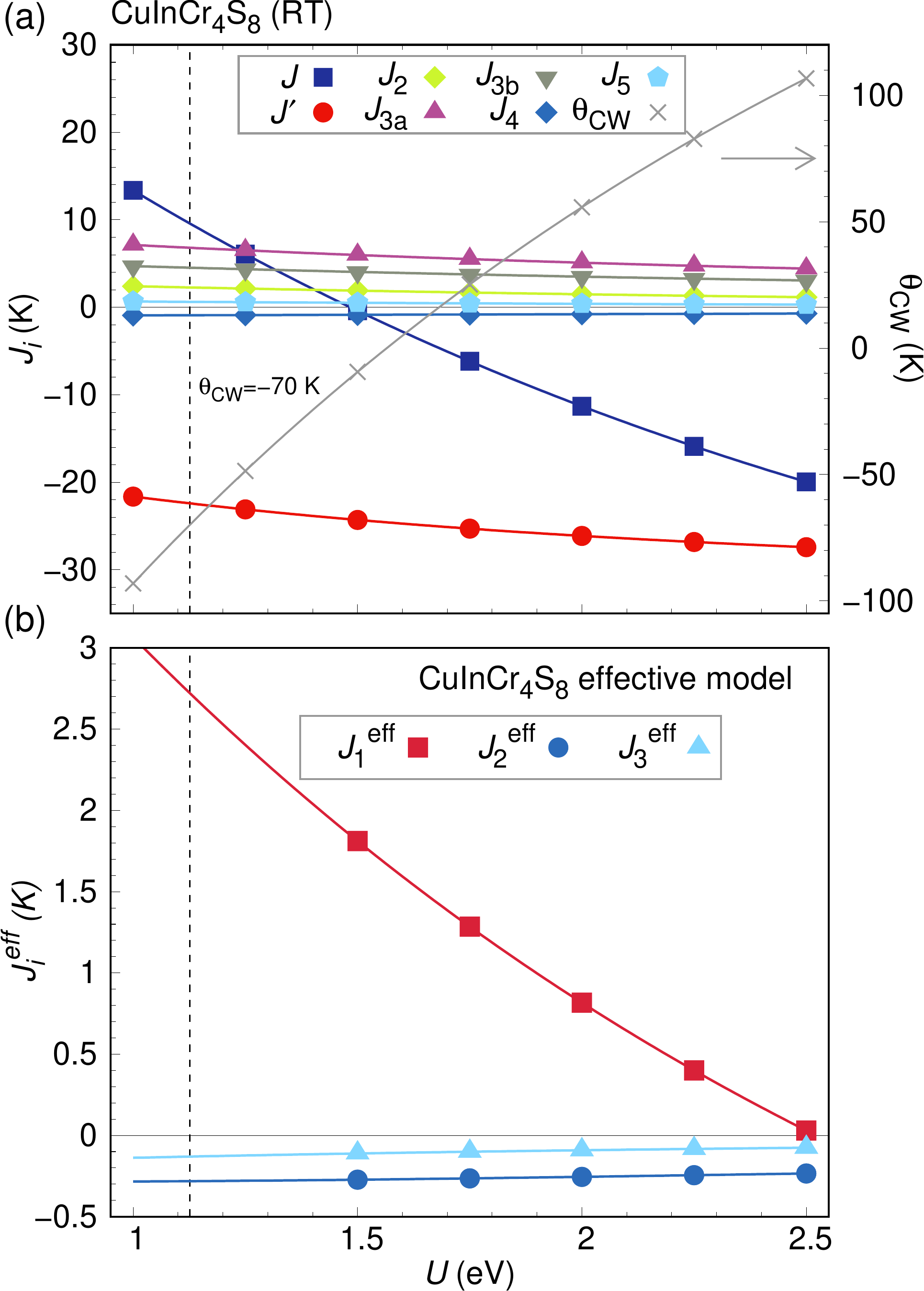}
\caption{
Exchange couplings of \ce{CuInCr4S8}, obtained for the room-temperature
structure, calculated within GGA\(+U\) at \(J_{\rm H}=0.72\)~eV using a
\(6\times6\times6\) \(k\)-point mesh in a supercell containing 16
Cr\(^{3+}\) sites.  Panels and symbols are as in
Fig.~\ref{fig:cas_exchange}.
}
\label{fig:cis_exchange}
\end{figure}

Figures~\ref{fig:cas_exchange} and~\ref{fig:cis_exchange} show two representative
examples, \ce{CuAlCr4S8} and \ce{CuInCr4S8}.  The upper panels show the
microscopic Cr--Cr exchange parameters as a function of \(U\).  The vertical line
marks the value of \(U\) fixed by matching the calculated Curie--Weiss
temperature to the measured one.  The lower panels show the corresponding
effective-fcc couplings obtained after coarse graining over Cr\(_4\) tetrahedra.

The two examples make clear why the coarse-grained description is useful.  The
microscopic exchanges vary strongly with \(U\), and several couplings contribute
at a comparable level.  After projecting the problem onto tetrahedral units,
however, the effective model becomes much more compact.  In both materials the
leading term is an antiferromagnetic \(J_1^{\rm eff}\) on the fcc lattice, while
the further-neighbor effective couplings are one to two orders of magnitude
smaller.  This is the sense in which the materials realize effective-fcc physics,
even when the microscopic exchange network is not itself a simple
nearest-neighbor model.

\subsection{Effective fcc Hamiltonian}
\label{sec:materials_fcc}

Because every \(J'\) is ferromagnetic and, in most compounds, larger in magnitude
than the remaining couplings, the four Cr spins of a large tetrahedron are
strongly locked together.  It is then natural to treat each large tetrahedron as
a rigid composite moment
\(\mathbf{T}_R=\sum_{i\in R}\mathbf{S}_i\), whose centers form an fcc lattice.
This substitution is illustrated in Fig.~\ref{fig:fcc_mapping}(b), and the
neighbor shells of the resulting fcc network in Fig.~\ref{fig:fcc_mapping}(c).
The effective model is
\begin{equation}
    H_{\rm eff}
    =
    \sum_n
    J_n^{\rm eff}
    \sum_{\langle RR'\rangle_n}
    \mathbf{T}_R\cdot\mathbf{T}_{R'} ,
    \label{eq:materials_effective_fcc}
\end{equation}
where \(R\) labels the Cr\(_4\) tetrahedral units and \(\langle RR'\rangle_n\)
denotes the \(n\)-th-neighbor shell on the fcc lattice.  The ratios
\(J_n^{\rm eff}/J_1^{\rm eff}\), rather than the absolute normalization of
\(\mathbf{T}_R\), determine where a material lies in the effective-fcc phase
diagram; for reference, four fully polarized Cr\(^{3+}\) moments give
\(|\mathbf{T}_R|=6\).

This effective-fcc model should be kept distinct from the spin-\(1/2\) singlet
pseudospin Hamiltonian of Sec.~\ref{sec:effective_pseudospin}.  There, the active
degree of freedom is the two-dimensional singlet manifold of an
antiferromagnetic tetrahedron, and pseudospin order describes a dimer or tetramer
pattern of bond energies.  Here the degrees of freedom are coarse-grained
magnetic moments of ferromagnetic Cr\(_4\) tetrahedra in \(S=3/2\) compounds.
The common element is the fcc arrangement of tetrahedral units.

The coarse graining can also be carried out analytically.  Replacing \(\mathbf{S}_i\to\mathbf{T}_R/4\) for every spin of a rigid tetrahedron
and summing the microscopic couplings that connect a given pair of tetrahedra, an
enumeration of the pyrochlore neighbor shells gives
\begin{subequations}
\label{eq:projection}
\begin{align}
    J_1^{\rm eff}
    &=
    \tfrac{1}{16}\left(J+4J_2+2J_{3a}+2J_{3b}+2J_4+4J_5\right),
    \\
    J_2^{\rm eff}&=\tfrac{1}{4}J_4,
    \qquad
    J_3^{\rm eff}=\tfrac{1}{8}J_5,
    \qquad
    J_{n\ge4}^{\rm eff}=0 .
\end{align}
\end{subequations}
The first line generalizes the nearest-neighbor cluster coupling used in
Refs.~\cite{Pokharel-2020,Gao-2021}, which truncates at \(J_{3b}\); evaluated on
Table~\ref{tab:materials_couplings} it reproduces the directly computed
\(J_1^{\rm eff}\) of Table~\ref{tab:materials_fcc_couplings} to within
\(3\)--\(25\%\) for every entry, which is a useful consistency check on the
supercell energy mapping.  Note that Eq.~\eqref{eq:projection} generates no
fourth-neighbor fcc coupling at all, simply because no microscopic Cr--Cr bond up
to \(J_5\) connects fourth-neighbor tetrahedra.  We return to this point below.

The effective exchanges, their dimensionless ratios, and the resulting phase
assignment are collected in Table~\ref{tab:materials_fcc_couplings}.  These
entries are not obtained from Eq.~\eqref{eq:projection}; they come from a second,
independent energy mapping in which Eq.~\eqref{eq:materials_effective_fcc},
truncated at the fourth fcc neighbor shell, is fitted directly to GGA\(+U\)
total energies of a tenfold supercell containing ten Cr\(_4\) tetrahedra, with the
four Cr spins of every tetrahedron chosen parallel so that each unit enters as a
single moment \(\mathbf{T}_R\).  The fit thus returns four numbers,
\(J_1^{\rm eff}\) to \(J_4^{\rm eff}\), and the agreement between its
\(J_1^{\rm eff}\) and the projection of Eq.~\eqref{eq:projection} is the
consistency check quoted above.  Several
compounds lie close to the nearest-neighbor antiferromagnetic fcc limit: for
\ce{LiGaCr4S8} at room temperature, \ce{LiInCr4S8}, and \ce{CuInCr4S8} at
\(30\)~K, both
\(|J_2^{\rm eff}/J_1^{\rm eff}|\) and \(|J_3^{\rm eff}/J_1^{\rm eff}|\) are below
\(0.03\).  Thus, despite substantial differences in their microscopic exchange
paths, these compounds are organized at the tetrahedral level by a dominant
antiferromagnetic nearest-neighbor fcc coupling.  The low-temperature structures
show that the effective-fcc parameters can nevertheless move appreciably with
small structural changes: \ce{LiGaCr4S8} shifts strongly toward positive
\(J_2^{\rm eff}\), while \ce{CuAlCr4S8} moves toward negative \(J_2^{\rm eff}\)
and negative \(J_3^{\rm eff}\); conversely, the room-temperature structure of
\ce{CuInCr4S8} carries appreciably larger negative \(J_2^{\rm eff}\) and
\(J_3^{\rm eff}\) than its \(30\)~K structure.

\begin{table*}[!t]
\caption{
Effective fcc couplings obtained by coarse graining over Cr\(_4\) tetrahedra,
calculated within GGA\(+U\) at \(J_{\rm H}=0.72\)~eV and quoted in Kelvin for
classical Cr\(^{3+}\) spins of length \(S=3/2\), without double counting of bonds.
A dash indicates a coupling that was not determined.  The dimensionless ratios are
the quantities that fix the position of a compound in
Fig.~\ref{fig:materials_fcc_phase_diagram}.  The last two columns give the
degeneracy-lifting combination \(s=J_2^{\rm eff}-4J_3^{\rm eff}\) of
Eq.~\eqref{eq:selector}, in units of \(J_1^{\rm eff}\), and the classical ordering
wave vector it selects.  Entries with \(|s|<0.035\,J_1^{\rm eff}\), marked with an
asterisk, lie inside the window in which thermal order-by-disorder rather than
exchange decides the ordering wave vector, and are then driven to \(X\); see
Eq.~\eqref{eq:obd_criterion}.  ``RT'' denotes room temperature.
}
\label{tab:materials_fcc_couplings}
\begin{ruledtabular}
\begin{tabular}{lccccccccc}
Material & \(T\) & \(U\) &
\(J_1^{\rm eff}\) & \(J_2^{\rm eff}\) & \(J_3^{\rm eff}\) & \(J_4^{\rm eff}\) &
\(J_2^{\rm eff}/J_1^{\rm eff}\) & \(J_3^{\rm eff}/J_1^{\rm eff}\) &
\(s/J_1^{\rm eff}\ \rightarrow\ \mathbf{q}_{\rm ord}\) \\
 & (K) & (eV) & (K) & (K) & (K) & (K) & & & \\
\hline
\ce{LiGaCr4S8} & RT & 1.769 & \(1.24\)  & \(-0.0350\) & \(0.0125\)   & \(0.0642\) & \(-0.028\) & \(0.010\)  & \(-0.069\ \rightarrow\ X\) \\
\ce{LiGaCr4S8} & 10 & 1.729 & \(0.826\) & \(0.215\)   & \(0.00494\)  & \(0.0270\) & \(0.260\)  & \(0.006\)  & \(+0.236\ \rightarrow\ W\) \\
\ce{LiInCr4S8} & RT & 1.677 & \(1.43\)  & \(-0.0372\) & \(0.00542\)  & \(0.0624\) & \(-0.026\) & \(0.004\)  & \(-0.041\ \rightarrow\ X\) \\
\ce{CuInCr4S8} & RT & 1.126 & \(2.72\)  & \(-0.2815\) & \(-0.1300\) & \(-\) & \(-0.103\) & \(-0.048\)  & \(+0.088\ \rightarrow\ W\) \\
\ce{CuInCr4S8} & 30 & 1.850 & \(3.33\)  & \(-0.0493\) & \(-0.0329\)  & \(0.106\)  & \(-0.015\) & \(-0.010\) & \(+0.025^{*}\rightarrow\ X\) \\
\ce{CuAlCr4S8} & RT & 1.633 & \(1.61\)  & \(-0.137\)  & \(-0.00927\) & \(0.0566\) & \(-0.085\) & \(-0.006\) & \(-0.062\ \rightarrow\ X\) \\
\ce{CuAlCr4S8} & 12 & 1.684 & \(1.73\)  & \(-0.372\)  & \(-0.248\)   & --         & \(-0.214\) & \(-0.143\) & \(+0.358\ \rightarrow\ W\) \\
\ce{CuGaCr4S8} & RT & 1.712 & \(2.78\)  & \(0.0663\)  & \(-0.0864\)  & \(0.145\)  & \(0.024\)  & \(-0.031\) & \(+0.148\ \rightarrow\ W\)
\end{tabular}
\end{ruledtabular}
\end{table*}

\subsection{Degeneracy lifting on the fcc soft line}
\label{sec:materials_selector}

Before placing the compounds in a phase diagram we identify which combination of
couplings actually controls the selection.  On the fcc lattice, the
nearest-neighbor antiferromagnet is degenerate along the entire soft line
\(\mathbf{q}=(1,\delta,0)\) of Eq.~\eqref{eq:softline}.  Evaluating the Fourier
transform of Eq.~\eqref{eq:materials_effective_fcc} along that line gives the
exact result
\begin{align}
    J^{\rm eff}(1,\delta,0)
    &=
    -4J_1^{\rm eff}+2J_2^{\rm eff}+8J_3^{\rm eff}-4J_4^{\rm eff}
    \nonumber\\
    &\quad
    +\,4\left(J_2^{\rm eff}-4J_3^{\rm eff}+4J_4^{\rm eff}\right)\cos^2(\pi\delta) .
    \label{eq:softline_energy}
\end{align}
The nearest-neighbor coupling drops out of the \(\delta\) dependence entirely, as
it must.  The degeneracy of the line is therefore lifted by the single
combination
\begin{equation}
    s \equiv J_2^{\rm eff}-4J_3^{\rm eff}+4J_4^{\rm eff} ,
    \label{eq:selector}
\end{equation}
and by nothing else: \(s<0\) selects \(\delta=0\), i.e.\ the Type-I state at
\(X=(1,0,0)\); \(s>0\) selects \(\delta=1/2\), i.e.\ the Type-III state at
\(W=(1,\tfrac{1}{2},0)\); and \(s=0\) leaves the entire line degenerate.  The
locus \(s=0\) is the only phase boundary of the classical fcc \(J_1\)--\(J_2\)--\(J_3\)
model that passes through the nearest-neighbor point, and it is the subextensively
degenerate manifold identified in Ref.~\cite{Balla2020}.  No incommensurate,
Type-II or ferromagnetic phase is a competitor anywhere in the parameter range
relevant here, since those phases require \(|J_2^{\rm eff}|\gtrsim0.3J_1^{\rm eff}\).

Although the selection is unambiguous, it is energetically very weak.  The energy
difference per tetrahedral moment between the two commensurate endpoints is
\(2|s|\,|\mathbf{T}|^2\), i.e.\ a fraction \(|s|/(2J_1^{\rm eff})\) of the
nearest-neighbor energy scale, and for the compounds of
Table~\ref{tab:materials_fcc_couplings} this fraction lies between \(0.4\%\) and
\(6\%\).  All of these materials therefore sit inside a nearly degenerate remnant
of the fcc soft-mode manifold, where the ordering wave vector is fixed by a
balance among small further-neighbor terms rather than by the dominant exchange.

Thermal order-by-disorder competes directly with this weak exchange selection,
and always in favor of \(X\).  Evaluating the harmonic magnon free
energy of the coplanar states along the soft line for the nearest-neighbor fcc
antiferromagnet, we find that \(\langle\ln\omega_{\mathbf{k}}\rangle\) increases
monotonically from \(\delta=0\) to \(\delta=1/2\), with
\(\Delta F(W\!\to\!X)=0.157\,T\) per site.  Comparing this with the exchange
splitting at the ordering temperature \(T_c\simeq0.446\,J_1^{\rm eff}|\mathbf{T}|^2\)
of the nearest-neighbor fcc antiferromagnet~\cite{Gvozdikova05a}, exchange
selection dominates only if
\begin{equation}
    |s| \gtrsim 0.035\,J_1^{\rm eff} .
    \label{eq:obd_criterion}
\end{equation}
The \(30\)~K entry of \ce{CuInCr4S8} falls below this threshold and is
therefore predicted to order at \(X\) irrespective of the sign of \(s\), by the
same order-by-disorder mechanism established for the ideal model in
Sec.~\ref{sec:classical_cmc}.  Its room-temperature entry, with
\(s=+0.088\,J_1^{\rm eff}\), lies outside the window on the Type-III side; since
the ordered state forms at \(T_N\simeq35\)~K, it is the low-temperature structure
that is relevant to the selection, and the contrast between the two entries
mainly illustrates how sensitively \(s\) responds to the modest changes in the
microscopic couplings between the two refinements
(Table~\ref{tab:materials_couplings}).

One caveat concerns \(J_4^{\rm eff}\).  Equation~\eqref{eq:selector} weights the
fourth-neighbor coupling by the same factor \(4\) as the third, so a coupling
that is negligible next to \(J_1^{\rm eff}\) need not be negligible for the
selection.  In several compounds of Table~\ref{tab:materials_fcc_couplings} one
has \(J_4^{\rm eff}\gtrsim|J_2^{\rm eff}|\), and retaining these values would move
every entry to the Type-III side.  We do not retain them.

The reason is not the size of the supercell.  The cell used here contains ten
formula units and spans about \(25\)~\AA{} in two directions, comfortably larger
than the \(\simeq 14\)~\AA{} separation of fourth-neighbor tetrahedral units.  It
is rather that the fit behind Table~\ref{tab:materials_fcc_couplings}
retains only the first four fcc shells, whereas the supercell energies also
contain contributions from further-neighbor effective couplings.  A supercell
of this size cannot resolve those couplings separately, and their weight is
absorbed into the fitted parameters, into \(J_4^{\rm eff}\) and, depending on
symmetry, also into the shorter-range couplings.  The number listed under
\(J_4^{\rm eff}\) in Table~\ref{tab:materials_fcc_couplings} therefore
contains further-neighbor contributions and is not the fourth-neighbor
coupling by itself.

Equation~\eqref{eq:projection} constrains exactly these further-neighbor contributions.  Enumerating
the Cr--Cr shells of the pyrochlore network, the longest bond retained in
Table~\ref{tab:materials_couplings} is \(J_5\), at \(\sqrt{14}\,a/4\simeq
0.935\,a\), whereas the shortest Cr--Cr distance between two fourth-neighbor
tetrahedra is \(3\sqrt{2}\,a/4\simeq 1.061\,a\).  No microscopic bond up to
\(J_5\) connects tetrahedral units beyond the third fcc shell, so the projection
gives \(J_n^{\rm eff}=0\) for every \(n\ge4\), and not merely for \(n=4\).  The
whole family of couplings that is folded into the fitted \(J_4^{\rm eff}\)
vanishes in the coarse-grained model.  We therefore set \(J_4^{\rm eff}=0\) when
evaluating Eq.~\eqref{eq:selector}, and \(s\) is quoted on that basis in
Table~\ref{tab:materials_fcc_couplings}.  The resulting Type-I assignment is also
the one consistent with the single compound whose magnetic structure has been
resolved experimentally.

This remains the largest quantitative uncertainty of the materials analysis.  It
would be removed only by an energy mapping that resolves the further-neighbor
effective couplings simultaneously, which is a considerably larger calculation
than the one performed here.

\begin{figure}[t]
    \centering
    \includegraphics[width=\columnwidth]{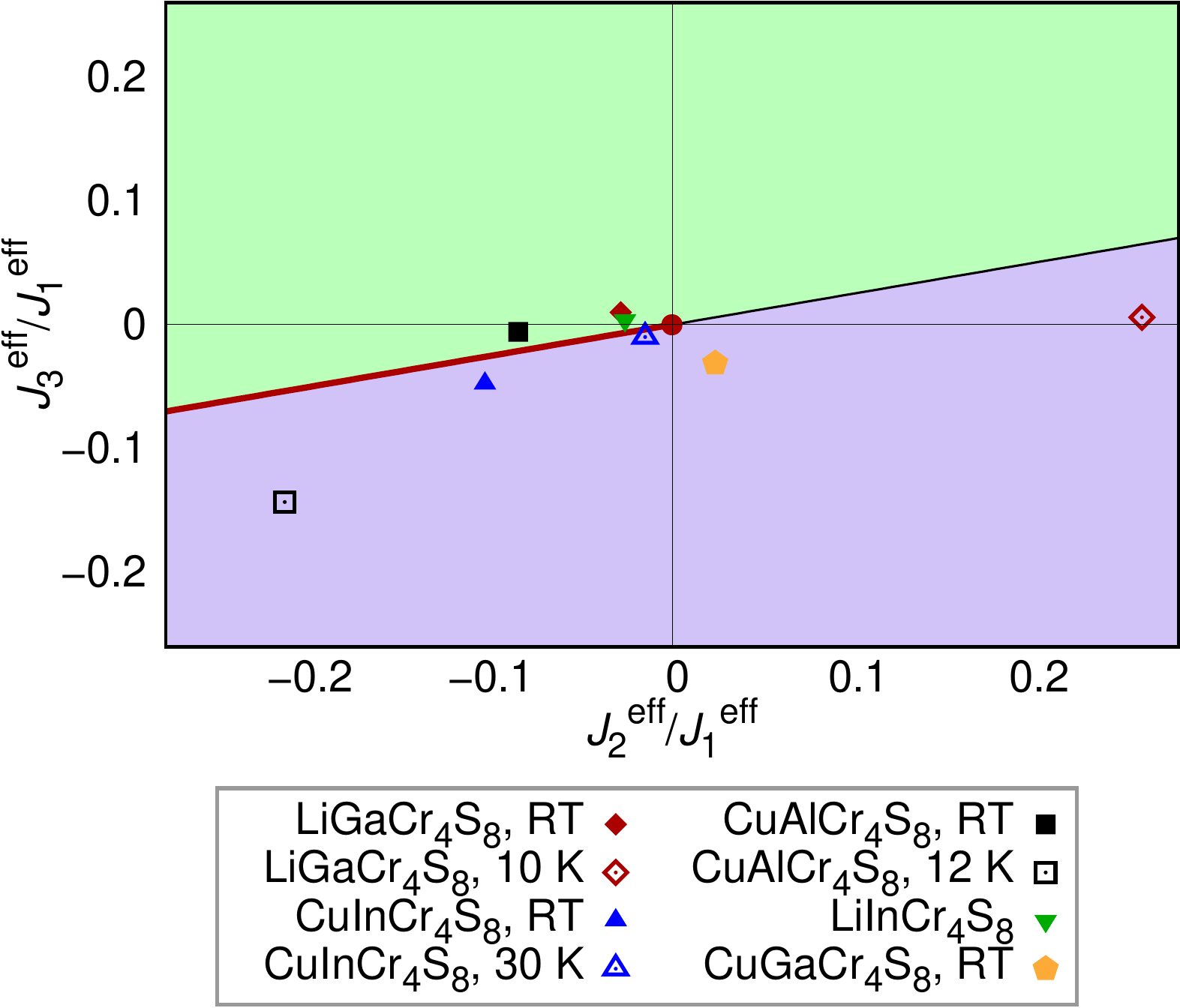}
    \caption{
    Breathing chromium thiospinels placed in the classical effective-fcc phase
    diagram using the ratios \(J_2^{\rm eff}/J_1^{\rm eff}\) and
    \(J_3^{\rm eff}/J_1^{\rm eff}\) of
    Table~\ref{tab:materials_fcc_couplings}.  The background phase diagram is
    that of the classical fcc Heisenberg model with antiferromagnetic
    nearest-neighbor exchange~\cite{Balla2020}; the thick line
    \(J_3^{\rm eff}=J_2^{\rm eff}/4\), i.e.\ \(s=0\) in
    Eq.~\eqref{eq:selector}, is the subextensively degenerate manifold that
    separates the Type-I (\(X\)) and Type-III (\(W\)) regions and is the only
    boundary emanating from the nearest-neighbor point at the origin.  All
    compounds lie within \(|s|\lesssim0.36\,J_1^{\rm eff}\) of that line.  The
    figure is a coarse-grained materials map and should not be read as a direct
    prediction of the spin-\(1/2\) tetramerized regime discussed in
    Sec.~\ref{sec:effective_pseudospin}.
    }
    \label{fig:materials_fcc_phase_diagram}
\end{figure}

\subsection{Comparison with the experimental literature}
\label{sec:materials_experiment}

We now compare these assignments with what is known experimentally.
Table~\ref{tab:materials_experiment} collects the relevant data for the breathing
chromium spinels, including the two oxides and the selenide for context.  Of the
five thiospinels, only one has a determined magnetic structure together with a
lattice that remains cubic; two order incommensurately, in both cases together
with a symmetry-lowering structural transition; and two have no determined
magnetic structure at all.  We take them in that order.

\begin{table*}[!t]
\caption{
Experimental status of the breathing chromium spinels.  \(T^\ast\) is the
temperature of the magnetic (or freezing) transition, \(\theta_{\rm CW}\) the
Curie--Weiss temperature, and \(\mathbf{k}\) the magnetic propagation vector in
the cubic setting.  ``LRO'' denotes long-range magnetic order and ``IC''
incommensurate.  Where two values of \(\theta_{\rm CW}\) are quoted, the
literature is in conflict.  The oxides and the selenide are listed for context;
the effective-fcc analysis of this section is carried out for the sulfides only.
}
\label{tab:materials_experiment}
\begin{ruledtabular}
\begin{tabular}{l l c r p{4.6cm} l}
Material & Low-\(T\) structure & \(T^{\ast}\)~(K) & \(\theta_{\rm CW}\)~(K) &
Magnetic ground state & Refs. \\
\hline
\ce{LiGaCr4O8} & \(I\bar{4}m2+F\bar{4}3m\) & 13.8,\,20 & \(-659\) &
LRO; \(\mathbf{k}=(0,0,1)\) and \((1,0,\tfrac12)\), \(m\simeq1.4\,\mu_B\) &
\cite{Okamoto13a,tanaka14-PRL113,Saha-2016,wawrzynczak17} \\[2pt]
\ce{LiInCr4O8} & \(I\bar{4}m2+F\bar{4}3m\) & 15.9,\,13 & \(-332\) &
quasi-LRO; \(\mathbf{k}=(1,0,\delta)\), \(\delta\simeq0.2\), \(\xi\simeq44\)~\AA{} &
\cite{Okamoto13a,nilsen15-PRB91,Nilsen-2025} \\[2pt]
\ce{LiGaCr4S8} & \(F\bar{4}3m\); no transition & 10.3--12 & \(+19.5\)/\(-20(10)\) &
\textbf{no LRO}; cluster-glass freezing &
\cite{Pokharel-2018,Pokharel-2020,okamoto18-JPSP87} \\[2pt]
\ce{LiInCr4S8} & \textbf{not determined} & 24 & \(+30(10)\) &
\textbf{not determined}; LRO inferred from bulk probes only &
\cite{okamoto18-JPSP87,Kanematsu-2020} \\[2pt]
\ce{CuInCr4S8} & \(F\bar{4}3m\) down to 2~K & 28--35 & \(-70(20)/-5\) &
LRO; \(\mathbf{k}=(1,0,0)\), Type-I, \(m=2.54(2)\,\mu_B\) &
\cite{Plumier-1971,Unger1975,Plumier-1977,okamoto18-JPSP87,Gao-2021,Gen-2022} \\[2pt]
\ce{CuAlCr4S8} & \(F\bar{4}3m\!\rightarrow\!Imm2\) & 20--21 & \(-68.7(9)\) &
LRO; IC cycloid, \(\mathbf{k}\simeq(0.89,0.11,0)\), \(m=2.20(3)\,\mu_B\) &
\cite{Sharma-2022,Gen-2024} \\[2pt]
\ce{CuGaCr4S8} & \(F\bar{4}3m\!\rightarrow\!Imm2\) & 31 & \(-103\) &
LRO; IC cycloid, \(\mathbf{k}\simeq(0.81,0.19,0)\), \(m=2.47(3)\,\mu_B\) &
\cite{Wilkinson-1976,Gen-2023,Gen-2024} \\[2pt]
\ce{CuInCr4Se8} & not determined & \(\sim14\) & --- &
\textbf{not determined}; spiral spin liquid predicted &
\cite{Pinch-1970,Duda-2008,ghosh19a}
\end{tabular}
\end{ruledtabular}
\end{table*}

\paragraph*{\ce{CuInCr4S8}.}
\ce{CuInCr4S8} provides the one direct test.  It is the only breathing chromium
thiospinel whose magnetic structure has been refined, and the only one whose
crystal symmetry is known to remain cubic \(F\bar{4}3m\) over the whole
temperature range \(2\)--\(300\)~K~\cite{Gao-2021}.  Neutron powder diffraction
finds long-range antiferromagnetic order below \(T_N\simeq35\)~K with propagation
vector \(\mathbf{k}=(1,0,0)\) and an ordered moment \(2.54(2)\,\mu_B\) per
Cr~\cite{Gao-2021}, that is, the Type-I state at the \(X\) point.  The same
experiment resolves a hierarchy of excitations in which each large tetrahedron
behaves as a rigid ferromagnetic \(S=6\) cluster on an fcc lattice, which is the
coarse graining of Eq.~\eqref{eq:materials_effective_fcc}.  Our \(30\)~K parameter set, the one appropriate to the ordered state, gives
\(s=+0.025\,J_1^{\rm eff}\): nominally in the \(W\) region of
Fig.~\ref{fig:materials_fcc_phase_diagram}, but inside the order-by-disorder
window of Eq.~\eqref{eq:obd_criterion}, so that thermal selection drives the
system to \(X\).  The effective-fcc mapping therefore gives
\(\mathbf{k}=(1,0,0)\), in agreement with experiment.  The room-temperature set
gives \(s=+0.088\,J_1^{\rm eff}\), outside the window on the Type-III side; as
discussed in Sec.~\ref{sec:materials_selector}, this reflects the sensitivity of
\(s\) to the structural refinement rather than a competing prediction for the
ordered state.  The agreement is not a trivial one: the
microscopic Hamiltonian of Table~\ref{tab:materials_couplings} has an
antiferromagnetic \(J\) and a ferromagnetic \(J'\) of comparable magnitude,
together with sizable third-neighbor terms, and a commensurate Type-I state is
not an obvious outcome of that Hamiltonian before the tetrahedral reorganization
is carried out.  The refined spin arrangement within the \(\mathbf{k}=(1,0,0)\)
cell remains ambiguous~\cite{Plumier-1971,Gao-2021}, but that ambiguity concerns
the orientation of the moments inside each tetrahedron rather than the
propagation vector, and lies below the resolution of the coarse-grained
description.

\paragraph*{\ce{LiGaCr4S8}.}
\ce{LiGaCr4S8} does not order.  Neutron scattering and \(\mu\)SR find only
spin freezing at \(T_f=12(2)\)~K, with no magnetic Bragg peaks and a
frequency-dependent susceptibility characteristic of a cluster
glass~\cite{Pokharel-2020}; high-resolution diffraction finds no symmetry lowering
down to \(10\)~K, only pronounced negative thermal expansion below
\(111(4)\)~K~\cite{Pokharel-2018}.  The effective-fcc parameters suggest a reason.
The room-temperature structure gives \(s=-0.069\,J_1^{\rm eff}\),
placing the compound in the Type-I region, whereas the \(10\)~K structure gives
\(s=+0.236\,J_1^{\rm eff}\), placing it well inside the Type-III region.  The two
structures differ only through the anomalous thermal contraction --- there is no
structural transition --- yet they fall on \emph{opposite} sides of the
degeneracy line, and both lie far enough from it that order-by-disorder cannot
decide between them.  Two candidate ordering wave vectors that exchange stability
as the lattice contracts, with no symmetry change to select one of them, are
unfavorable conditions for long-range order, and glassy freezing of correlated
clusters is a plausible outcome.  This gives a microscopic content to the
``cluster frustration'' reported for this compound~\cite{Pokharel-2020}.  We add
that the sign of \(\theta_{\rm CW}\) for \ce{LiGaCr4S8} is disputed in the
literature~\cite{Pokharel-2018,okamoto18-JPSP87}, which propagates into the choice
of \(U\).  The values quoted here are anchored to the negative determination,
\(\theta_{\rm CW}=-20(10)\)~K of Ref.~\cite{okamoto18-JPSP87}, as in
Ref.~\cite{ghosh19a}.  The conclusion above does not depend on this choice, since
both parameter sets place the compound far from the degeneracy line, but a
redetermination of \(\theta_{\rm CW}\) would be useful.

\paragraph*{\ce{CuAlCr4S8} and \ce{CuGaCr4S8}.}
These two compounds order into incommensurate cycloids, with
\(\mathbf{k}\simeq(0.89,0.11,0)\) and \((0.81,0.19,0)\) in the cubic setting,
below first-order transitions at \(21\) and \(31\)~K
respectively~\cite{Gen-2024}.  Neither wave vector lies on the fcc soft line
\((1,\delta,0)\), so the coarse-grained cubic model of
Eq.~\eqref{eq:materials_effective_fcc} does not reproduce them.  It does,
however, indicate why these two compounds fall outside its range of validity.
Both order simultaneously with a symmetry-lowering structural transition
\(F\bar{4}3m\rightarrow Imm2\)~\cite{Gen-2024}, which partially relieves the
breathing distortion and splits the nearest-neighbor bonds, whereas the
rigid-cluster coarse graining assumes cubic symmetry and equivalent tetrahedra
and is therefore inapplicable below \(T^\ast\).  The one compound in the family
that does \emph{not} distort, \ce{CuInCr4S8}, is also the one whose order the
mapping reproduces.  The size of the effect follows the distortion: the
orthorhombic distortion is larger for \(M=\)~Ga than for \(M=\)~Al, and so is the
incommensurability (\(0.19\) against \(0.11\)).  In the effective-fcc language the
observed states are long-wavelength modulations of the \(X\)-point Type-I
structure, as expected when a nearly degenerate soft-mode manifold is perturbed
by a weak symmetry-breaking magnetoelastic coupling of the kind analyzed in
Refs.~\cite{Aoyama-2019,Aoyama-2021}.  A quantitative treatment would require the
effective-fcc couplings to be recomputed in the \(Imm2\) structure with the
spin--lattice coupling retained, which lies beyond the scope of the present work.

\ce{CuAlCr4S8} is also the compound whose low-temperature parameter set is the
most extreme in Table~\ref{tab:materials_fcc_couplings}, with
\(s=+0.358\,J_1^{\rm eff}\), so the coarse-grained analysis identifies it as the
largest departure from the nearest-neighbor fcc limit even before the structural
transition is taken into account.  The low-temperature structural refinements of
these two compounds are, moreover, not fully consistent across the literature:
Ref.~\cite{Sharma-2022}
refines \ce{CuAlCr4S8} at \(12\)~K in \(F\bar{4}3m\) and reports no transition,
whereas Ref.~\cite{Gen-2024} refines it at \(4\)~K in \(Imm2\).  The distortion is
very small (pseudo-cubic \(\gamma=89.83^\circ\)), which plausibly accounts for the
discrepancy.  In either case the \(12\)~K parameter set of
Table~\ref{tab:materials_couplings} was computed from a cubic refinement, and
describes a cubic structure just below \(T^\ast\) rather than the true
ground-state structure.

\paragraph*{\ce{LiInCr4S8}: a prediction.}
\ce{LiInCr4S8} shows a sharp, first-order-like anomaly in the heat capacity at
\(24\)~K, a discontinuous drop of the susceptibility, and a volume contraction of
\(\sim700\)~ppm across the transition, all of which point to magnetic
order~\cite{okamoto18-JPSP87,Kanematsu-2020}.  Its magnetic structure, however, is
completely unknown: no neutron diffraction, \(\mu\)SR or NMR measurement of this
compound has been reported, and no low-temperature crystal structure has been
refined.  It is therefore the most direct available test of the effective-fcc
description.

Our room-temperature parameter set gives
\(J_2^{\rm eff}/J_1^{\rm eff}=-0.026\),
\(J_3^{\rm eff}/J_1^{\rm eff}=+0.004\), hence \(s=-0.041\,J_1^{\rm eff}\), and
places \ce{LiInCr4S8} firmly in the Type-I region of
Fig.~\ref{fig:materials_fcc_phase_diagram}, with exchange selection and thermal
order-by-disorder acting in the same direction.  We therefore predict that, if
cubic symmetry survives the \(24\)~K transition, \ce{LiInCr4S8} orders into the
collinear Type-I antiferromagnet with
\begin{equation}
    \mathbf{k}=X=(1,0,0)
\end{equation}
and cubic-symmetry-related wave vectors, with the four Cr moments of each large
tetrahedron ferromagnetically aligned and successive \((100)\) planes of
tetrahedra antiparallel.  The ordered moment should be close to, but somewhat
below, the full \(3\,\mu_B\) per Cr, as in \ce{CuInCr4S8}.  \ce{LiInCr4S8} is the
member of the family for which this description should work best: its microscopic
\(J\) is essentially zero while \(J'\) is strongly ferromagnetic, so of the
compounds considered here it comes closest to a set of rigid ferromagnetic
tetrahedra on an fcc lattice.  Should the transition instead involve a symmetry
lowering, as the sharp specific-heat anomaly may indicate, we would expect
\ce{LiInCr4S8} to follow \ce{CuAlCr4S8} and \ce{CuGaCr4S8} into an incommensurate
cycloid.  A single powder neutron-diffraction measurement would settle which of
the two occurs: the Type-I state gives resolution-limited magnetic Bragg peaks at
\((1,0,0)\)-type positions, the cycloid a pair of satellites displaced along
\(\langle1\bar{1}0\rangle\).

\paragraph*{Oxides and selenide.}
The two oxides fall outside the mixed-sign regime studied here, since both
nearest-neighbor couplings are antiferromagnetic~\cite{ghosh19a}, and both
undergo magnetostructural transitions with phase coexistence, so their reported
propagation vectors, \(\mathbf{k}=(0,0,1)\) and \((1,0,\tfrac{1}{2})\) for
\ce{LiGaCr4O8}~\cite{Saha-2016} and \(\mathbf{k}=(1,0,\delta)\) with
\(\delta\simeq0.2\) for \ce{LiInCr4O8}~\cite{nilsen15-PRB91,Nilsen-2025}, are not
directly comparable with the effective-fcc analysis.  We note, however, that the
\ce{LiInCr4O8} propagation vector has the form \((1,\delta,0)\) of the fcc soft
line, Eq.~\eqref{eq:softline}, with a short correlation length that
Ref.~\cite{Nilsen-2025} attributes to residual frustration in the distorted
structure.  The selenide \ce{CuInCr4Se8} has both nearest-neighbor couplings
ferromagnetic and was predicted in Ref.~\cite{ghosh19a} to realize a spiral spin
liquid with a spherical degeneracy surface, which is a different regime; its
magnetic ground state remains experimentally uncharacterized~\cite{Duda-2008}.

\subsection{Relation to the model results}
\label{sec:materials_relation}

The effective-fcc materials map also gives a practical interpretation of the
finite-temperature signatures discussed in Sec.~\ref{sec:hte_dynamics}.  In a
neutron-scattering experiment on a chromium thiospinel, the measured response is
the spin response of microscopic Cr moments.  It is not a direct measurement of
Eq.~\eqref{eq:materials_effective_fcc}.  Even so, the model calculations identify
momentum-space features that should remain useful in comparing different
compounds.

The first feature is diffuse scattering centered near the \(X\)-type wave vectors
of the fcc antiferromagnet.  In the ideal model this structure appears in the
classical SCGA, in the nonmagnetic pf-FRG susceptibility, in the DMRG structure
factor, and in the finite-temperature HTE response.  Its appearance in a
thiospinel would be a direct indication that the correlations are controlled by an
effective fcc antiferromagnetic manifold of tetrahedral units.  The one existing
measurement of this kind, the diffuse powder data of Ref.~\cite{Plumier-1977} on
\ce{CuInCr4S8}, is consistent with this picture~\cite{ghosh19a}.  The compound has in addition a complex ultrahigh-field phase diagram, including a wide half-magnetization plateau~\cite{Gen-2020,Gen-2022}, which lies outside the scope of the present zero-field analysis.

The second feature is the sharpening of this response toward Type-I correlations.
In the ideal model this corresponds to the \(\mathrm{AF}(1,0,0)\) state selected by
order-by-disorder and recovered as the leading magnetic instability at stronger
mixed-sign coupling.  In the materials setting, the same sharpening would indicate
motion toward the Type-I region of the effective-fcc phase diagram.

The third feature is the distinction between effective-fcc correlations and a true
ferromagnetic response.  Several microscopic exchange networks contain
ferromagnetic bonds, and in every compound of
Table~\ref{tab:materials_couplings} the dominant nearest-neighbor coupling
\(J'\) is ferromagnetic.  This does not by itself imply a ferromagnetic spin
structure factor.  In the mixed-sign effective-fcc regime, the relevant response
is centered near \(X\)-type wave vectors, whereas a true ferromagnet gives
dominant zone-center weight.  This distinction is especially important when
interpreting broad diffuse scattering at finite temperature, and it is directly
illustrated by the FM column of Fig.~\ref{fig:hte_suscep}.

The connection to the tetramerized nonmagnetic regime should be made with more
care.  The tetramer physics discussed in Secs.~\ref{sec:dmrg}
and~\ref{sec:effective_pseudospin} is a spin-\(1/2\), strong-breathing,
singlet-sector phenomenon.  The Cr\(^{3+}\) thiospinels have \(S=3/2\), and they
are not expected to realize this limit directly.  Their relevance here is
different: they show that real breathing-pyrochlore materials can naturally enter
an effective-fcc regime through tetrahedral coarse graining, and that the
resulting near degeneracy of the fcc soft-mode manifold controls which state they
select.  They are therefore natural systems in which to look for the classical
and finite-temperature aspects of the theory, especially \(X\)-centered diffuse
scattering and its evolution with exchange hierarchy, temperature, and structural
distortion.

\section{Conclusions and outlook}
\label{sec:conclusions}

We have studied the nearest-neighbor Heisenberg model on the breathing pyrochlore
lattice in the mixed ferro-antiferromagnetic regime, where one tetrahedral
sublattice is antiferromagnetic and the other ferromagnetic.  This regime differs
in an essential way from the antiferromagnetic breathing pyrochlore.  Its
classical ground-state manifold is organized by an effective fcc antiferromagnet
formed by tetrahedral units, rather than by the usual pyrochlore Coulomb
constraint.  The main question addressed in this work is how this effective-fcc
manifold is modified by thermal and quantum fluctuations, and whether the
resulting picture survives contact with real materials.

In the classical model, finite-temperature fluctuations resolve the mixed-sign
manifold by order-by-disorder.  Classical Monte Carlo simulations show a strongly
first-order transition into the collinear Type-I antiferromagnet with ordering
wave vector \(X=(1,0,0)\).  This selection appears consistently in the heat
capacity, in the growth of the AF1 order parameter and the suppression of AF3
with increasing system size, and in the equal-time structure factor.  The SCGA
results give the same physical picture from the paramagnetic side: the mixed-sign
regime develops square-ring diffuse scattering, reflecting the
soft-mode structure inherited from the effective fcc antiferromagnet.  This
establishes the classical reference point for the rest of the paper.

Quantum fluctuations modify this outcome.  The pf-FRG calculations for both \(S=1/2\) and \(S=1\) show
that a finite nonmagnetic region appears near the decoupled
antiferromagnetic-tetrahedron limit when ferromagnetic inter-tetrahedron coupling
is introduced, extending to \((J_B/J_A)_c\simeq-0.61(5)\) for \(S=1/2\) and to
\((J_B/J_A)_c\simeq-0.08(2)\) for \(S=1\).  At
stronger mixed-sign coupling, the leading magnetic instability occurs at the same
\(X=(1,0,0)\) wave vector selected in the classical problem.  Quantum
fluctuations therefore do not favor an unrelated competing magnetic order.
Instead they postpone the onset of the Type-I state, leaving a nonmagnetic regime
between isolated antiferromagnetic tetrahedra and the \(X\)-ordered phase.

The DMRG results show that this nonmagnetic regime is not a featureless
paramagnet.  Its real-space correlations display nearly ideal tetramer patterns
on the antiferromagnetic tetrahedra: four bonds carry strong antiferromagnetic
correlations close to \(-1/2\), while the remaining pair of opposite bonds carries
ferromagnetic correlations close to \(+1/4\).  At the same time, the spin
structure factor retains broad maxima at the same \(X\)-type wave vectors that
control the neighboring ordered phase.  The quantum nonmagnetic regime thus keeps
the momentum-space memory of the effective-fcc manifold while avoiding dipolar
order, by reorganizing the local tetrahedral degrees of freedom into tetramerized
singlet textures.

The strong-breathing pseudospin theory explains why this tetramer pattern is
selected.  Starting from isolated antiferromagnetic spin-\(1/2\) tetrahedra, the
low-energy Hilbert space consists of a two-dimensional singlet doublet on each
tetrahedron.  Degenerate perturbation theory generates, at third order in the
inter-tetrahedron coupling, an effective pseudospin Hamiltonian on the fcc lattice
of tetrahedron centers, Eq.~\eqref{eq:model_hamiltonian}.  Only half of the
elementary fcc triangles contribute to it: for those whose three tetrahedra share
a common \(B\) tetrahedron the third-order matrix element vanishes identically,
for the permutation-symmetry reason given in Sec.~\ref{sec:third_order}.  The sign
of the resulting cubic term then decides the outcome.  For the usual
antiferromagnetic inter-tetrahedron perturbation one recovers the partially
dimerized state discussed by Tsunetsugu; for the mixed-sign case studied here the
sign is reversed, and the effective Hamiltonian favors uniform tetramer
orientations.
Exact diagonalization of the 32-site pseudospin model supports this
interpretation: the finite-cluster ground state is well described as the symmetric
superposition of the three uniform tetramer states, and the corresponding
threefold degeneracy is already visible in the exact spectrum of three coupled
tetrahedra.

We also studied the finite-temperature spin response using a dynamic
high-temperature expansion, which provides the connection to scattering
experiments.  At weak mixed-sign coupling, the dynamical structure factor still
carries the local magnetic excitation scales of antiferromagnetic tetrahedra.  As
the ferromagnetic inter-tetrahedron coupling is increased, these local features
broaden and lose their isolation, while low-energy spectral weight grows near the
\(X\)-type wave vectors of the effective-fcc manifold.  The equal-time structure
factor gives the same message: the nonmagnetic regime is broad and
non-Bragg-like, but its correlations are already organized around the \(X\)
points.  The tetramer order itself is not expected to appear as a sharp branch in
the dipolar dynamical structure factor, since it belongs to the singlet
bond-energy sector.  Its presence is instead inferred from local tetrahedral
correlations, from the absence of conventional dipolar order, and from the
persistence of soft \(X\)-centered spin correlations.

The breathing chromium thiospinels provide a materials context for these results.
After coarse graining over Cr\(_4\) tetrahedra, all six compounds considered lie
close to an effective nearest-neighbor antiferromagnetic fcc model.  We showed
that the degeneracy of the fcc soft-mode line is lifted by a single combination
of further-neighbor effective couplings,
\(s=J_2^{\rm eff}-4J_3^{\rm eff}+4J_4^{\rm eff}\), and that \(|s|\) is small in
every compound, so that these materials sit inside a nearly degenerate remnant of
the fcc manifold.  This accounts for the range of behavior observed across the
family.  \ce{CuInCr4S8},
the only member of the family whose magnetic structure has been refined and whose
lattice remains cubic to \(2\)~K, orders at \(\mathbf{k}=(1,0,0)\), as the mapping
requires: its low-temperature structure lies inside the order-by-disorder window of
Eq.~\eqref{eq:obd_criterion}, so that thermal selection fixes the wave vector.  \ce{LiGaCr4S8}, whose room-temperature and low-temperature
structures fall on opposite sides of the degeneracy line although no symmetry
change intervenes, does not order at all, consistent with the cluster-glass
freezing observed experimentally.  \ce{CuAlCr4S8} and \ce{CuGaCr4S8}, the two
compounds that undergo a magnetostructural transition to \(Imm2\), order into
incommensurate cycloids that the rigid cubic coarse graining cannot describe, and
the incommensurability follows the size of the orthorhombic distortion.  For
\ce{LiInCr4S8}, whose \(24\)~K transition has never been characterized
microscopically, the mapping predicts Type-I order at \(\mathbf{k}=(1,0,0)\).
The effective-fcc description is thus more than a formal rewriting of the ideal
model: it makes testable statements, and its range of validity --- materials that
remain cubic --- can be stated.

Several questions remain open.  First, the static tetramer order should be
established or ruled out in the thermodynamic limit.  Since the tetramer order
parameter lives in the bond-energy sector, future numerical work should compute
equal-time tetramer--tetramer and dimer--dimer correlation functions
on larger clusters.  Such observables would distinguish more sharply between
true long-range tetramer order, short-range tetramer correlations, and a
fluctuating singlet regime.  A finite-size analysis of the pseudospin structure
factor at the ordering wave vector of the tetramer state would be the natural
first step.

Second, it would be useful to connect the strong-breathing pseudospin description
more quantitatively to the full microscopic model.  The third-order effective
Hamiltonian explains why tetramer correlations appear close to \(J_B=0^{-}\), but
it cannot locate the transition into the \(X=(1,0,0)\) ordered phase.
Higher-order perturbation theory, larger-cluster exact diagonalization of the
pseudospin model, and variational wave functions built within the tetrahedral
singlet manifold could clarify how the tetramerized regime gives way to dipolar
Type-I order.  A related question is whether the near degeneracy of the
pseudospin-ferromagnetic manifold seen on the 32-site cluster survives in the
thermodynamic limit, or is lifted into a genuine three-state Potts-like
transition.

Third, independently of whether the tetramer order is long ranged, the
dynamics of the singlet sector remains largely unexplored.  The
high-temperature expansion gives access to intermediate-temperature dipolar spin
dynamics, but not to the low-temperature dynamics of the tetramer manifold
itself, whose excitations are pseudospin flips between tetramer orientations
and carry no dipole moment.  A calculation of the frequency-resolved
bond-energy response, or a finite-temperature treatment of the effective
pseudospin model, would be a natural next step, and could clarify whether
tetramer fluctuations have observable signatures in Raman scattering or in other
higher-order probes that couple to the bond-energy sector.

The effective-fcc analysis should also be extended to the distorted phases.
Two of the compounds studied here order only together with a
\(F\bar{4}3m\rightarrow Imm2\) transition, and a coarse graining performed in the
orthorhombic structure, retaining the spin--lattice coupling in the spirit of
Refs.~\cite{Aoyama-2019,Aoyama-2021}, would test whether the observed
incommensurate cycloids emerge from the same near-degenerate soft-mode manifold.
The sensitivity of the selector \(s\) to the fourth-neighbor effective coupling,
noted in Sec.~\ref{sec:materials_selector}, should be resolved at the same time.
Enlarging the supercell alone will not settle it, since the obstacle is the
simultaneous resolution of \(J_4^{\rm eff}\) and the further-neighbor effective
couplings that are folded into it, rather than the range accessible to the
mapping.

On the experimental side, powder
neutron diffraction on \ce{LiInCr4S8} below \(24\)~K would test the prediction
made above, and with it the effective-fcc description.  The analysis also
suggests diagnostics common to the whole family: diffuse \(X\)-centered
scattering, its sharpening toward Type-I correlations, and its distinction from a
zone-center ferromagnetic response.  Single crystals, which do not yet exist for
any breathing chromium thiospinel, would allow these to be mapped out in
reciprocal space.  For quantitative comparison with individual compounds,
further-neighbor couplings, spin-lattice effects, and possible anisotropies will
have to be included.

Mixed-sign breathing pyrochlores are therefore not an interpolation between
antiferromagnetic pyrochlore physics and ferromagnetism.  Local tetrahedral
physics, effective-fcc frustration, thermal order-by-disorder, quantum
suppression of dipolar order, and singlet tetramer formation all appear within
the same model, and the near degeneracy that drives order-by-disorder in the
ideal model can be followed into the magnetic ground states of a real family of
materials.  They are, in this sense, a useful platform for studying how classical
frustrated manifolds are reshaped by quantum fluctuations, and how local singlet
degrees of freedom coexist with strong signatures of nearby magnetic order.

\section*{Acknowledgments}
Y.I., K.P., and J.R. performed part of this work at the Aspen Center for Physics, which is supported by National Science Foundation (NSF) Grant No. PHY-2210452, with the participation of Y.I. and K.P. supported by the Simons Foundation (1161654, Troyer).  Y.I. and K.P. also worked at the Kavli Institute for Theoretical Physics, Santa Barbara, supported by NSF Grant No. PHY-2309135, during the programs ``A New Spin on Quantum Magnets'' (summer 2023) and ``Correlated Gapless Quantum Matter'' (spring 2024).  A.R., J.R., J.G.R., K.P., H.O.J., and L.J. acknowledge IIT Madras for Visiting Faculty Fellow positions under the Institute of Eminence (IoE) program, during which this work was initiated and completed.  A.R., Y.I., L.J., K.P., and H.O.J. thank the International Centre for Theoretical Sciences (ICTS), Bengaluru, India, for hospitality during the program ``Frustrated Metals and Insulators'' (Code No. ICTS/frumi2022/9), and Y.I. and H.O.J. the University of Bordeaux/CNRS during a work visit in September 2023.  Y.I. acknowledges support from the ICTP through the Associates Programme, from the Simons Foundation through Grant No. 284558FY19, and from IIT Madras through the IoE program for establishing QuCenDiEM (Project No. SP22231244CPETWOQCDHOC).  Y.I., A.R., and L.J. acknowledge support from the Indo French Centre for the Promotion of Advanced Research (CEFIPRA) through Project No. SP26270224PHCEFI008874.  L.J. acknowledges financial support from Grants No. ANR-18-CE30-0011-01 and No. ANR-23-CE30-0038-01.  S.C. acknowledges support from the Anusandhan National Research Foundation (ANRF), India, in the form of a Junior Research Fellowship via the Prime Minister Early Career Research Grant Scheme ANRF/ECRG/2024/001198/PMS.  K.P. was supported by the Hungarian National Research, Development and Innovation Office (NKFIH) under Grant No.~K-142652.  I.H. was supported by the NKFIH through Grant No.~FK142985 and by the J\'anos Bolyai Research Scholarship of the Hungarian Academy of Sciences.  I.H. and J.R. were supported by the Alexander von Humboldt Foundation in the framework of the Research Group Linkage Programme funded by the German Federal Ministry of Research, Technology and Space.  H.O.J. acknowledges support through JSPS KAKENHI Grant No.~25K08460.  Y.I. and S.C. acknowledge the use of the computing resources at HPCE, IIT Madras, and I.H. of the Noctua2 cluster at the Paderborn Center for Parallel Computing (PC$^2$).

%---------------------------------------------------------------------------------
\appendix

%---------------------------------------------------------------------------------
\section{Pseudofermion functional renormalization group}
\label{app:pffrg}

In this appendix we summarize the pseudofermion functional
renormalization group (pf-FRG) scheme used in Sec.~\ref{sec:pffrg_results}.
The method has been applied extensively to frustrated quantum spin models
in two and three dimensions~\cite{
Reuther2010,Reuther2011,Reuther2011_2,Reuther2011_3,Ronny2014,
Yasir2016,yasir2016_2,Finn2016,Reuther2017,Finn2018,Finn2018_2,
Finn2019,Yasir2019,Dominik2020,Finn2021,Vincent2022,Yasir2023,
fukui2023,Lasse2023,Lasse2024,Chern2024}.  We use it here for the
nearest-neighbor breathing-pyrochlore Heisenberg model for both
\(S=1/2\) and \(S=1\).  A detailed account of the spin-\(S\) formulation
and its numerical implementation can be found in Ref.~\cite{Tobi2024}.

\subsection{Pseudofermion representation and RG flow}
\label{app:pffrg_formalism}

pf-FRG is formulated for a general bilinear spin Hamiltonian of the form
\begin{equation}
    \hat H
    =
    \sum_{i<j}\sum_{\mu,\nu}
    J_{ij}^{\mu\nu}
    \hat S_i^\mu \hat S_j^\nu ,
    \label{eq:pffrg_general_hamiltonian}
\end{equation}
where \(\mu,\nu\in\{x,y,z\}\).  In the present work the interactions are
spin-rotation invariant, so that \(J_{ij}^{\mu\nu}=J_{ij}\delta_{\mu\nu}\),
with \(J_{ij}=J_A\) or \(J_B\) depending on whether the bond belongs to
an \(A\)- or \(B\)-tetrahedron.

For \(S=1/2\), the spin operators are represented in terms of Abrikosov
pseudofermions~\cite{Abrikosov1965},
\begin{equation}
    \hat S_i^\mu
    =
    \frac{1}{2}
    \sum_{\alpha,\beta}
    \hat f_{i\alpha}^{\dagger}
    \sigma^\mu_{\alpha\beta}
    \hat f_{i\beta},
    \label{eq:pffrg_spin_representation}
\end{equation}
where \(\sigma^\mu\) are the Pauli matrices and
\(\alpha,\beta\in\{\uparrow,\downarrow\}\).  The physical spin Hilbert space corresponds to
the singly occupied sector,
\begin{equation}
    \hat n_i=\sum_\alpha \hat f^\dagger_{i\alpha}\hat f_{i\alpha}=1 .
    \label{eq:pffrg_constraint}
\end{equation}
In practice, pf-FRG works in the enlarged pseudofermion Hilbert space and
Eq.~\eqref{eq:pffrg_constraint} is not imposed exactly at every step of
the flow.  Two facts make this benign here.  First, the unphysical
sectors \(\hat n_i=0,2\) carry no magnetic moment, so they act as
nonmagnetic vacancies rather than as spurious magnetic degrees of
freedom.  Second, at \(T=0\) the particle-hole symmetry of the pf-FRG
equations enforces \(\langle \hat n_i\rangle=1\) on
average~\cite{Reuther2010}; an additional level-repulsion term can be
used to sharpen this, but it is known to have little effect on the
quantities computed here.

For \(S>1/2\) we use the spin-\(S\) extension of
Refs.~\cite{Reuther2017,Tobi2024}, in which the spin operator is
represented by \(2S\) pseudofermion flavors,
\begin{equation}
    \hat S_i^\mu
    =
    \frac{1}{2}
    \sum_{\kappa=1}^{2S}
    \sum_{\alpha,\beta}
    \hat f_{i\kappa\alpha}^{\dagger}
    \sigma^\mu_{\alpha\beta}
    \hat f_{i\kappa\beta},
    \label{eq:pffrg_spinS}
\end{equation}
subject to \(\sum_{\kappa\alpha}\hat f^\dagger_{i\kappa\alpha}\hat
f_{i\kappa\alpha}=2S\) and to the projection onto the fully symmetric
(maximal-spin) representation of the \(2S\) flavors.  This projection is not
enforced exactly in the flow; unphysical states of the enlarged flavor space can
in principle contribute, and the extent to which they do is discussed in
Ref.~\cite{Reuther2017}.  We do not employ a large-\(N\) control parameter, so
the \(S=1\) results should be read with the same caveats as the \(S=1/2\) ones,
together with this additional one.

Substitution of Eq.~\eqref{eq:pffrg_spin_representation} into
Eq.~\eqref{eq:pffrg_general_hamiltonian} converts the spin Hamiltonian
into an interacting pseudofermion problem with quartic vertices.  Since
there is no kinetic hopping term, the bare pseudofermion propagator is
local in real space and has the form
\begin{equation}
    G_0(i\omega)
    =
    \frac{1}{i\omega}.
    \label{eq:pffrg_bare_prop}
\end{equation}
The absence of a kinetic energy makes the problem strongly coupled from
the outset.  The purpose of pf-FRG is to resum classes of diagrams in a
self-consistent way by introducing an infrared cutoff \(\Lambda\) into
the Matsubara-frequency propagator~\cite{WETTERICH1993,kopietz2010,Metzner2012}.
For the sharp cutoff used in the present calculations,
\begin{equation}
    G_0(i\omega)
    \longrightarrow
    G_0^\Lambda(i\omega)
    =
    \frac{\Theta(|\omega|-\Lambda)}{i\omega}.
    \label{eq:pffrg_cutoff_prop}
\end{equation}
The flow starts at \(\Lambda=\infty\), where all fluctuations are
suppressed, and proceeds toward \(\Lambda=0\), where the full interacting
problem is recovered.  Smooth cutoff functions have also been used in
recent pf-FRG implementations~\cite{kiese2022,thoenniss2020}; here we use
the sharp-regulator implementation of \textsc{SpinParser}~\cite{Finn_SP,Finn_SP_2}.

The flowing self-energy and two-particle vertex are denoted by
\(\Sigma^\Lambda\) and \(\Gamma^\Lambda\).  With the cutoff included,
the full propagator is
\begin{equation}
    G^\Lambda(i\omega)
    =
    \frac{\Theta(|\omega|-\Lambda)}
    {i\omega-\Sigma^\Lambda(i\omega)} .
    \label{eq:pffrg_full_prop}
\end{equation}
The exact functional RG hierarchy contains coupled flow equations for
all one-particle irreducible \(n\)-particle vertices.  The flow of the
\(n\)-particle vertex depends on vertices up to order \(n+1\), leading to
an infinite hierarchy.  In pf-FRG this hierarchy is truncated at the
two-particle vertex, while retaining part of the feedback from the
three-particle vertex through the Katanin substitution.  In compact
multi-index notation, with \(1\equiv(i_1,\omega_1,\alpha_1)\), the
one-loop flow has the schematic structure
\begin{subequations}
\label{eq:pffrg_flow_schematic}
\begin{align}
    \frac{d}{d\Lambda}\Sigma^\Lambda(1',1)
    &=
    -\frac{1}{2\pi}
    \sum_{2,2'}
    \Gamma^\Lambda(1',2';1,2)\,
    S^\Lambda(2,2'),
    \label{eq:pffrg_self_energy_flow}
    \\
    \frac{d}{d\Lambda}\Gamma^\Lambda(1',2';1,2)
    &=
    \Phi_{\rm pp}^\Lambda
    +
    \Phi_{\rm ph,d}^\Lambda
    +
    \Phi_{\rm ph,cr}^\Lambda .
    \label{eq:pffrg_vertex_flow}
\end{align}
\end{subequations}
Here \(\Phi_{\rm pp}^\Lambda\), \(\Phi_{\rm ph,d}^\Lambda\), and
\(\Phi_{\rm ph,cr}^\Lambda\) denote the particle-particle, direct
particle-hole, and crossed particle-hole one-loop contributions,
respectively.  Each consists of two flowing two-particle vertices joined by a loop
containing one full and one single-scale propagator, symmetrized as
\(S^\Lambda G^\Lambda+G^\Lambda S^\Lambda\) (equivalently
\(-\tfrac{d}{d\Lambda}[G^\Lambda G^\Lambda]\) once the Katanin
substitution is made), with the particle-particle channel carrying an
additional factor \(1/2\).  The explicit channel expressions, including
all signs and symmetry factors, are standard and are given, for the
implementation used here, in Refs.~\cite{Finn2019,Tobi2024}.

Throughout this appendix we use the sign convention of
\textsc{SpinParser}~\cite{Finn_SP}, in which the single-scale propagator is
\begin{equation}
    S^\Lambda
    =
    G^\Lambda
    \frac{d}{d\Lambda}
    \left(G_0^\Lambda\right)^{-1}
    G^\Lambda
    =
    -\left.\frac{d}{d\Lambda}G^\Lambda\right|_{\Sigma^\Lambda\ {\rm fixed}} .
    \label{eq:pffrg_single_scale}
\end{equation}
For the sharp regulator of Eq.~\eqref{eq:pffrg_cutoff_prop} this
derivative is evaluated using Morris' lemma, which gives
\begin{equation}
    S^\Lambda(i\omega)
    =
    \,\frac{\delta(|\omega|-\Lambda)}
    {i\omega-\Sigma^\Lambda(i\omega)} ,
    \label{eq:pffrg_single_scale_sharp}
\end{equation}
the \(\Theta\delta\) ambiguity being resolved in the standard way.  In
the Katanin truncation the single-scale propagator is replaced by the
full cutoff derivative of the dressed propagator,
\begin{equation}
    S^\Lambda
    \longrightarrow
    S^\Lambda_{\rm Kat}
    =
    -\frac{d}{d\Lambda}G^\Lambda
    =
    S^\Lambda
    -
    \left(G^\Lambda\right)^2
    \frac{d}{d\Lambda}\Sigma^\Lambda .
    \label{eq:pffrg_katanin}
\end{equation}
This substitution feeds self-energy corrections back into the vertex
flow and incorporates an important subset of three-particle-vertex
contributions.  It is crucial in frustrated magnets, where a bare
one-loop truncation tends to overemphasize magnetic ordering tendencies
and may fail to describe magnetically disordered regimes
reliably~\cite{Reuther2010}.  Note that the overall sign of \(S^\Lambda\) in
Eqs.~\eqref{eq:pffrg_single_scale}--\eqref{eq:pffrg_katanin} is opposite to the
convention of Ref.~\cite{Reuther2010}; the flow equations are of course
unaffected, since \(S^\Lambda\) enters them only in the combinations quoted
above.

In practice the initial conditions are imposed at a large but finite
\(\Lambda_{\max}\gg\max_{ij}|J_{ij}|\), where
\begin{equation}
    \Sigma^{\Lambda=\infty}=0,
    \qquad
    \Gamma^{\Lambda=\infty}=\Gamma_{\rm bare},
\end{equation}
where \(\Gamma_{\rm bare}\) is fixed by the exchange couplings in
Eq.~\eqref{eq:pffrg_general_hamiltonian}.  The coupled flow equations
are then integrated down to the lowest accessible cutoff.

\subsection{Static susceptibility}
\label{app:pffrg_observable}

The principal observable used in this work is the cutoff-dependent static
spin susceptibility,
\begin{equation}
    \chi_{ij}^{zz,\Lambda}
    =
    \int_0^\infty d\tau\,
    \left\langle
    T_\tau S_i^z(\tau)S_j^z(0)
    \right\rangle_\Lambda ,
    \label{eq:pffrg_realspace_suscep}
\end{equation}
where \(T_\tau\) denotes imaginary-time ordering.  Its momentum-space
form is
\begin{equation}
    \chi^{zz,\Lambda}(\mathbf{k})
    =
    \frac{1}{N}
    \sum_{ij}
    \chi_{ij}^{zz,\Lambda}
    e^{i\mathbf{k}\cdot(\mathbf{r}_i-\mathbf{r}_j)} .
    \label{eq:pffrg_momentum_suscep}
\end{equation}
For spin-rotation-invariant models, the \(zz\) component is
representative of the full spin response.

The momentum dependence of \(\chi^{zz,\Lambda}(\mathbf{k})\) gives the
dominant spin-correlation profile at the RG scale \(\Lambda\).  During
the flow we monitor the maximum susceptibility,
\begin{equation}
    \chi_{\rm max}^{\Lambda}
    =
    \max_{\mathbf{k}}
    \chi^{zz,\Lambda}(\mathbf{k}) .
\end{equation}
A smooth flow of \(\chi_{\rm max}^{\Lambda}\) down to the lowest
accessible cutoff is taken as evidence against conventional dipolar
magnetic order within the resolution of the calculation.  By contrast,
a kink, cusp, or breakdown of the flow at a finite scale \(\Lambda_c\)
signals an instability toward magnetic order.  The momentum at which
\(\chi^{zz,\Lambda}(\mathbf{k})\) is maximal just above \(\Lambda_c\) is
then identified as the ordering wave vector.

This criterion should be interpreted with the usual caution.  A smooth
pf-FRG flow does not by itself determine the nature of the nonmagnetic
state.  It rules out, within the method's resolution, conventional
dipolar magnetic order and provides the dominant two-spin correlation
profile.  It does not directly distinguish, for example, a featureless
paramagnet from a valence-bond, plaquette, tetramerized, or topologically
ordered state.  This is why the main text supplements the pf-FRG phase
diagram with DMRG and the strong-breathing pseudospin analysis.

\subsection{Numerical implementation}
\label{app:pffrg_numerics}

The pf-FRG calculations were performed using the \textsc{SpinParser}
package~\cite{Finn_SP,Finn_SP_2}.  The Matsubara-frequency dependence of
the vertices is discretized on a logarithmic mesh symmetric about
\(\omega=0\).  We use \(N_\omega=64\) positive frequencies, giving \(2N_\omega\)
frequencies in total, spanning
\(\omega_{\min}=10^{-5}\,\bar{J}\) to \(\omega_{\max}=50\,\bar{J}\), with
\begin{equation}
    \omega_n
    =
    \omega_{\min}
    \left(
    \frac{\omega_{\max}}{\omega_{\min}}
    \right)^{\frac{n}{N_\omega-1}},
    \qquad
    n=0,\ldots,N_\omega-1 .
    \label{eq:pffrg_freq_mesh}
\end{equation}
In the present calculations the real-space range of the two-particle
vertex is truncated by setting vertex components to zero when the two
sites involved are separated by more than
\(L_{\rm cut}=8\)
nearest-neighbor bonds.  For the pyrochlore network this retains, besides the
reference site, \(1028\) correlated sites, reduced by the lattice symmetries to
\(202\) symmetry-inequivalent ones; the corresponding maximum real-space range is
\(2\sqrt{2}\,a\), that is, eight nearest-neighbor Cr--Cr distances.  The RG cutoff
is decreased geometrically,
\begin{equation}
    \Lambda_{n+1}=b\Lambda_n,
    \qquad b=0.98,
\end{equation}
from
\begin{equation}
    \Lambda_{\max}=50\,\bar{J}
    \quad \text{to} \quad
    \Lambda_{\min}=10^{-4}\,\bar{J} .
\end{equation}
All energies are measured in units of the overall exchange scale
\(\bar{J}\) of Eq.~\eqref{eq:polar_parametrization}, which is set to
unity.

The coupled flow equations are integrated with the explicit Euler stepping of
\textsc{SpinParser}, in which the vertices at \(\Lambda-\delta\Lambda\) are
obtained by linear extrapolation from their values and derivatives at
\(\Lambda\).  The internal frequency integrals are evaluated with the trapezoidal
rule on the same logarithmic mesh that carries the vertices, and a constant
extrapolation is used for frequency arguments falling outside the mesh, so that
the tail beyond \(\omega_{\max}\) contributes at its boundary value.

The breakdown scale \(\Lambda_c\) is identified by inspection of the RG flow,
as the cutoff at which \(\chi^{zz,\Lambda}_{\max}\) develops a kink or a cusp.
We do not extract it from a numerical derivative of the flow: the sharp
frequency regulator of Eq.~\eqref{eq:pffrg_cutoff_prop} makes
\(d\chi^{zz,\Lambda}_{\max}/d\Lambda\) too noisy to locate the breakdown
reliably.  The uncertainty quoted on the phase boundary,
Eq.~\eqref{eq:theta_c}, correspondingly reflects the spread in where the kink
can be placed by eye across the flows computed on the \(\theta\) grid, and is
the dominant source of error on \(\theta_c\).

The same frequency and cutoff discretization was used for \(S=1/2\) and
\(S=1\).  For \(S=1\) the \textsc{SpinParser} normalization constant, which
rescales all exchange couplings and defaults to \(2S\), was varied between
\(1.0\) and its default value; the phase boundary in
Fig.~\ref{fig:PD}(c) is unchanged, and in particular \((J_B/J_A)_c\) is
independent of this choice, as it must be for a ratio of couplings.

The finite frequency mesh, the real-space truncation range, and the
choice of cutoff grid set the numerical resolution of the pf-FRG flow.
They can slightly shift the apparent breakdown scale and broaden the
momentum-space susceptibility profile.  The phase assignments in the
main text are therefore based not on a single numerical signature, but
on the combined behavior of the RG flow and the momentum-resolved
susceptibility.
%---------------------------------------------------------------------------------

\section{Mean-field analysis and correlators of the pseudospin model}
\label{sec:appendix_MF}

This appendix collects three supplementary results used in
Sec.~\ref{sec:effective_pseudospin}.  First, we discuss the mean-field
minimization of the third-order pseudospin Hamiltonian,
Eq.~\eqref{eq:model_hamiltonian}, over product states in the tetrahedral
singlet manifold.  Second, we derive the correlator identities used to
interpret the exact-diagonalization ground state of the effective model.
Third, we analyze the first excited state within the
pseudospin-ferromagnetic manifold.

\subsection{Product-state mean-field analysis}

We consider product states built from the equatorial pseudospin
coherent state
\begin{equation}
    |\varphi\rangle
    =
    \frac{|+\rangle+e^{i\varphi}|-\rangle}{\sqrt{2}},
    \label{eq:appendix_theta_state}
\end{equation}
where \(|+\rangle\) and \(|-\rangle\) are the two chiral singlets of an
isolated antiferromagnetic tetrahedron, defined in
Eq.~\eqref{eq:chiral_singlet_basis}.  These equatorial states are
time-reversal invariant and have
\(\langle \tau^z\rangle=0\).  Their transverse pseudospin direction
specifies how the six bond energies are distributed within the
tetrahedron.

The three angles
\begin{equation}
    \varphi=\pi,\quad \frac{\pi}{3},\quad \frac{5\pi}{3}
\end{equation}
correspond to the three dimer-pair singlets, while
\begin{equation}
    \varphi=0,\quad \frac{2\pi}{3},\quad \frac{4\pi}{3}
\end{equation}
correspond to the three tetramer singlets.  In a tetramer singlet, four
bonds of the tetrahedron carry antiferromagnetic correlations and the
remaining two opposite bonds carry ferromagnetic correlations.  This is
the local pattern discussed in Secs.~\ref{sec:dmrg} and
\ref{sec:effective_pseudospin}.

For a four-sublattice product state on the fcc lattice of
\(A\)-tetrahedron centers,
\begin{equation}
    |\Psi_{\rm MF}\rangle
    =
    \prod_{n=1}^{4}\ \prod_{\mathbf{R}\in n}|\varphi_n\rangle_{\mathbf{R}},
    \label{eq:mf_product_state}
\end{equation}
where \(n=1,\dots,4\) labels the four cubic-cell sublattices of the fcc
lattice of \(A\)-tetrahedron centers, we evaluate the expectation value
of Eq.~\eqref{eq:model_hamiltonian}.  This gives an energy functional of
the four angles \(\varphi_1,\dots,\varphi_4\), which we minimize.

For \(J_B>0\), corresponding to antiferromagnetic inter-tetrahedron
coupling in the strong-breathing expansion, the product-state
minimization reproduces Tsunetsugu's mean-field solution.  Three of the
four fcc sublattices select dimer-pair angles, while the fourth
sublattice remains free at this level.  Representative solutions are
listed in Table~\ref{tab:MF_AFM}.  The undetermined angle reflects the
residual mean-field degeneracy of the antiferromagnetic singlet-sector
Hamiltonian.  In Tsunetsugu's analysis this remaining freedom is lifted
only after further fluctuation effects are included.

\begin{table}[bt]
\caption{
Four-sublattice product-state minima of the effective pseudospin
Hamiltonian for \(J_B>0\).  This is the antiferromagnetic
strong-breathing case studied by Tsunetsugu.  Three sublattices select
dimer-pair angles, while the fourth angle remains free at the
mean-field level.
}
\label{tab:MF_AFM}
\begin{ruledtabular}
\begin{tabular}{cccc}
\(\varphi_1\) & \(\varphi_2\) & \(\varphi_3\) & \(\varphi_4\) \\
\hline
\(\varphi\) & \(\pi\) & \(\pi/3\) & \(5\pi/3\)\\
\(\pi\) & \(\varphi\) & \(5\pi/3\) & \(\pi/3\)\\
\(\pi/3\) & \(5\pi/3\) & \(\varphi\) & \(\pi\)\\
\(5\pi/3\) & \(\pi/3\) & \(\pi\) & \(\varphi\)
\end{tabular}
\end{ruledtabular}
\end{table}

\begin{figure}[bt]
    \centering
    \includegraphics[width=0.8\linewidth]{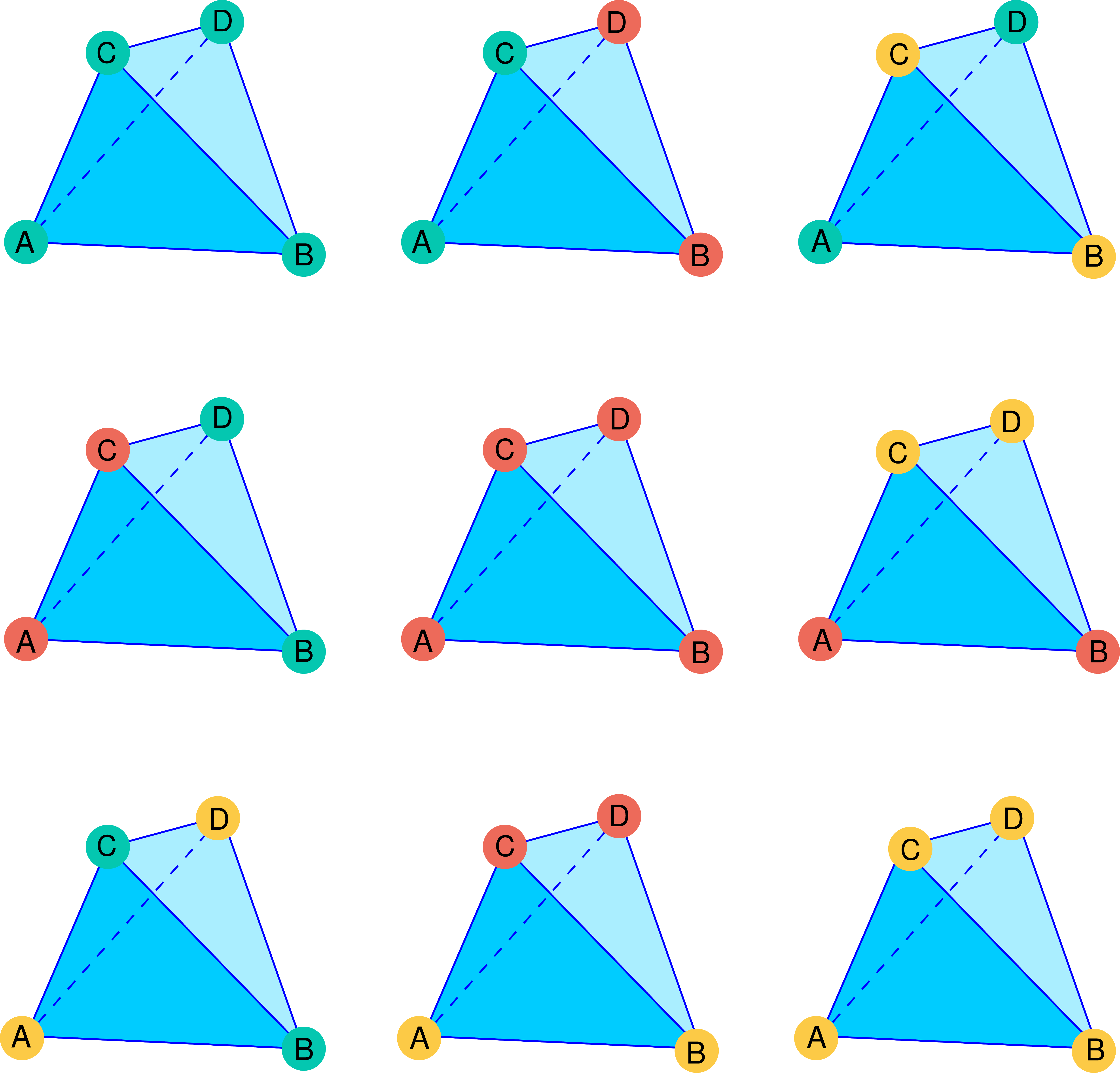}
    \caption{
    Product-state minima for \(J_B<0\) on an elementary fcc
    supertetrahedron.  Colors denote the three tetramer angles: green
    for \(0\), red for \(2\pi/3\), and yellow for \(4\pi/3\).  The
    uniform configurations are among the nine local mean-field minima.
    Exact diagonalization of the full 32-site pseudospin model selects
    the pseudospin-ferromagnetic sector corresponding to the uniform
    tetramer states.
    }
    \label{fig:FM_MFs}
\end{figure}

For the mixed-sign problem considered in the main text, \(J_B<0\).
Since the degeneracy-lifting term in the strong-breathing expansion is
third order in \(J_B\), this reverses the sign of the effective
Hamiltonian relative to the antiferromagnetic perturbation.  The
product-state minima are then shifted from dimer-pair angles to
tetramer angles.  On a single fcc supertetrahedron, the minimization
gives the nine solutions listed in Table~\ref{tab:MF_FM} and shown
schematically in Fig.~\ref{fig:FM_MFs}.  Three of these are uniform
tetramer states.  The other six are nonuniform four-sublattice
tetramer patterns, in which the tetramer orientation changes between
sublattices.

The four-site mean-field minimization should not be overinterpreted.
It identifies the local tendencies of the effective Hamiltonian, but it
does not by itself determine the ground state of the full quantum
problem.  The exact diagonalization of the 32-site pseudospin model,
discussed in Sec.~\ref{sec:effective_pseudospin}, provides the sharper
diagnostic.  It selects the pseudospin-ferromagnetic sector and
identifies the finite-cluster ground state as the symmetric
superposition of the three uniform tetramer orientations.

\begin{table}[bt]
\caption{
Four-sublattice product-state minima of the effective pseudospin
Hamiltonian for \(J_B<0\), appropriate to the mixed-sign
strong-breathing perturbation.  All angles belong to the tetramer set
\(\{0,2\pi/3,4\pi/3\}\).  The first, fifth, and ninth rows are uniform
tetramer states.  The remaining rows are nonuniform four-sublattice
tetramer patterns that are degenerate at the level of a single
supertetrahedron product-state minimization.
}
\label{tab:MF_FM}
\begin{ruledtabular}
\begin{tabular}{cccc}
\(\varphi_1\) & \(\varphi_2\) & \(\varphi_3\) & \(\varphi_4\) \\
\hline
\(0\) & \(0\) & \(0\) & \(0\)\\
\(0\) & \(2\pi/3\) & \(0\) & \(2\pi/3\)\\
\(0\) & \(4\pi/3\) & \(4\pi/3\) & \(0\)\\
\(2\pi/3\) & \(0\) & \(2\pi/3\) & \(0\)\\
\(2\pi/3\) & \(2\pi/3\) & \(2\pi/3\) & \(2\pi/3\)\\
\(2\pi/3\) & \(2\pi/3\) & \(4\pi/3\) & \(4\pi/3\)\\
\(4\pi/3\) & \(0\) & \(0\) & \(4\pi/3\)\\
\(4\pi/3\) & \(4\pi/3\) & \(2\pi/3\) & \(2\pi/3\)\\
\(4\pi/3\) & \(4\pi/3\) & \(4\pi/3\) & \(4\pi/3\)
\end{tabular}
\end{ruledtabular}
\end{table}

\subsection{Correlators for symmetric superpositions}

We now derive the correlators used in Sec.~\ref{sec:effective_pseudospin}
to identify the ED ground state.  The operators in this subsection are
pseudospin operators \(\tau^x,\tau^y,\tau^z\), not microscopic spin
operators.

Consider the symmetric superposition
\begin{equation}
    |\Psi_{\varphi,n}\rangle
    =
    \frac{1}{\sqrt{n}}
    \sum_{k=0}^{n-1}
    \prod_{j}
    \left|
    \varphi+\frac{2\pi k}{n}
    \right\rangle_j ,
    \label{eq:appendix_cat_state}
\end{equation}
where all pseudospins are aligned in each product component, and the
\(n\) components are equally spaced on the equator of the Bloch sphere.
For a single coherent state \(|\varphi\rangle\),
\begin{equation}
    \langle \tau^x\rangle=\frac{1}{2}\cos\varphi,
    \qquad
    \langle \tau^y\rangle=\frac{1}{2}\sin\varphi,
    \qquad
    \langle \tau^z\rangle=0 .
    \label{eq:single_tau_expectation}
\end{equation}
For two distinct sites \(i\neq j\), the two-pseudospin correlators in
the state \(|\Psi_{\varphi,n}\rangle\) are obtained by averaging over
the \(n\) equatorial directions:
\begin{subequations}
\label{eq:appendix_twobody}
\begin{align}
    \langle \tau_i^x\tau_j^x\rangle
    &=
    \frac{1}{4n}
    \sum_{k=0}^{n-1}
    \cos^2
    \left(\varphi+\frac{2\pi k}{n}\right),
    \\
    \langle \tau_i^y\tau_j^y\rangle
    &=
    \frac{1}{4n}
    \sum_{k=0}^{n-1}
    \sin^2
    \left(\varphi+\frac{2\pi k}{n}\right),
    \\
    \langle \tau_i^x\tau_j^y\rangle
    &=
    \frac{1}{4n}
    \sum_{k=0}^{n-1}
    \cos
    \left(\varphi+\frac{2\pi k}{n}\right)
    \sin
    \left(\varphi+\frac{2\pi k}{n}\right).
\end{align}
\end{subequations}
Using the elementary sums over roots of unity, this gives
\begin{subequations}
\label{eq:appendix_twobody_result}
\begin{align}
    \langle \tau_i^x\tau_j^x\rangle
    &=
    \begin{cases}
    \dfrac{1}{4}\cos^2\varphi, & n=1,2,\\[0.4em]
    \dfrac{1}{8}, & n>2,
    \end{cases}
    \\
    \langle \tau_i^y\tau_j^y\rangle
    &=
    \begin{cases}
    \dfrac{1}{4}\sin^2\varphi, & n=1,2,\\[0.4em]
    \dfrac{1}{8}, & n>2,
    \end{cases}
    \\
    \langle \tau_i^x\tau_j^y\rangle
    &=
    \begin{cases}
    \dfrac{1}{8}\sin 2\varphi, & n=1,2,\\[0.4em]
    0, & n>2 .
    \end{cases}
\end{align}
\end{subequations}
Thus the ED values
\(\langle\tau^x\tau^x\rangle\simeq
\langle\tau^y\tau^y\rangle\simeq1/8\) and
\(\langle\tau^x\tau^y\rangle\simeq0\) show that the finite-cluster
ground state is not a single equatorial product state.  They are
consistent with a symmetric superposition of more than two equatorial
orientations.

The two-body correlators alone do not distinguish a threefold
superposition from a higher-fold one.  For this purpose we use
three-pseudospin correlators.  For three distinct sites,
\begin{align}
    \langle \tau_i^x\tau_j^x\tau_k^x\rangle
    &=
    \frac{1}{8n}
    \sum_{m=0}^{n-1}
    \cos^3
    \left(\varphi+\frac{2\pi m}{n}\right),
    \label{eq:appendix_tauxxx_sum}
    \\
    \langle \tau_i^x\tau_j^x\tau_k^y\rangle
    &=
    \frac{1}{8n}
    \sum_{m=0}^{n-1}
    \cos^2
    \left(\varphi+\frac{2\pi m}{n}\right)
    \sin
    \left(\varphi+\frac{2\pi m}{n}\right).
    \label{eq:appendix_tauxxy_sum}
\end{align}
Using
\begin{equation}
    \cos^3\phi=\frac{3\cos\phi+\cos3\phi}{4},
\end{equation}
and
\begin{equation}
    \cos^2\phi\,\sin\phi
    =
    \frac{\sin\phi+\sin3\phi}{4},
\end{equation}
one obtains
\begin{subequations}
\label{eq:appendix_threebody}
\begin{align}
    \langle \tau_i^x\tau_j^x\tau_k^x\rangle
    &=
    \begin{cases}
    \dfrac{1}{8}\cos^3\varphi, & n=1,\\[0.4em]
    \dfrac{1}{32}\cos 3\varphi, & n=3,\\[0.4em]
    0, & n\neq 1,3,
    \end{cases}
    \\
    \langle \tau_i^x\tau_j^x\tau_k^y\rangle
    &=
    \begin{cases}
    \dfrac{1}{8}\cos^2\varphi\,\sin\varphi, & n=1,\\[0.4em]
    \dfrac{1}{32}\sin 3\varphi, & n=3,\\[0.4em]
    0, & n\neq 1,3 .
    \end{cases}
\end{align}
\end{subequations}
The ED ground state satisfies
\begin{equation}
    \langle\tau^x\tau^x\tau^x\rangle\simeq \frac{1}{32},
    \qquad
    \langle\tau^x\tau^x\tau^y\rangle\simeq 0 .
\end{equation}
Combining this with the two-body correlators gives
\begin{equation}
    n=3,\qquad
    \varphi=0\quad {\rm mod}\quad \frac{2\pi}{3}.
\end{equation}
The finite-cluster ground state is therefore naturally identified as
the symmetric superposition
\begin{equation}
    |\Psi_{\rm tet}\rangle
    =
    \frac{
    |\Theta_0\rangle+
    |\Theta_{2\pi/3}\rangle+
    |\Theta_{4\pi/3}\rangle
    }{\sqrt{3}},
    \label{eq:appendix_tetramer_cat}
\end{equation}
where
\begin{equation}
    |\Theta_{\varphi}\rangle=\prod_j|\varphi\rangle_j .
\end{equation}
In the thermodynamic limit, one of these three components may be
selected spontaneously.  In the finite periodic cluster, the exact
eigenstate preserves the symmetry and appears as the equal-weight
superposition in Eq.~\eqref{eq:appendix_tetramer_cat}.

\subsection{First excited state in the pseudospin-ferromagnetic manifold}
\label{app:first_excited_pseudospin}

The ground-state correlators discussed above identify the finite-cluster
ground state as the symmetric superposition of the three uniform
tetramer orientations.  On a finite periodic cluster, however, one does
not expect an exact symmetry-broken state.  Instead, the states obtained
by forming different linear combinations of the three uniform tetramer
configurations should appear as a small, nearly degenerate multiplet,
with splittings that are finite-size effects.

It is therefore useful to examine the first excited state of the
effective pseudospin Hamiltonian.  Let
\begin{equation}
    |\Theta_0\rangle,\qquad
    |\Theta_{2\pi/3}\rangle,\qquad
    |\Theta_{4\pi/3}\rangle
\end{equation}
denote the three uniform pseudospin-ferromagnetic tetramer product
states.  The symmetric combination,
\begin{equation}
    |\Psi_{\rm tet}\rangle
    =
    \frac{
    |\Theta_0\rangle+
    |\Theta_{2\pi/3}\rangle+
    |\Theta_{4\pi/3}\rangle
    }{\sqrt{3}},
\end{equation}
describes the ED ground state to good accuracy, as shown in
Sec.~\ref{sec:effective_pseudospin}.  The two-dimensional subspace
orthogonal to this state is spanned, for example, by
\begin{subequations}
\label{eq:first_excited_basis}
\begin{align}
    |\Psi_{1,1}\rangle
    &=
    \frac{
    |\Theta_{2\pi/3}\rangle-
    |\Theta_{4\pi/3}\rangle
    }{\sqrt{2}},
    \\
    |\Psi_{1,2}\rangle
    &=
    \frac{
    -2|\Theta_0\rangle+
    |\Theta_{2\pi/3}\rangle+
    |\Theta_{4\pi/3}\rangle
    }{\sqrt{6}} .
\end{align}
\end{subequations}
More generally, a state in this subspace can be written as
\begin{align}
    |\Psi_1^{\alpha\beta}\rangle
    & =
    \sqrt{\alpha}\,|\Psi_{1,1}\rangle
    +
    \sqrt{1-\alpha}\,e^{i\beta}|\Psi_{1,2}\rangle , 
    \label{eq:first_excited_alpha}
    \\
    \nonumber\text{with } & \alpha\in [0,1],\,\beta\in[0,2\pi)
\end{align}
which is normalized as written, since
\(|\Theta_0\rangle,|\Theta_{2\pi/3}\rangle,|\Theta_{4\pi/3}\rangle\) are
orthogonal up to corrections of order \(2^{-N}\) on an \(N\)-site cluster
since $\langle\varphi|\varphi'\rangle=\frac{1}{2}(1+e^{i(\varphi'-\varphi)})$ has modulus 1/2 for $\varphi'-\varphi=\pm2\pi/3$.  This parametrization is not meant to introduce a new
variational theory; it is only a convenient way of comparing the ED first
excited state with linear combinations inside the same three-state
tetramer manifold.  We note that if the effective Hamiltonian retains the
exact threefold symmetry, the first excited level is itself a doublet,
\(\alpha\) and $\beta$ then merely labels the particular eigenvector returned by the
diagonalization rather than a physical parameter. The diagonalization indeed returns two degenerate states at $E_1 - E_0 \sim 10^{-4}$, split from one another by $\sim10^{-13}$, we label them $e_1$ and $e_2$.

\begin{figure}[bt]
    \centering
    \includegraphics[width=\columnwidth]{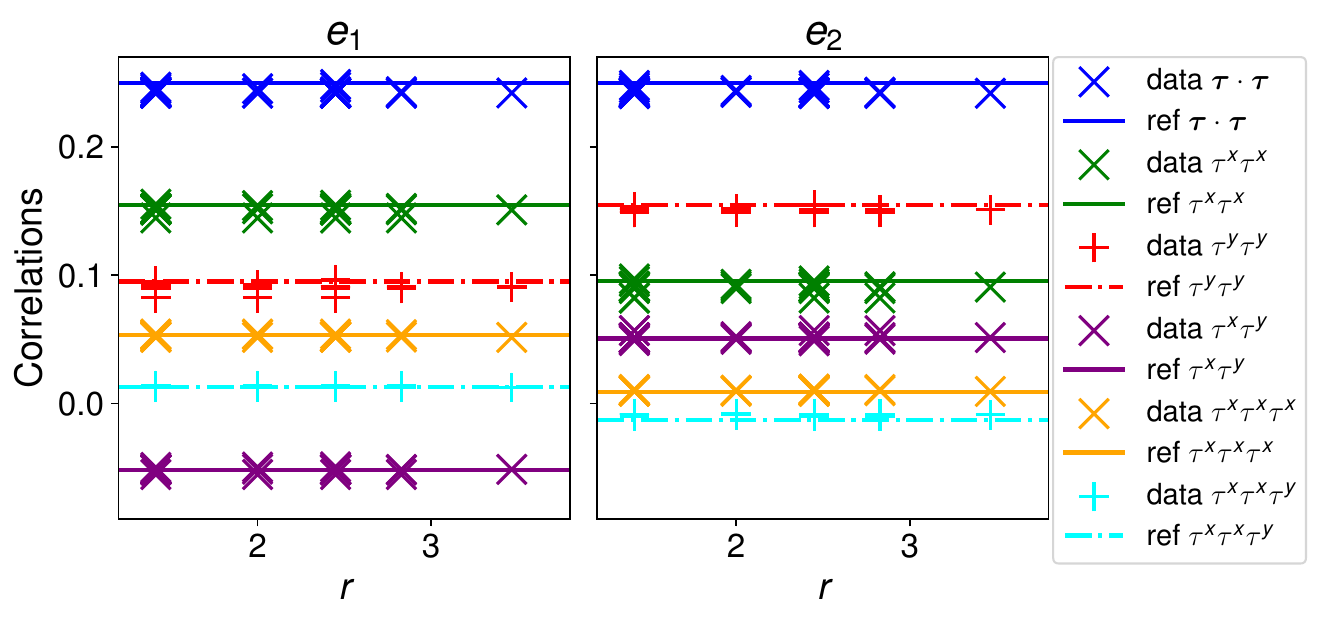}
    \caption{
    Two- and three-pseudospin correlations in the two members of the first excited doublet
    of the 32-site effective fcc model.  The horizontal lines show the
    values expected for the state \(|\Psi_1^{\alpha,\beta}\rangle\) in
    Eq.~\eqref{eq:first_excited_alpha}, with \(\alpha=0.26,\,\beta=0.11\pi\) for $e_1$ and \(\alpha=0.74,\,\beta=1.13\pi\) for $e_2$.  The
    correlations are consistent with a state in the two-dimensional
    subspace orthogonal to the symmetric ground-state combination of
    the three uniform tetramer orientations.  
    }
    \label{fig:all_correlations_plot_cross_e}
\end{figure}

For such a state, the two-pseudospin correlations remain essentially
ferromagnetic,
\begin{equation}
    \langle \bm{\tau}_i\cdot\bm{\tau}_j\rangle_1^{\alpha,\beta}
    =
    \frac{1}{4},
\end{equation}
while the individual components depend on the particular linear
combination.  For the parametrization in Eq.~\eqref{eq:first_excited_alpha},
one obtains
\begin{subequations}
\label{eq:first_excited_correlators}
\begin{align}
    \langle \tau^x_i\tau^x_j\rangle_1^{\alpha,\beta}
    &=
    \frac{3-2\alpha}{16},
    \\
    \langle \tau^y_i\tau^y_j\rangle_1^{\alpha,\beta}
    &=
    \frac{1+2\alpha}{16},
    \\
    \langle \tau^x_i\tau^y_j\rangle_1^{\alpha,\beta}
    &=
    -\frac{\sqrt{\alpha(1-\alpha)}\cos{\beta}}{8},
    \\
    \langle \tau^x_i\tau^x_j\tau^x_k\rangle_1^{\alpha,\beta}
    &=
    \frac{10-12\alpha}{128},
    \\
    \langle \tau^x_i\tau^x_j\tau^y_k\rangle_1^{\alpha,\beta}
    &=
    \frac{\sqrt{\alpha(1-\alpha)}\cos{\beta}}{32} .
\end{align}
\end{subequations}
Each correlator depends on \(\beta\) only through \(\cos\beta\), so a fit to them determines \(\beta\) only up to \(\beta\to-\beta\).
The two members of the doublet are, moreover, not independent.  
For two states of the form of Eq.~\eqref{eq:first_excited_alpha} in the same subspace,
\begin{equation}
    \langle\Psi_1^{\alpha\beta}|\Psi_1^{\alpha'\beta'}\rangle
    =
    \sqrt{\alpha\alpha'}
    +
    \sqrt{(1-\alpha)(1-\alpha')}\;e^{i(\beta'-\beta)} ,
    \label{eq:doublet_overlap}
\end{equation}
and since the first term is real and non-negative the two contributions cancel only if they have equal modulus and opposite phase.  
Equal modulus gives \(\alpha\alpha'=(1-\alpha)(1-\alpha')=1-\alpha-\alpha'+\alpha\alpha'\), hence \(\alpha'=1-\alpha\); opposite phase gives \(e^{i(\beta'-\beta)}=-1\), hence \(\beta'=\beta+\pi\). 
$e_1$ is well described by \(\alpha_1=0.26\), \(\beta_1=0.11\pi\) this should fix $e_2$ with \(\alpha_2=1-\alpha_1=0.74\) and \(\beta_2=\beta_1+\pi=1.11\pi\).  
Fitting the two members independently confirms this: \(\alpha_2=0.74\) against \(1-\alpha_1=0.74\), and \(\beta_2=1.13\pi\) against \(\beta_1+\pi=1.11\pi\), where the branch quoted for \(\beta_2\) is the one consistent with orthogonality, as illustrated in Fig.~\ref{fig:all_correlations_plot_cross_e}. 
Numerically, we find the values collected in Table~\ref{tab:first_excited_ed_values}.
\begin{table}[bt]
\caption{
Two- and three-pseudospin correlations of the two members
\(e_1\) and \(e_2\) of the first excited doublet of the
32-site effective fcc model, compared with
Eq.~\eqref{eq:first_excited_correlators} fitted independently to each
member, giving \(\alpha_1=0.26\), \(\beta_1=0.11\pi\) and
\(\alpha_2=0.74\), \(\beta_2=1.13\pi\).
}
\label{tab:first_excited_ed_values}
\begin{ruledtabular}
\begin{tabular}{ccccc}
 & \multicolumn{2}{c}{\(e_1\)} & \multicolumn{2}{c}{\(e_2\)}\\
 & ED & fit & ED & fit\\
\hline
\(\langle\bm{\tau}_i\cdot\bm{\tau}_j\rangle\) & \(0.2429\) & \(0.2500\) & \(0.2429\) & \(0.2500\)\\
\(\langle\tau^x_i\tau^x_j\rangle\) & \(0.1509\) & \(0.1549\) & \(0.0903\) & \(0.0952\)\\
\(\langle\tau^y_i\tau^y_j\rangle\) & \(0.0903\) & \(0.0951\) & \(0.1509\) & \(0.1548\)\\
\(\langle\tau^x_i\tau^y_j\rangle\) & \(-0.0517\) & \(-0.0518\) & \(0.0517\) & \(0.0507\)\\
\(\langle\tau^x_i\tau^x_j\tau^x_k\rangle\) & \(0.0525\) & \(0.0536\) & \(0.0102\) & \(0.0089\)\\
\(\langle\tau^x_i\tau^x_j\tau^y_k\rangle\) & \(0.0132\) & \(0.0129\) & \(-0.0088\) & \(-0.0127\)
\end{tabular}
\end{ruledtabular}
\end{table}

The residual differences are small and comparable to the discrepancy between $\langle \bm{\tau}_i\cdot\bm{\tau}_j\rangle_1^{\rm ED}$ and its saturated value which reflects the dressing of the ideal tetramer product states by quantum fluctuations in a finite size cluster. The agreement is sufficient to identify the first excited state as another combination of the same three uniform tetramer configurations that built the ground state.
This analysis supports the interpretation of the low-energy spectrum as
a finite-size remnant of a threefold tetramer symmetry-breaking
manifold.  The first excited state is therefore not a separate competing
phase, but another linear combination of
the same three uniform tetramer configurations which become degenerate
with the ground state in the thermodynamic symmetry-broken limit. 

\begin{figure}[t]
    \centering
    \includegraphics[width=\columnwidth]{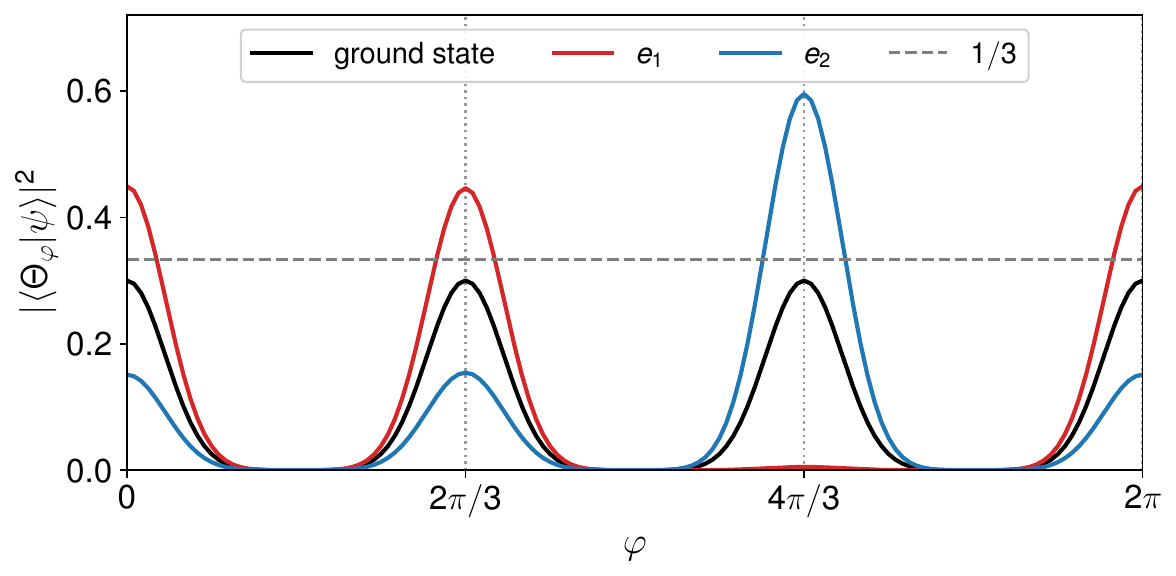}
    \caption{
    Overlap \(|\langle\Theta_\varphi|\psi\rangle|^2\) of the ground state and of the two members of the first excited doublet with the uniform product state \(|\Theta_\varphi\rangle\), as \(\varphi\) runs over the equator.  
    All three peak at the tetramer angles \(\varphi=0,2\pi/3,4\pi/3\); the ground state weights them equally, the doublet members do not.}
    \label{fig:overlap_theta}
\end{figure}

This can be further captured by looking at the overlap of the ground state and first excited state with the product state $|\Theta_\varphi\rangle$ which shows a peak in the overlap for $\varphi=0,2\pi/3$ and $4\pi/3$, the tetramer product states.
The ground state has equal weight \(|\langle\Theta_\varphi|\Psi_{\rm ED}\rangle|^2=0.300\) on all three, while the two members of the doublet weight them unequally, as shown in Fig.~\ref{fig:overlap_theta}.

%------------------------------------------------------------------------

\section{Details of the dynamic high-temperature expansion}
\label{app:dynhte}

All Dyn-HTE data shown here are based on the $10$th-order expansion of the
Matsubara correlator, which provides access to expansions of the first five
even-frequency moments of the dynamic structure factor.  For the $2n$-th moment,
the approach yields an expansion to order $10-2n$.  As described in detail in
Ref.~\cite{Burkard-2026-PRL}, the dynamic structure factor is then reconstructed
using a continued-fraction expansion with a termination condition that linearly
extrapolates the continued-fraction parameters.

At infinite temperature, all available moments are exact and are therefore
used to reconstruct the dynamic structure factor.  At lower temperatures,
however, only the $u$-Pad\'e approximants of Ref.~\cite{Burkard-2026-PRL}, with
parameter $f=0.35$, for the first three available moments are used.  This is
because, as the temperature decreases, the quality of the approximation of the
higher moments deteriorates more quickly than that of the lower moments.

%%%%%%%%%%%%%%%%%%%%%%%%%%%%%%%%%%%%
\bibliography{biblio.bib}
\end{document}